\documentclass[%
 reprint,
superscriptaddress,
nofootinbib,
 amsmath,amssymb,
 aps,
a4paper,twocolumn, floatfix
]{revtex4-2}
\usepackage{dsfont}
\usepackage{graphicx}
\usepackage{dcolumn}
\usepackage{bm}
\usepackage[hidelinks]{hyperref}
\usepackage{caption}
\usepackage{subcaption}
\hypersetup{
    colorlinks,
    linkcolor={red!50!black},
    citecolor={blue!50!black},
    urlcolor={blue!80!black}
}
\usepackage{comment}

\newcommand{\ket}[1]{|#1\rangle}

\newcommand{\shiftleft}[2]{\makebox[0pt][r]{\makebox[#1][l]{#2}}}
\newcommand{\fs}[1]{\textcolor{red}{\textit{#1}}}

\usepackage{tikz}
\usetikzlibrary{matrix,backgrounds,quantikz}
\pgfdeclarelayer{myback}
\pgfsetlayers{myback,background,main}

\tikzset{mycolor/.style = {line width=1bp,color=#1}}%
\tikzset{myfillcolor/.style = {draw,fill=#1}}%

\NewDocumentCommand{\highlight}{O{blue!40} m m}{%
\draw[mycolor=#1] (#2.north west)rectangle (#3.south east);
}
\usepackage{booktabs}
\usepackage{multirow}

\newcommand{\Tr}{\text{Tr}}
\usepackage{mathtools}

\newcommand\myeq{\stackrel{\mathclap{\normalfont\mbox{\tiny N=5}}}{=}}

\newtheorem{definition}{Definition}

\begin{document}

\preprint{APS/123-QED}

\title{A General Approach to Dropout in Quantum Neural Networks}
\author{Francesco Scala}
 \email{francesco.scala01@ateneopv.it}
 \affiliation{%
    Dipartimento di Fisica, Universit\`a di Pavia (Italy)
}%
\author{Andrea Ceschini}%
 \email{andrea.ceschini@uniroma1.it}
 \affiliation{%
   Dipartimento di Ingegneria dell'Informazione, Elettronica e Telecomunicazioni\\
Università degli Studi di Roma "La Sapienza"  (Italy)\\
}
\author{Massimo Panella}%
\email{massimo.panella@uniroma1.it}
 \affiliation{%
   Dipartimento di Ingegneria dell'Informazione, Elettronica e Telecomunicazioni\\
Università degli Studi di Roma "La Sapienza"  (Italy)\\
}
\author{Dario Gerace}
 \email{dario.gerace@unipv.it}
 \affiliation{%
    Dipartimento di Fisica, Universit\`a di Pavia (Italy)
}%
%


\date{\today}

\begin{abstract}
In classical Machine Learning, ``overfitting'' is the phenomenon occurring when a given model learns the training data excessively well, and it thus performs poorly on unseen data. A commonly employed technique in Machine Learning is the so called ``dropout'', which prevents computational units from becoming too specialized, hence reducing the risk of overfitting. With the advent of Quantum Neural Networks as learning models, overfitting might soon become an issue, owing to the increasing depth of quantum circuits as well as multiple embedding of classical features, which are employed to give the computational nonlinearity. Here  we present a generalized approach to apply the dropout technique in Quantum Neural Network models, defining and analysing different quantum dropout strategies to avoid overfitting and achieve a high level of generalization. Our study allows to envision the power of quantum dropout in enabling generalization, providing useful guidelines on determining the maximal dropout probability for a given model, based on overparametrization theory. It also highlights how quantum dropout does not impact the features of the Quantum Neural Networks model, such as expressibility and entanglement. All these conclusions are supported by extensive numerical simulations, and may pave the way to efficiently employing deep Quantum Machine Learning models based on state-of-the-art Quantum Neural Networks.
\end{abstract}

\maketitle

\section{Introduction\label{sec:intro}}
Machine Learning (ML) has been rapidly emerging in the last few years as a most promising computational model to analyze and extract insights from large and complex datasets, improve decision-making processes, and automate different tasks across a wide range of industrial processes \cite{tysonMachineLearning2020,dhall2020machine}. ML models generally require high flexibility, with lots of trainable parameters in order to learn the underlying function in a supervised fashion. However, being able to learn with low in-sample error is not enough: it is also desirable to have a model that is capable of high \emph{generalization}, meaning that it is able to provide good predictions on previously unseen data \cite{amanpreetReviewSupervised2016}.

Overfitting is a common issue in ML when dealing with highly expressive models \cite{hawkins2004problem,ying2019overview}. It is a phenomenon that occurs when a model is trained too well on the training data, and as a result, performs poorly on new, unseen data; if the model's performance on the testing set is significantly worse than on the training set, it probably indicates overfitting. This happens because the model has learnt the noise in the training data, rather than the underlying pattern that is generalizable to new data. When a learning model has a high number of parameters relative to the amount of training data, then it is likely to produce overfiting: the model becomes too complex for the amount of data it has been trained on, resulting in a lack of generalization.

Deep Neural Networks (DNNs) are powerful neural network models that are able to employ a large number of parameters, which allows to well approximate complex functions and achieve high accuracy in training. However, the high complexity of these models can also lead to overfitting the training data. To mitigate overfitting, regularization methods such as the ``dropout'' are widely employed in the Deep Learning (DL) community. Dropout is a technique that acts by randomly dropping either neurons or connections within a DNN during the training phase, to block the information flow and prevent units from becoming too specialized, thus reducing the risk of overfitting~\cite{srivastava_dropout_2014}.

Recently, an emerging fiels combining ML with the principles of quantum computing has been emerging, Quantum machine learning (QML). The goal of QML is to leverage the unique properties of quantum systems, such as superposition and entanglement, to improve the performance of ML and DL algorithms. The potential advantages of QML include faster training times~\cite{huang2022quantum}, improved generalization~\cite{caro_generalization_2022}, and the ability to solve problems that are intractable for classical algorithms~\cite{biamonte2017quantum,sajjan2022quantum}. Even though prototypes of real quantum computers are available nowadays, these are still of limited size and strongly affected by noise, posing an additional challenge to researchers in the field and giving rise to the so-called Noisy Intermediate Scale Quantum (NISQ) era~\cite{Linke2017, Preskill_2018,Bharti_2022}. Variational Quantum Algorithms (VQA)~\cite{Cerezo_2021VQA} are a particular QML approach that demonstrated to be very efficient in dealing with such noisy hardware by exploiting the combination of Parametrized Quantum Circuits (PQC), often referred also as Quantum Neural Networks (QNNs)~\cite{Mangini_2021, schuld2021effect, Abbas_2021,Nguyen2022equivariant}, and an optimization process run on a classical computer~\cite{Peruzzo_2014, Benedetti2019, Tacchino2021variational, Sim2019, Hubregtsen2021, ceschini2022hybrid, scala2022quantum, leone2022practical, Ballarin_2023}.

The search for high generalization capability also applies to QNN models~\cite{banchi2021generalization, caro_encoding-dependent_2021, caro_generalization_2022}, in which a quantum model can be trained to analyse classical or quantum data. Since quantum mechanics is intrinsically linear, in order to nonlinearly analyse classical data and enhance the expressive power of QML models, the data re-uploading technique was introduced~\cite{perez2020data}. Further efforts on the relationship between classical data encoding and generalization performances suggest that overfitting may be caused by the repeated encoding of classical data~\cite{banchi2021generalization,caro_encoding-dependent_2021}. 
Although overfitting may also occur in deep QML models, little has been said about how to practically deal with it, so far.
If on the one hand training deep overparametrized quantum models~\cite{Haug_2021, Larocca_2023, peters_generalization_2022, garcia2023effects} greatly helps to avoid barren plateaus~\cite{arrasmith2020effect, Patti_2021, marrero2021entanglement, holmes2022connecting,Arrasmith_2022equivalence, thanasilp_exponential_2022}, on the other hand, the presence of multiple data re-uploading and redundancy in the parameters may lead to overfitting the data. Some bounds for the scaling of generalization error are given in Ref.~\cite{caro_generalization_2022}, but their application to deep QML models has not been explored, yet. Moreover, a recent work~\cite{gilfuster2023understanding}  reports on the limitations of uniform generalization bounds applied to QML models.

The seminal idea of implementing dropout to address the overfitting problem in QNNs is given in Ref.~\cite{verdon2018universal}, where the presence or absence of a unitary operation is controlled by an ancillary register. More recently, it has been proposed to randomly select and measure one of the qubits and set it aside for a certain number of optimization steps during the training of PQCs~\cite{schuld_circuit-centric_2020}, which is a sort of dropout technique. Nevertheless, this type of dropout regularization does not seem to solve the overfitting issue.
Recently, entangling dropout was proposed to address the problem of overfitting in deep QNNs~\cite{kobayashi_overfitting_2022}: some entangling gates (specifically, controlled-NOT ones) are randomly removed during the training phase in order to decrease the expressibility of the QML model. In order to clarify possible misunderstandings on the relationship between the present work with others already present in the literature, let us point out that the terminology `quantum dropout' has been employed also in the context of Quantum Approximate Optimization Algorithms~\cite{wang_2023_qdrop}, which consists in a completely different technique. In addition,  \emph{classical} dropout has been used to avoid overfitting in hybrid Quantum Convolutional NNs~\cite{Chen_2022_QCNN}, which is not what we mean as quantum dropout.

In this paper, we present the first 
extensive theoretical assessment of the dropout technique applied to QNNs (i.e., \emph{quantum dropout}) inspired by the comparison between classical and quantum models, in which the role of artificial neurons is implemented by rotation gates, whereas entangling gates work as inter-neural connections. Taking into account this analogy, rotations are randomly removed, in addition to some entangling gates depending on the chosen dropout method. Additionally, starting from the overparametrization theory, we provide guidelines about how to determine the maximal dropout probability for a given model.
Our results indicate that all the proposed quantum dropout strategies are effective in mitigating overfitting, especially when parametrized quantum gates are dropped. 
Moreover, we examine the use of parameter rescaling for quantum dropout strategies, a potentially useful strategy that has not been addressed in the previous literature. Interestingly, unlike classical dropout, quantum dropout does not benefit from parameter rescaling. As a consequence, we analyse the learning performances without it.
Finally, we analyse the behaviour of genuine quantum features related to the QNN when quantum dropout is applied. The main conclusion is that quantum dropout does not reduce expressibility~\cite{Sim2019} or entanglement~\cite{beckey2021concentrable} when the QNN is overparametrized. These findings have significant implications for the effective implementation of deep QNN models and provide a promising foundation for the development of more efficient strategies for QNNs training.

The paper is structured as follows. Section~\ref{sec:preliminaries} provides a comprehensive overview of QNNs, overparametrization and dropout. Section~\ref{sec:quantum dropout} presents a general introduction to dropout for QNNs and examines the proposed strategies. Section~\ref{sec:results} delivers and discusses the main results of numerical experiments on quantum dropout, together with an analysis of entanglement, expressibility and parameters scaling with QNNs. Conclusions are drawn in Sec.~\ref{sec:conclusions}. Lastly, the employed Methods are illustrated in Section~\ref{sec:methods}.

\begin{figure*}
    \centering
    \includegraphics[trim={1cm 2cm 1cm 2cm},clip,width=\textwidth]{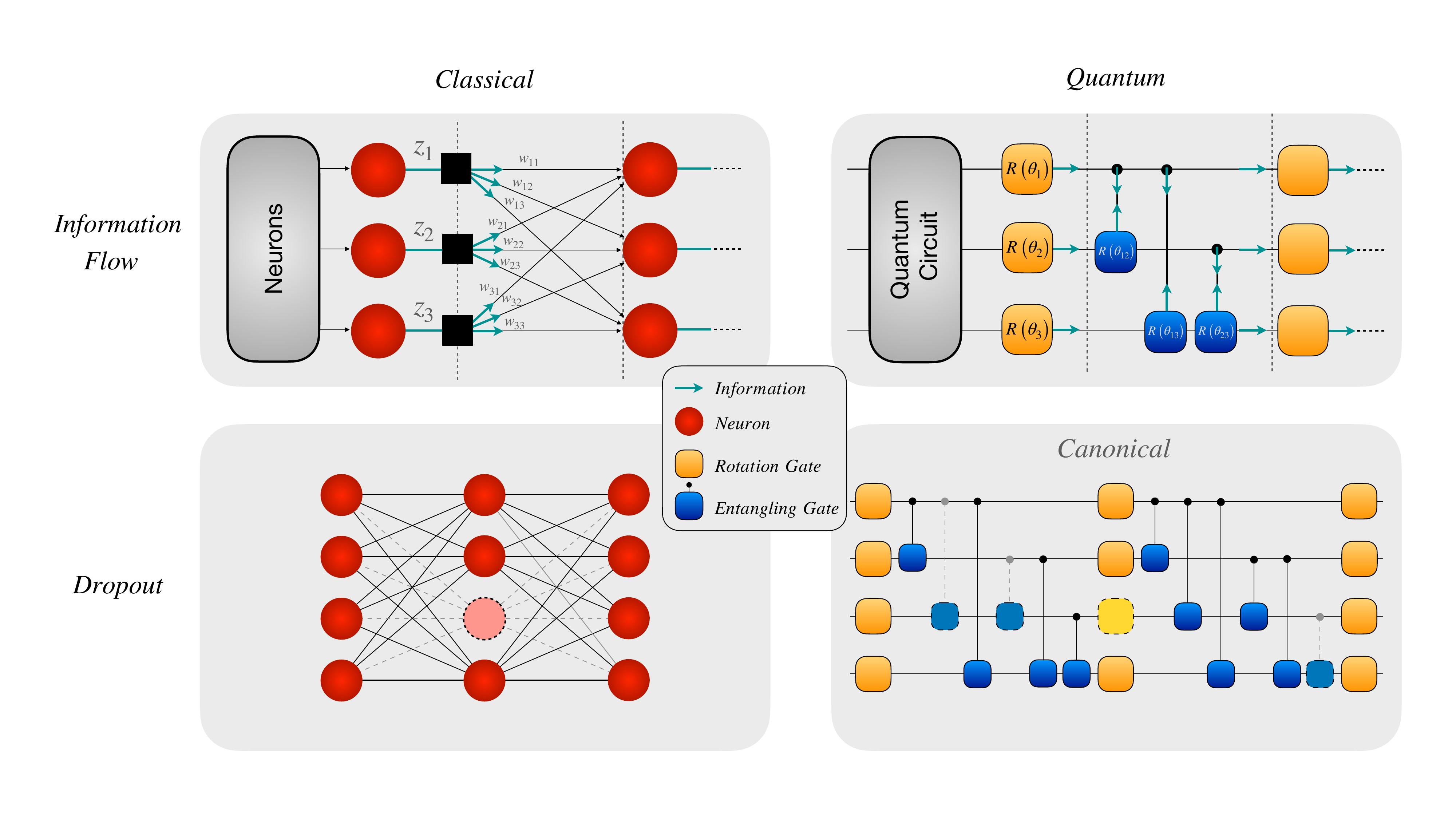}
    \caption{Illustration of the similarities between classical and quantum NNs. Quantum rotational gates take the role of artificial neurons, and entangling gates work as connections between them. Neurons process the incoming information (from a previous layer of neurons/quantum circuit) and then send it to all the neurons of the next layers throughout their connections; for the QNN, this is just a schematic representation; in fact, if the state is entangled, a rotation on one qubit will affect also others. In the classical NN, when one neuron is dropped, it is removed together with all its connections, whereas the canonical quantum dropout leaves some entangling gates unaltered, to avoid a strong impact on the quantum learning process.}
    \label{fig:cl_vs_quant}
\end{figure*}

\section{Technical Background}
\label{sec:preliminaries}
\subsection{Quantum Neural Networks }
\label{subsec:QNNs}
Due to their similarities with classical NNs, PQCs employed within VQAs are often referred to as QNNs~\cite{Mangini_2021}. They are composed of a data encoding stage, a parametrized ansatz, and a measurement operation at the end to retrieve the result followed by a classical update of the parameters. Following this scheme, QNNs can learn to perform various tasks, such as regression and classification.

Classical patterns ${\mathbf{x} \in \mathbb{R}^n}$ are encoded into a $N$ qubit quantum circuit via a quantum feature map, which corresponds to a unitary matrix $S(\mathbf{x})$ that maps $\mathbf{x}$ to a $2^N$-dimensional Hilbert space. The embedding is then followed by the variational ansatz $W(\boldsymbol{\theta})$. The latter is composed of layers of parametrized single-qubit rotation gates and two-qubits entangling gates: rotation gates are employed to manipulate the quantum state by addressing single qubits independently, while layers of entangling gates allow for the creation of multipartite entanglement~\cite{Guhne_2009ent} throughout the whole state vector (see, e.g., Fig.~\ref{fig:QNN circuit} as an example). This can be summarized as
\begin{equation}		
\label{eq:QNN}
	W(\boldsymbol{\theta})S(\mathbf{x})\ket{\mathbf{0}} = W(\boldsymbol{\theta})\ket{\phi(\mathbf{x})} \,,
\end{equation}
where $\ket{\mathbf{0}}=\ket{0}^{\otimes N}$.

Data encoding and ansatz unitaries may be applied repeatedly in a data re-uploading fashion to realise more expressive models~\cite{perez2020data}. In this regard, given $L$ data re-uploading layers, the final unitary describing the evolution of the quantum state is formally expressed as
\begin{equation}	
\label{eq:data re-uploading}
	U_L(\mathbf{x};\boldsymbol{\theta}) = \prod_{l=L}^1{U_l(\mathbf{x};\boldsymbol{\theta})} = \prod_{l=L}^1{W(\boldsymbol{\theta}^{(l)})S(\mathbf{x})}\,,
\end{equation}
where $\boldsymbol{\theta}^{(l)}$ refers to the parameters in the $l$-th layer of the quantum circuit.

At the end of the quantum circuit, a measurement with respect to an observable is performed to retrieve the outcome of the algorithm: 
\begin{equation}	
\label{eq:prediction}
 f_L(\mathbf{x}; \boldsymbol{\theta}) = \bra{\mathbf{0}}U_L^{\dagger}(\mathbf{x};\boldsymbol{\theta}) \hat{O} U_L(\mathbf{x};\boldsymbol{\theta})\ket{\mathbf{0}} \,,  
 \end{equation}
where $\hat{O}$ is the measured operator. In this work, we measure in the Pauli-Z basis one or more qubits depending on the task. When employed in VQAs, from $f_L(\mathbf{x}; \boldsymbol{\theta})$ a classical optimizer estimates the cost function to update the parameters.

The data re-uploading strategy affects the ability of QNNs to approximate functions. In fact, as shown in Ref.~\cite{schuld2021effect}, a quantum model $f_L(\mathbf{x}; \boldsymbol{\theta})$ can be expressed as a partial Fourier series in the data, and by repeating the data encoding stage quantum models can access a wider range of frequency spectra growing as $L$. On the other hand, if QNN  depth increases too much, this procedure will be prone to overfitting of the training data due to their enhanced expressiveness~\cite{peters_generalization_2022}. The issue of overfitting is discussed in Sec.~\ref{subsec:overparam}, where we also discuss its link to  overparametrization~\cite{Larocca_2023}.

\subsection{Overparametrization and overfitting}
\label{subsec:overparam}

In classical ML, a model is said to be overparametrized when the number of trainable parameters, $M$, exceeds the number of training samples. This is often the case for deep learning models, where the number of trainable parameters is usually way larger than the size of the training dataset. Differently, in QML such a term has taken slightly different meanings depending on the context in which it has been  employed~\cite{Haug_2021,Larocca_2023, peters_generalization_2022, garcia2023effects}. In this work, we apply the definitions provided in Refs.~\cite{Haug_2021,Larocca_2023}: a QNN is said to be overparametrized when the \emph{average parameter dimension} (over the whole training set) saturates its maximum achievable value, $D_{max}=2^{N+1}-2$. This implies that the QNN is fully capable of representing every possible quantum state on $N$ qubits. As a consequence, adding more trainable parameters will only increase the \emph{redundancy}. In App.~\ref{appendix: overparam} we provide the definitions of the mentioned quantities and we verify that we are working in the overparametrized regime, indeed. 

In order to strengthen the presence of overfitting, we also employ QNNs with multiple data encoding, both in parallel on different qubits and with the data re-uploading technique~\cite{banchi2021generalization, caro_encoding-dependent_2021, caro_generalization_2022}, respectively.

\subsection{Dropout}
\label{subsec:dropout}

Dropout, which was first proposed in Ref.~\cite{srivastava_dropout_2014}, is a widespread technique to prevent overfitting in classical DL models. It consists in temporarily removing neurons from the network, along with all their incoming and outgoing connections, during the training phase. Each unit is dropped randomly and independently, usually according to a fixed probability $p$. Dropout is repeated at each iteration and corresponds to sampling a thinned network from the original one. Thus, training a $n$-units neural network with dropout can be seen as training a collection of $2^n$ thinned networks with extensive weight sharing, where each thinned network gets trained very rarely, namely with probability equal to $1/2^n$.

At test time, one employs the original neural network with scaled weights with respect to the ones obtained from the training. If a unit is dropped with probability $p$ during training, the final weight of that unit is multiplied by $1-p$ at test time. This amounts to averaging the behaviour of $2^n$ thinned neural networks when using the original one.

In addition to the characterization of dropout as an ensemble technique, one can also try to catch its functioning by analysing what happens during training. When applying dropout the information flow is hindered, hence each hidden neuron is forced to learn to work with a randomly chosen group of other units. This should make each unit more robust and help it develop useful features on its own without relying on the surrounding units to correct its mistakes (co-adapting) and, on the other hand, it should also avoid the hyperspecialization of the units. However, the hidden units within a layer will still learn to do different things from each other.

\begin{table*}[tbp]
    \centering
    \begin{tabular}{lccp{75mm}}
        \toprule
        Name & Gates to drop & $p_G$ & Rule \\\midrule
        Canonical & $R_G$, $E_G$ & $p_R$  & Drop a single $R_G$, all previous $E_G$s with target on $R_G$, all next $E_G$s with control on $R_G$.\\
        Canonical-forward & $R_G$, $E_G$ & $p_R$ & Drop a single $R_G$, all next $E_G$s with control on $R_G$.  \\
        Rotation & $R_G$ & $p_R$ &  Drop a single $R_G$. \\
        Entangling & $E_G$ & $p_E$ &  Drop a single $E_G$. \\
        Independent & $R_G$, $E_G$ & $p_R$, $p_E$ &  Drop a single $R_G$ and a single $E_G$. \\
        \bottomrule
    \end{tabular}
    \caption{Quantum dropout strategies. Rotation and entangling gates are represented as $R_G$ and $E_G$, respectively. $p_G$ is employed together with $p_L$ to obtain the dropping probability $p=p_Gp_L$.}
    \label{tab:drop_strategies}
\end{table*}

\section{Quantum dropout}
\label{sec:quantum dropout}

\subsection{General approach}
Since overfitting usually appears when a model is overparametrized, we present the quantum dropout technique in the context of overparametrized QNNs. 
By comparing classical and quantum NNs, one can associate quantum single-qubit rotation gates to neurons and entangling gates to connections between neurons, see Fig~\ref{fig:cl_vs_quant}. In this view, quantum dropout consists in randomly removing parametrized unitaries in an overparametrized QNN during the training phase. More precisely, herein dropping a gate means that we substitute it with an identity gate. 
We argue that it is crucial to drop parametrized gates (or groups of gates including some parametrized ones) in order to fully benefit from dropout, since it prevents the QNN from becoming too reliant on specific parameters, thereby reducing the risk of overfitting. By randomly removing these parametrized unitaries, the QNN is forced to rely on a wider range of parameters and learn more robust features.

The general scheme of quantum dropout at each training step, inspired by~\cite{kobayashi_overfitting_2022}, can be roughly summarized as follows:

\begin{enumerate}
    \item Randomly select QNN layers according to the layer dropout ratio ($p_L$), i.e., the probability of choosing a layer to which the dropout is applied.
    \item Remove (groups of) gates in the selected layers based on a probability defined as the gate dropout ratio ($p_G$). This probability determines the likelihood of a quantum gate of being dropped.
    \item Compute the gradient of the cost function for the dropout circuit and update the parameters accordingly.
    \item Iterate the above procedure until the termination criterion is met.
\end{enumerate}

Usually, in classical ML  one applies dropout only to certain selected layers with probability $p$. The selection of the layers is often related to the particular design of the NN under usage, since it is very common to build part of the model with specific purposes. Conversely, in a quantum setting we have a QNN composed of identical repeated layers and this is why we randomly sample the layers where to apply quantum dropout.

Given the layer dropout rate $p_L$ and the gate dropout rate $p_G$, the probability $p$ that a (group of) gate(s) is dropped in a layer can be calculated with the conditioned probability law:
\begin{equation}
    \label{eq:probability}
p=p(A\cap B)=p(A|B)p(B)=p_Gp_L
\end{equation}
where $A$ is the selection of a specific (group of) gate(s) and $B$ is the selection of a specific layer. 

Before applying quantum dropout, one must always consider how many parameters are needed to preserve the full capabilities of the QNN under usage, in order to choose appropriate values of $p$. If the probability of dropout is set too high, the QNN will leave the overparametrization regime, and this will have direct consequences on its expressibility and entanglement produced.
In this work, as a rule of thumb for checking the maximal allowed number of dropped parameters we employ the following inequality:
\begin{equation}
    \label{eq:max_drop_params}
    M^{drop}_{max} \leq M-D_{max}
\end{equation}
where $M$ is the total number of parameters, $D_{max}$ is the maximal parameter dimension (setting the minimum number of parameters required to fully explore the Hilbert space as explained in Sec.~\ref{subsec:overparam}). One can then obtain $p^{drop}_{max} = M^{drop}_{max}/M$. The reader can find more details in Appendix~\ref{appendix: overparam}; we show the effects of dropout on genuine quantum feature in Sec.~\ref{subsec:expr and ent} and in Appendices~\ref{appendix:expressib}~and~\ref{appendix: multip ent}.

\subsection{Proposed strategies}
\label{subsec:strategies}
The quantum dropout scheme described above can be applied in different fashions depending on the dropped unitary operations. We hereby present at first the quantum dropout approach that embodies the more intuitive and practical essence of dropout, then we define alternative quantum dropout techniques. All these are summarized in Tab.~\ref{tab:drop_strategies}.

Following the classical concept of dropout, a certain neuron (single-qubit rotation) is selected according to the drop probability $p$ and it is removed together with all its connections (entangling gates). Hence, we define the \emph{canonical quantum dropout} where all the previous entangling gates, having the single-rotated qubit as a target, and all the next ones, having it as a control, are dropped. To exactly reproduce the classical dropout in a quantum circuit, one should remove all the entangling gates connecting the rotated qubit with the other qubits. However, this is not practical in a quantum setting, since it would deeply change the entire quantum state. For this reason, we preferred to define the canonical dropout rather than an exact quantum counterpart of classical dropout. Besides, quantum dropout with PQC as QNNs will always be intrinsically different from its classical counterpart, because we are acting multiple times on the same qubits, evolving the quantum state. This implies a temporal connection that is never removed by the dropout, even if we were to drop all the entangling gates liked to a single rotation.

Exploiting this fact, one can think of a purely quantum \emph{partial dropout}, which has no classical analogue. More in detail, one may want to remove only the entangling gates previous (\emph{canonical-backward}) or subsequent (\emph{canonical-forward}) to the selected rotation gate.\footnote{Since these two are conceptually equivalent, in this study we will only work with canonical-forward.}

It is also possible to only drop single-qubit rotations or entangling gates: we call these approaches \emph{rotation dropout} and \emph{entangling dropout}, respectively. The latter approach was previously defined in~\cite{kobayashi_overfitting_2022} and is included in our quantum dropout framework as a special case. In fact, it can be considered as an improper dropout since it fits more the definition of a randomized quantum version of the pruning technique~\cite{pruning} (applied only during training), which consists in removing connections in a NN according to their level of importance.
In addition, the employment of CNOTs as entangling gates would result in having equally-weighted connections, whereas controlled rotation would allow for different weighting of the connections. For this reason, in this work, we employ two different QNN models: the first one, proposed in~\cite{kobayashi_overfitting_2022}, has CNOTs as entangling gates and is utilised for regression, while for classification we choose a QNN with parametrized entangling gates. More details about the QNN structures can be found in Appendix~\ref{appendix:QNN circs}.

Ultimately, one can remove both the rotation and the entangling gates independently after having selected the layer (\emph{independent dropout}). In this last case, we will have two separate drop gate ratios, i.e. one for the rotation ($p_R$) and one for the entangling ($p_E$) gates; the dropout probability applied to each type of gate will be $p=p_Lp_G$, where $p_G=p_R$ for the rotation gates and $p_G=p_E$ for the entangling gates.


\section{Results}
\label{sec:results}
In this Section, we apply quantum dropout in all its different flavours to avoid overfitting in both regression and classification tasks. We performed three different experiments to evaluate the goodness of the proposed quantum dropout methods in a noiseless simulation setting. The first two experiments concerned regression tasks, while the third was about a binary classification problem. Interestingly, unlike classical dropout, quantum dropout does not benefit from parameter rescaling and, consequently, we analyse the learning performances without it. In fact, in Sec.~\ref{subsec:params scaling} we show evidence that scaling the parameters at best gives the same performance of not rescaling at all.

More in detail, we address the regression on a dataset generated by a sinusoidal function, which is a common and straightforward benchmark for neural models and then we test the performances of the dropout on another synthetic dataset generated by the modulus function (reported in Appendix~\ref{appendix:experiment2}). 
Since our general approach includes as a subcase the entangling dropout~\cite{kobayashi_overfitting_2022}, a performance evaluation of general quantum dropout on the same datasets was appropriate. Finally, in the third experiment, we tackle a more challenging problem related to the binary classification of the Moon dataset retrieved on the scikit-learn library, which has a non-trivial geometric arrangement of data points. We want to investigate the behaviour of different dropout strategies in a non-linear classification setting to demonstrate the robustness of our proposed methodologies experimentally. Afterwards, we discuss the relationship between quantum dropout and features of the QNN circuit like expressibility and entanglement~\cite{Sim2019, Ballarin_2023}. 


\begin{figure*}
     \centering
     \begin{subfigure}[b]{0.49\textwidth}
         \centering
         \includegraphics[trim={0.3cm 0cm 1.2cm 1cm},clip,width=\textwidth]{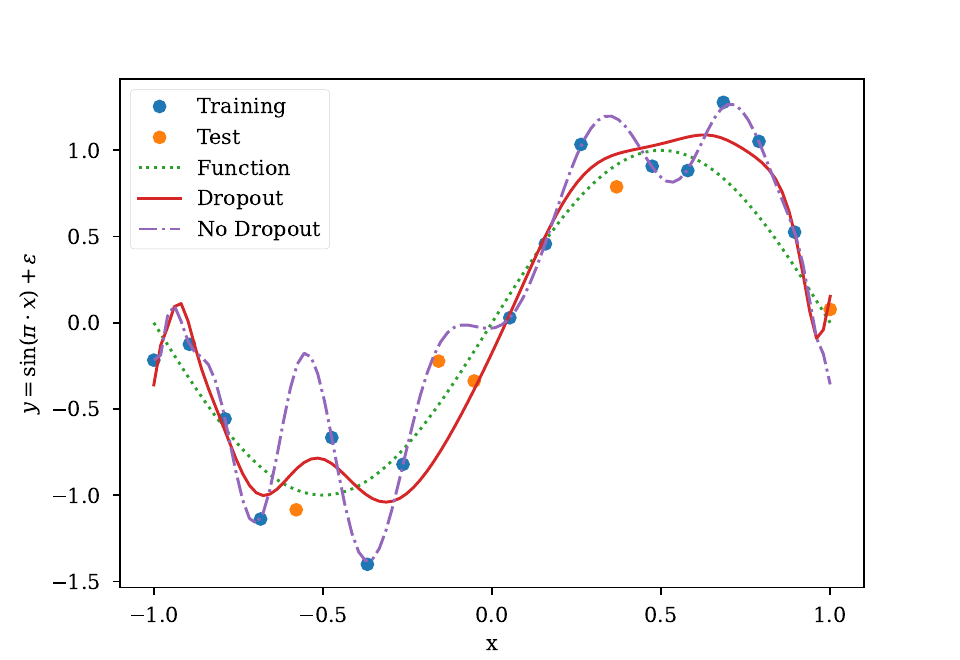}
         \caption{}
         \label{fig:y equals x}
     \end{subfigure}
     \begin{subfigure}[b]{0.49\textwidth}
         \centering
         \includegraphics[trim={0.3cm 0.cm 1.2cm 1cm},clip,width=\textwidth]{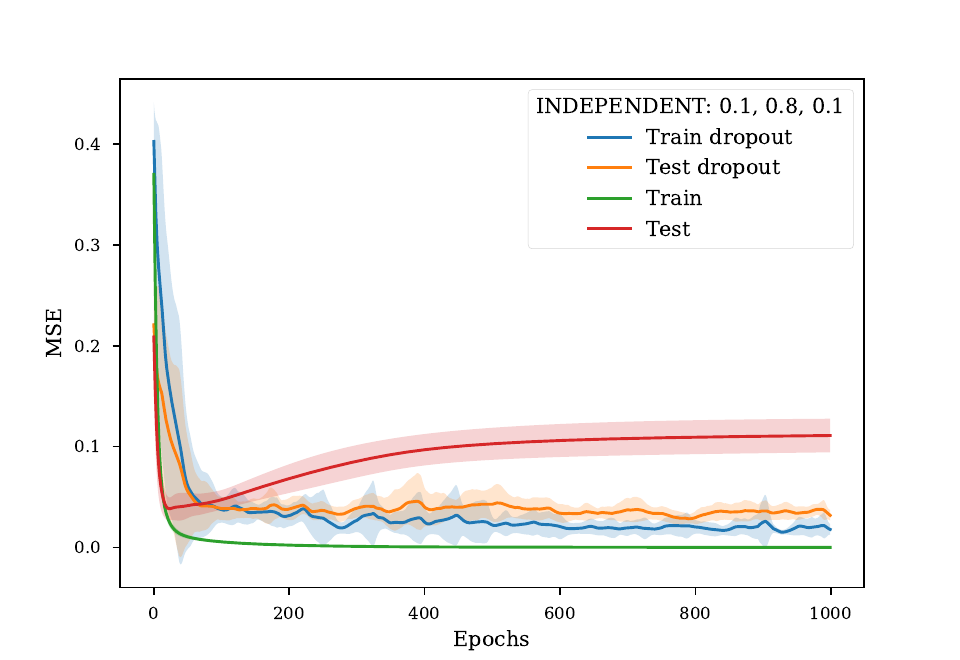}
         \caption{}
         \label{fig:five over x}
     \end{subfigure}
        \caption{\textbf{Regression} The plots illustrate the performances of an overparametrized model employed in a regression task of the \texttt{sin} function. \textbf{(a)} The model without dropout overfits the noisy data by predicting exactly each of them, whereas with dropout the approximate function is smoother. \textbf{(b)} The trend of the loss function when training without dropout shows an increasing prediction error typical of an overfitting model, while with dropout the prediction error does not increase. The standard deviation over 10 different runs is shown as a shadow.}
        \label{fig:no_drop}
\end{figure*}

\subsection{Regression of sinusoidal function}
\label{subsec:experiment1}
\begin{figure}[]
    \includegraphics[trim={0.cm 0.4cm 0.cm .cm}, clip, width=0.5\textwidth]{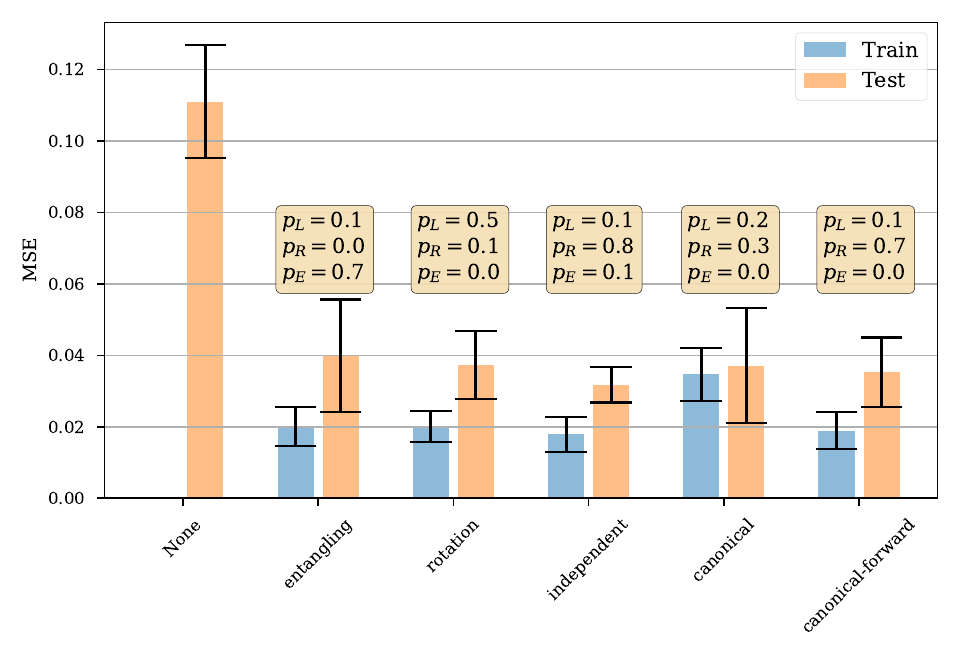}
    \caption{\textbf{Regression} Bar chart comparing the final average performances of all the dropout strategies on the \texttt{sin} dataset with their respective optimal hyperparameters. The standard deviation is taken over 10 different runs.}
    \label{fig:histogram sin function}
\end{figure}

In this experiment, we tackle in detail the \texttt{sin} regression problem. The analytical expression of the function under exam is the following:
\begin{equation}
    y = \sin(\pi x) + \epsilon\,,
\end{equation}
where $x \in [-1,1]$ and $\epsilon$ is an additive white Gaussian noise with amplitude equal to 0.4, zero mean and a standard deviation of 0.5. An extensive analysis of the results is illustrated in Fig.~\ref{fig:histogram sin function}; for each dropout model, the best configuration of $p_L$, $p_R$ and $p_E$ is reported, where $p_L$ refers to the layer dropout rate, $p_R$ to the rotation dropout rate and $p_E$ to the entangling dropout rate. The Mean Squared Error (MSE) obtained by the model without dropout is almost 0 in the training set but it is remarkably high in the test set, which is a clear indicator of overfitting. On the contrary, all quantum dropout techniques do not exhibit signs of overfitting in their bests configurations: the errors on the training set and test set are comparable to each other and way lower than the test error of the overfitting model without dropout, highlighting the effectiveness of these techniques. Independent dropout on average (over 10 different runs) achieves the best MSE of 0.032 on the test set, followed by canonical-forward and rotation dropout, and is $71.4\%$ lower than the one achieved by the model without dropout. It is not surprising that independent obtains the best performances, since the three hyperparameters ($p_L$, $p_R$, $p_E$) search is more exhaustive (and costly) with respect to the other methods leading to a higher-quality dropout. The test error obtained by entangling dropout is $25.3\%$ higher than the one of independent, although it is still comparable. However, the standard deviation of entangling dropout in the test set is large and suggests that this approach may be less robust to random initializations. This may be related to the fact that these gates are not parametrized, in fact also canonical dropout has a considerable standard deviation. 
For the regression QNN eq.~\eqref{eq:max_drop_params} allows to determine that about 60\% of the parameters can be removed while still having an overparametrized model.
Considering the optimal drop ratios in Fig.~\ref{fig:histogram sin function}, the best dropout probabilities are within the range of 5\% to 8\%. Such a result indicates that small dropout ratios are enough for quantum circuits to avoid overfitting while maintaining the ability to effectively learn from the data.

Taking as a reference the independent dropout, namely the best-performing approach among the ones proposed, a further analysis of the results is reported in Fig.~\ref{fig:no_drop}. Fig.~\ref{fig:no_drop}a shows how independent dropout actually mitigates overfitting and makes the prediction of the underlying \texttt{sin} function way smoother than the one performed by the overfitting QNN, which in contrast is highly affected by noise. The benefits of applying dropout are also demonstrated by the convergence of the train and test losses in Fig.~\ref{fig:no_drop}b: the ones of the model with dropout almost converge to the same MSE value after 1000 epochs, whereas the train and test losses of the model without dropout exhibit clear signs of overfitting and have very distinct average MSE values of $1.32\cdot 10^{-5}$ and $1.11\cdot 10^{-1}$ respectively, which differs by four orders of magnitude. Moreover, the test loss of the model without dropout has also a higher standard deviation compared to the other cases, meaning that it is more sensitive to changes in the input data and may indicate instability in the model.

\subsection{Regression of the modulus function}

Here, we briefly discuss the \texttt{module} regression problem while relegating a more detailed analysis with the relative plots in Appendix~\ref{appendix:experiment2}. The function under study is:
\begin{equation}
    y = \lvert x \rvert -\frac{1}{2} + \epsilon\,,
\end{equation}
where $x \in \mathbb{R}$ and $\epsilon$ is still an additive white Gaussian noise with amplitude equal to 0.3, zero mean and a standard deviation of 0.5. 
Using quantum dropout helps to achieve better generalization also for this kind of function. In this case, a pronounced standard deviation in the generalization performances is obtained, and this can be attributed to the Gibbs phenomenon~\cite{introductionFourier}.
The model without dropout overfits with almost 0 training MSE and high test error. Conversely, all quantum dropout techniques demonstrate low and comparable errors, both on the training and test sets, indicating their effectiveness in preventing overfitting compared to the model without dropout. The Independent dropout model still gets the best performance in terms of test error. The standard deviation of all quantum dropout models is lower than the standard deviation of the overfitted model after 10 runs, indicating that the quantum dropout models have better generalization performance.
As in the \texttt{sin} function experiment, also here the gate drop ratios are very low, typically between 6\% and 16\%.


\begin{figure*}
     \centering
     \begin{subfigure}[b]{0.49\textwidth}
         \centering
         \includegraphics[trim={0.5cm 0cm 1.2cm 1cm},clip,width=\textwidth]{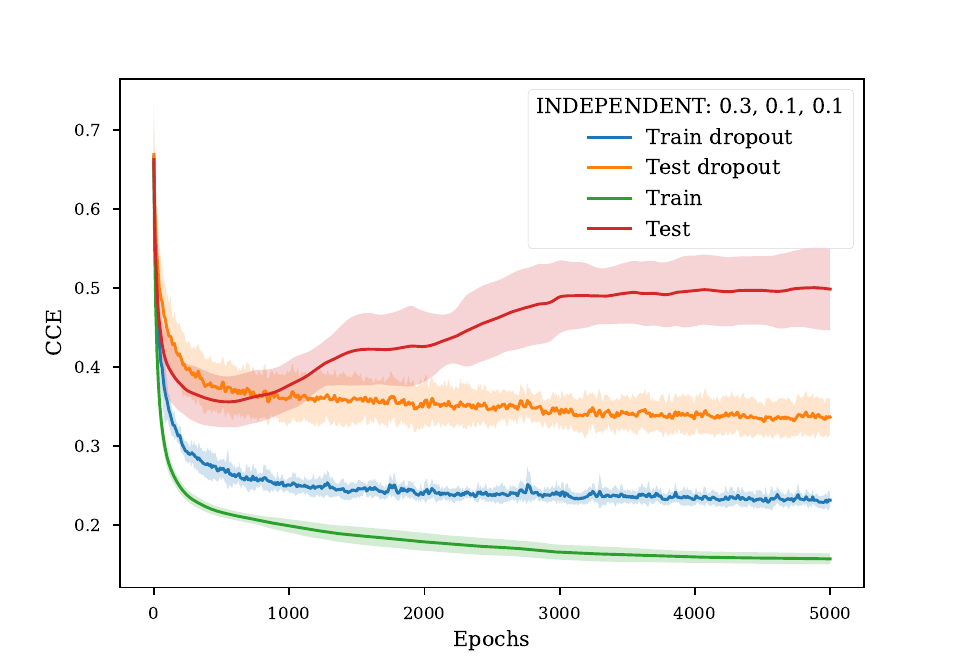}
         \caption{}
     \end{subfigure}
     \begin{subfigure}[b]{0.49\textwidth}
         \centering
         \includegraphics[trim={0.5cm 0cm 1.2cm 1cm},clip,width=\textwidth]{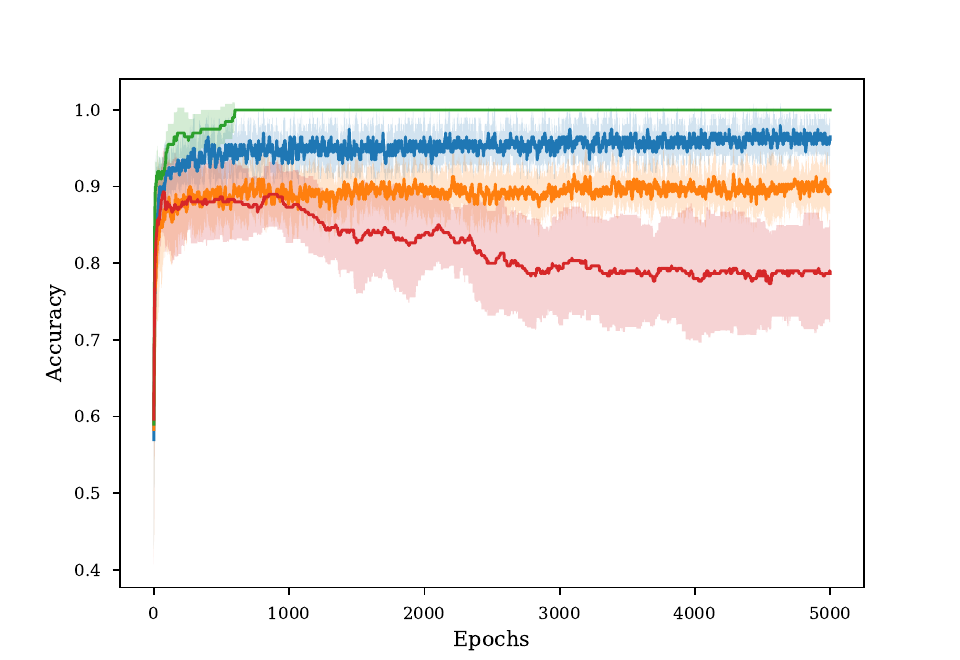}
         \caption{}
     \end{subfigure}
        \caption{\textbf{Classification} The plots illustrate the performances of the second QNN employed in a classification task of the Moons dataset. \textbf{(a)} Standard training shows an increasing prediction error, expressed as CCE, typical of an overfitting model, while with dropout the prediction error keeps decreasing. \textbf{(b)} A similar behaviour can be found in the accuracy of the predictions. The standard deviation over 10 different runs is shown as a shadow.}
        \label{fig:loss_acc_moons}
\end{figure*}

\begin{figure}[!ht]
    \includegraphics[trim={0.cm 0.45cm 0.cm 0.cm}, clip, width=0.5\textwidth]{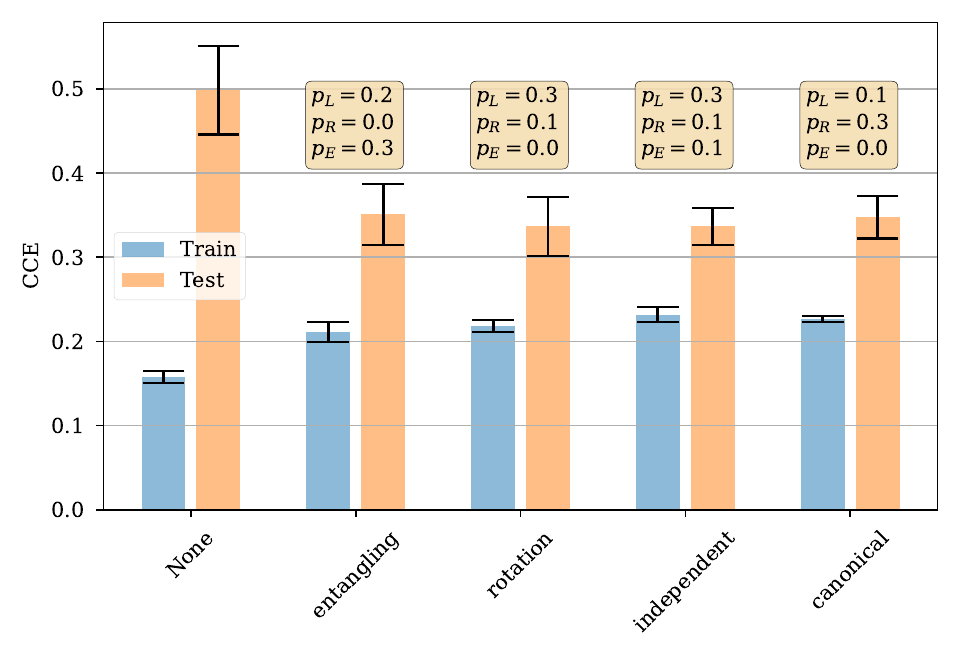}
    \caption{\textbf{Classification} Bar chart comparing the final average performances of all the dropout strategies on the Moons dataset with their respective optimal hyperparameters. The standard deviation is taken over 10 different runs.}
    \label{fig:histogram moons}
\end{figure}

\subsection{Classification}
\label{subsec:classif}
We now address a more difficult problem, i.e., the binary classification task of the Moons dataset with the addition of a further Gaussian noise with amplitude equal to 1, zero mean and a standard deviation of 0.2. In order to show that quantum dropout actually works for different QNN models, we hereby make use of a different QNN architecture with respect to the one employed for the regression, described in detail in Appendix~\ref{appendix:QNN circs}. The main difference lies in the kind of entangling gates employed, which are parametrized and consequently, all the dropout strategies removing entangling gates will directly have an impact on trainable parameters. In addition, this architecture is composed of 20 layers where only one variational sublayer per each data-reuploading is present. This last artifice helps achieve a high level of overfitting with a non-trivial dataset.

Following the same steps done in the previous sections, we report in Fig.~\ref{fig:histogram moons} a comparison of the generalisation capabilities of the model trained with all the dropout strategies in their optimal configuration (best $p_L$, $p_R$, $p_E$ are displayed) with respect to the standard training procedure. In addition, we show the different performances, expressed as Categorical Cross Entropy (CCE) and Accuracy, between the QNN with the best quantum dropout technique and the one without dropout in Fig.~\ref{fig:loss_acc_moons}. 

Similarly to the regression case, Fig.~\ref{fig:histogram moons} reveals that dropout succeeds in removing overfitting in all the employed fashions with comparable performances. Moreover, in this case, we notice good convergence for all the strategies, witnessed by small standard deviation values with respect to the mean value of the loss function (this holds also for accuracy, as a consequence). It has to be noticed that the canonical-forward technique does coincide with the canonical one since the QNN employed in this problem has only one variational sublayer per each layer (see Appendix~\ref{appendix:QNN circs}) and this is why in Fig.~\ref{fig:histogram moons} we only show canonical dropout performance. Interestingly, the optimal drop probabilities here are smaller than in regression problems, ranging from 1\% to 6\%. Independent dropout again achieves the best performance, diminishing the test loss of about $32.5\%$, whereas the entangling dropout is slightly worse than the others being the only one reducing the test CCE less than $30\%$. The hierarchies are the same also for what concerns the standard deviation: independent dropout has the tiniest one, whereas entangling presents the largest standard deviation. We stress that if on the one hand independent dropout is able to achieve the best performances, on the other hand, it requires a much more intense parameter search with respect to the other strategies.

From Figs.~\ref{fig:loss_acc_moons}a-b one can really understand the effectiveness of quantum dropout in removing overfitting: both the test loss and Accuracy smoothly follow the trend of their respective training counterpart with low standard deviation. Conversely, in the case without dropout the test loss and Accuracy depart from the training ones after 1000 epochs with high standard deviation. Here we report these trends only for the overall best technique, but all the other dropout configurations (in terms of $p_L$ and $p_G$) displayed in Fig.~\ref{fig:histogram moons} show completely analogous behaviours.

\subsection{Entanglement and expressibility}
\label{subsec:expr and ent}

\begin{figure*}
     \centering
     \begin{subfigure}[b]{0.49\textwidth}
         \centering
         \includegraphics[trim={0.cm 0.2cm 1.2cm 1cm},clip,width=\textwidth]{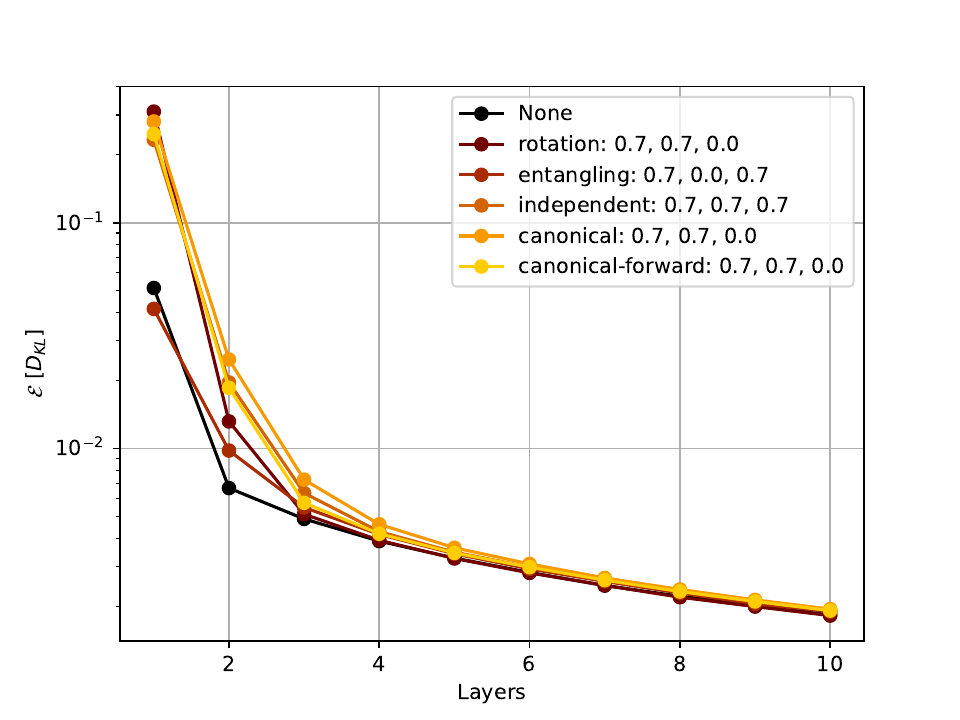}
         \caption{}
     \end{subfigure}
     \hfill
     \begin{subfigure}[b]{0.49\textwidth}
         \centering
         \includegraphics[trim={0.cm 0.2cm 1.2cm 1cm},clip,width=\textwidth]{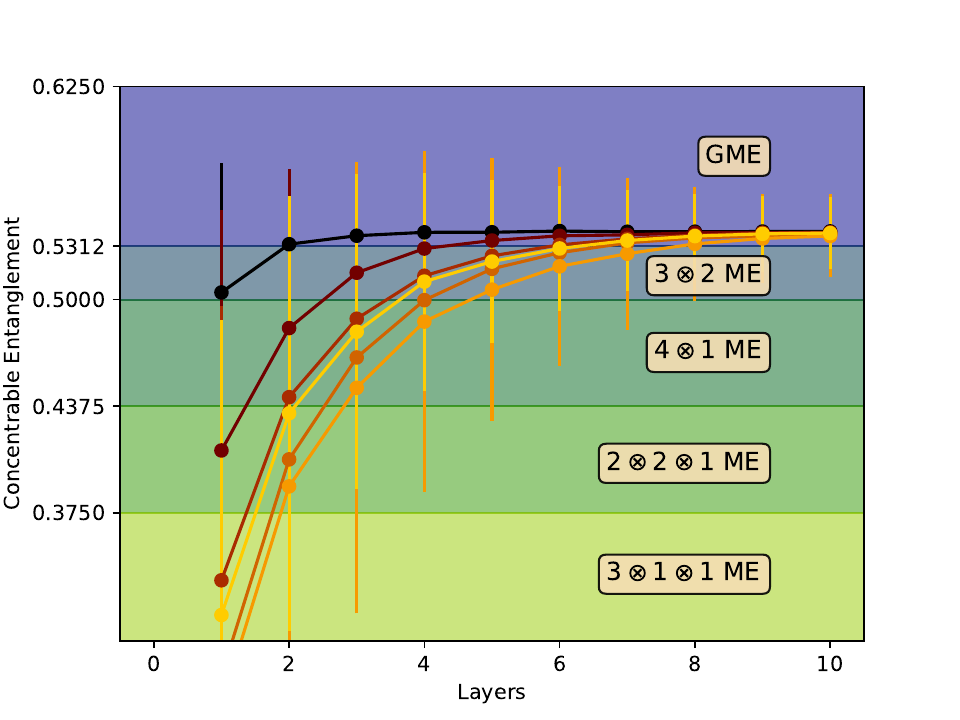}
         \caption{}
     \end{subfigure}
       \caption{\textbf{(a)} Expressibility (in log scale) of QNN w.r.t. the employed layers. After a few layers, the expressibility is approximately the same for all dropout strategies. \textbf{(b)}
       Degree of concentrable entanglement (CE) produced on average by the QNN w.r.t. the employed layers. The CE soon reaches a plateau, witnessing the presence of a non-trivial amount of GME in the QNN. Even in the case of high drop probability ($\sim 50\%$), the mean entanglement produced by the QNN with 10 layers is the same for different dropout strategies.}
        \label{fig:expressib+ent}
\end{figure*}

The expressibility and entangling capability~\cite{Sim2019, Ballarin_2023} of a QNN do depend on the structure of the chosen ansatz and, in particular, this is strongly related to the kind and number of rotations and entangling gates of which it is composed. The here proposed quantum dropout techniques randomly remove some gates in the circuit during the training process of overparametrized QNNs. In the following, differently from what conjectured in~\cite{kobayashi_overfitting_2022}, we show that expressibility and entangling capability are not affected by dropout even for high values of dropout ratio (see Figs.~\ref{fig:expressib+ent}) for the overparametrized QNN proposed in their work and employed in the regression tasks here. This ensures that expressibility and entanglement are not lowered also in the regime with small dropout rates, i.e. removing few gates, in which one usually operates. Expressibility measures the circuit’s ability to produce states that are well representative of the Hilbert space and to quantify it we follow the guidelines drawn in~\cite{Sim2019}. The entangling capability, instead, is quantified as the mean concentrable entanglement~\cite{beckey2021concentrable} produced, which is a measure of multipartite entanglement directly computed on a quantum computer. In addition, for the sake of completeness, we report in the appendix a similar analysis for the other QNN used in the classification problem confirming the same findings also for another QNN architecture.

The QNN model we are working with has elevated expressibility (see Appendix~\ref{appendix:expressib}) and in addition is highly overparametrized (see Appendix~\ref{appendix: overparam}). In this regard, applying quantum dropout around the optimal operating regime is very unlikely to reduce the expressibility directly. In fact, the drop ratios in all the possible dropout strategies presented in Fig.~\ref{fig:histogram sin function} are not very high, implying that only a few gates are dropped in each epoch of the training phase. On the opposite, to decrease the expressibility one should affect the parameter dimension $D$ (defined in Sec.~\ref{subsec:overparam}), making the QNN underparametrized, but using eq.~\eqref{eq:max_drop_params} with $M=150$, more than $M-D_{max}=88$ rotation gates should be removed, recalling that $D_{max}=2^{N+1}-2$ is the maximal parameter dimension and accounts for the minimal number of parameters required to explore the whole Hilbert space. This would correspond to about 60\% of the parameters of the model with 10 layers. From Fig.~\ref{fig:expressib+ent}a, one can capture that a 50\% dropout does not have an impact on the expressibility of the model, with exception made for the cases with less than 5 layers; above this threshold the model is overparametrized and the expressibility has the same value as the model without dropout.

In~\cite{kobayashi_overfitting_2022} it is argued that quantum dropout outperforms $L_1$ and $L_2$ regularization due to its ability to limit the expressibility of the model. They also suggest that regularization allows full model expressibility, consequently necessitating a higher number of training epochs to reach an optimal solution. Here we show that these assertions are false. To start with, our findings indicate that quantum dropout does not curtail the model's expressibility, as indicated in Fig.~\ref{fig:expressib+ent}a. On the contrary, it introduces a form of structured randomness into the quantum model, effectively helping it avoid overfitting without limiting its ability to represent complex functions. Secondly, $L_1$ and $L_2$ regularization methods work by adding a penalty term to the loss function, which encourages the model to keep its parameters small, as highlighted in~\cite{kobayashi_overfitting_2022} Sec.~3.4. Therefore, adopting this approach during the training of a QNN may be very restrictive as it does not allow the full exploration of the Hilbert space~\cite{holmes2022connecting}: by coercing parameter values to zero, regularization may limit the expressibility of the quantum model, which is exactly the opposite of what the authors in ~\cite{kobayashi_overfitting_2022} suggested.



The regression QNN architecture with 10 layers produces an average level of multipartite entanglement which is slightly above the Genuine Multipartite Entanglement (GME) threshold (the same holds for the classification QNN as reported in Appendix~\ref{appendix: multip ent}), see Fig.~\ref{fig:expressib+ent}b. Even in the case of entanglement applying dropout with a probability of about 50\% does not reduce its average level at the depth utilised for the learning task. The final average value of entanglement reached with 10 layers is the one expected for random states (Haar states) and this witnesses again that the QNN has high expressibility and that quantum dropout does not hinder expressibility. For both expressibility and entanglement, we estimated average and standard deviation from 15000 states. See Appendices~\ref{appendix:expressib} and~\ref{appendix: multip ent} for further details on these calculations. 

So, a natural question will be: why does the quantum dropout work if it does not modify the QNN features on average? To understand this we have to interpret the quantum dropout as sampling a different sub-architecture from the original QNN at each training step, leading to training a whole ensemble at the same time. Just like classical dropout, this prevents hidden units (rotations in the QNN) from relying too much on their neighbourhoods. However, the rotations belonging to the same layer (or sub-layer) still have different roles in changing the state. Consequently, quantum dropout does not work well with excessively high drop ratios because, in that case, we will have very different models producing different results at each epoch and this will definitely not help the optimization process. 


\subsection{Parameters scaling}
\label{subsec:params scaling}


To completely match classical and quantum dropout, one should scale the weights in the layers where dropout was applied by a quantity $s=1-p$, i.e. the probability of the computational units being active during training~\cite{srivastava_dropout_2014}. This ensures that the expected output of a computational unit during inference is the same as its expected output during training, i.e., it guarantees that the average output is the same for both training and inference. In classical NNs, the scaling helps in maintaining the statistical properties of the network and ensures that the network produces consistent and reliable results during inference.

\begin{figure*}
     \centering
     \begin{subfigure}[b]{0.49\textwidth}
         \centering
         \includegraphics[trim={0.cm 0.4cm 0.cm 0.cm},clip,width=\textwidth]{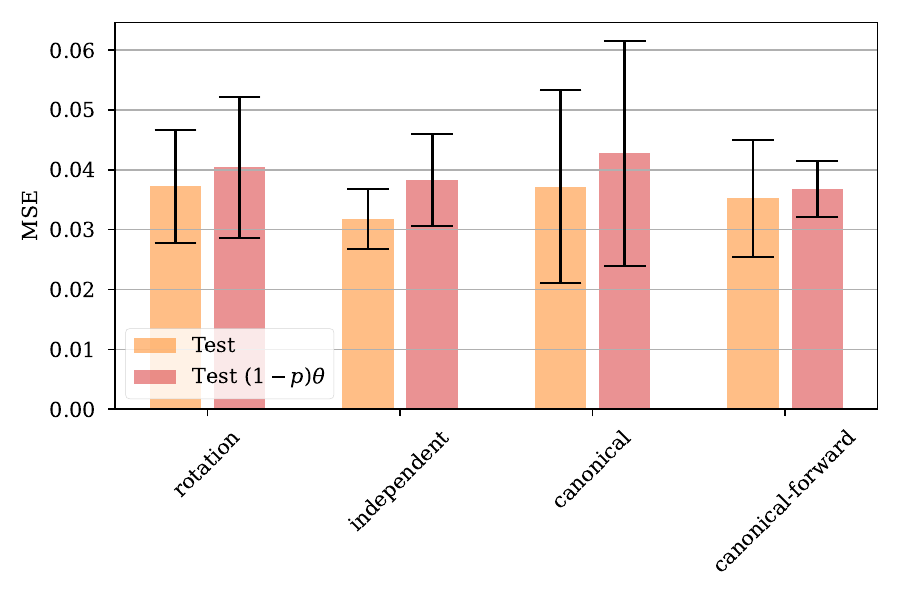}
         \caption{\texttt{sin}}
     \end{subfigure}
     \begin{subfigure}[b]{0.49\textwidth}
         \centering
         \includegraphics[trim={0cm 0.5cm 0cm 0cm},clip,width=\textwidth]{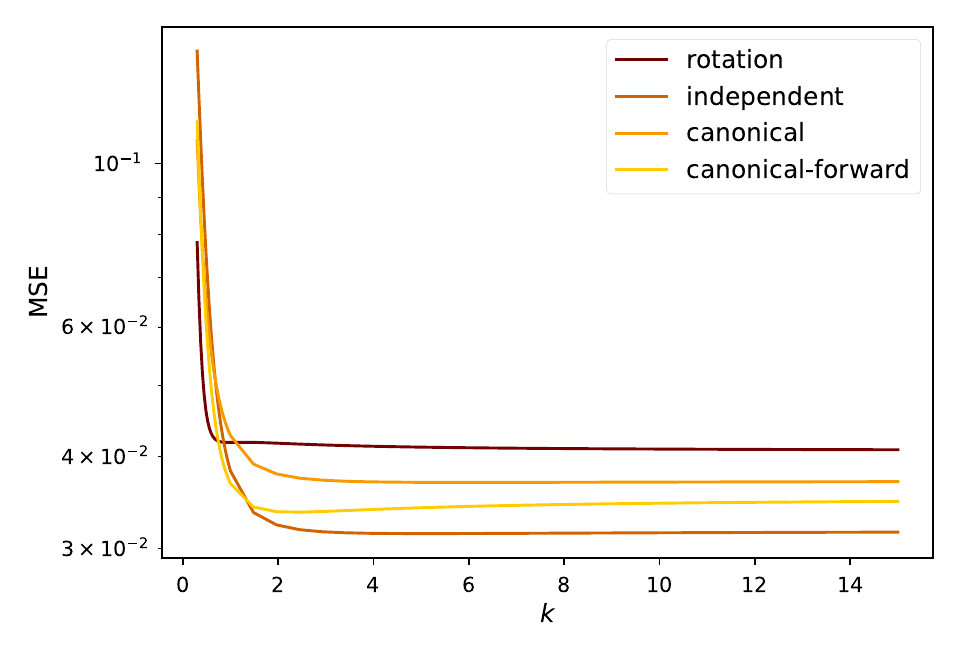}
         \caption{\texttt{sin}}
     \end{subfigure}
     \begin{subfigure}[b]{0.49\textwidth}
         \centering
         \includegraphics[trim={0.cm 0cm 0.cm 0.cm},clip,width=\textwidth]{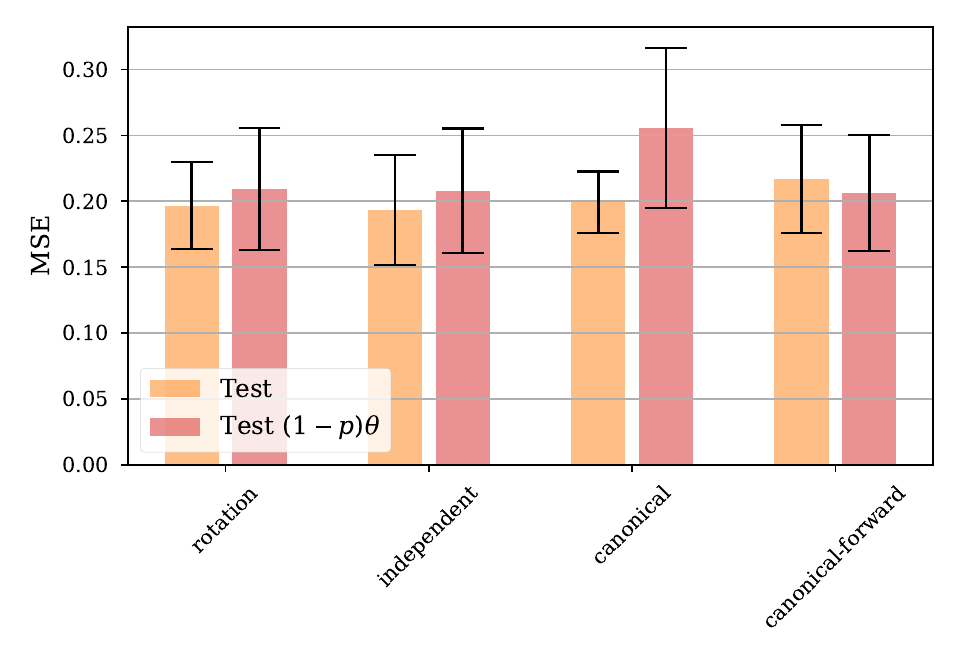}
         \caption{\texttt{module}}
     \end{subfigure}
     \begin{subfigure}[b]{0.49\textwidth}
         \centering
         \includegraphics[trim={0cm 0cm 0cm 0cm},clip,width=\textwidth]{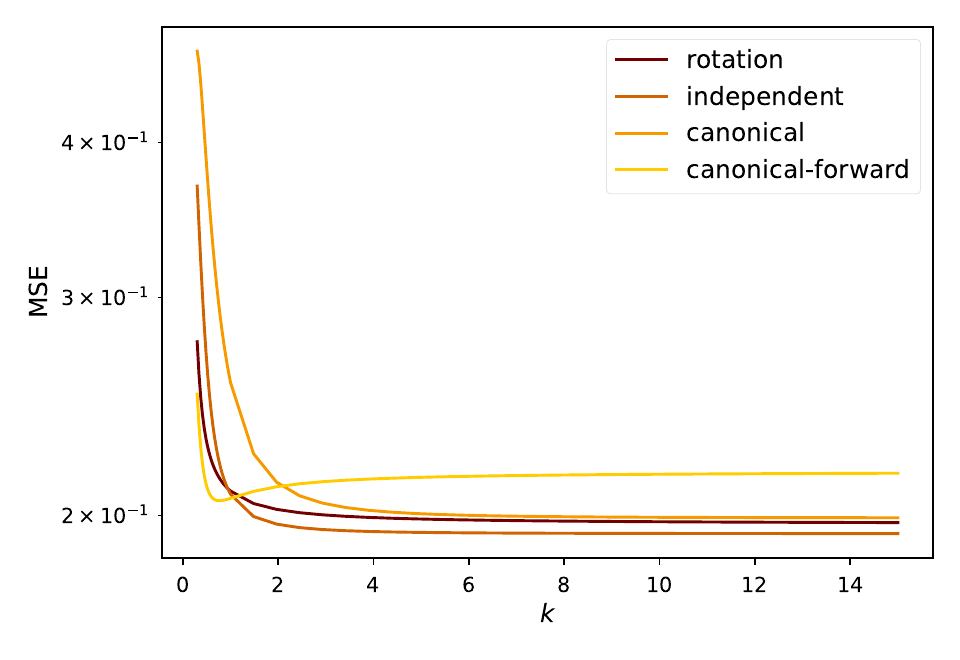}
         \caption{\texttt{module}}
     \end{subfigure}
     \begin{subfigure}[b]{0.49\textwidth}
         \centering
         \includegraphics[trim={0.cm 0.4cm 0cm 0cm},clip,width=\textwidth]{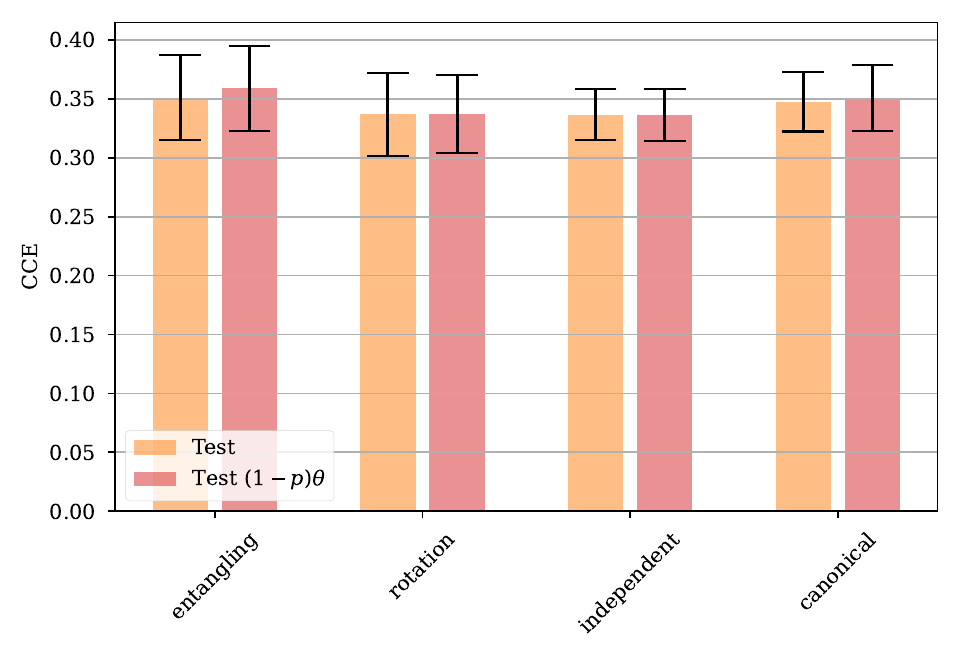}
         \caption{\texttt{moons}}
     \end{subfigure}
     \begin{subfigure}[b]{0.49\textwidth}
         \centering
         \includegraphics[trim={0.cm 0.3cm 0cm 0cm},clip,width=\textwidth]{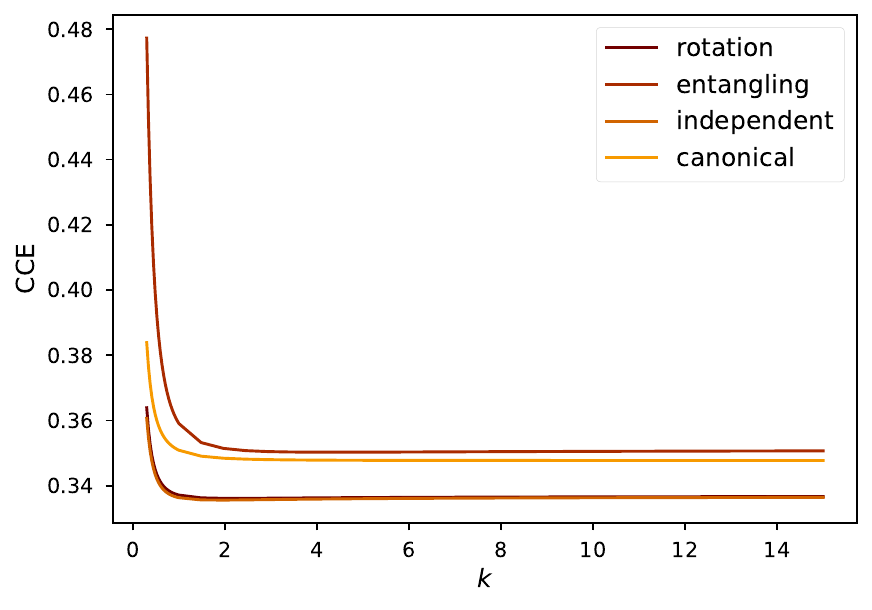}
         \caption{\texttt{moons}}
     \end{subfigure}
       \caption{\textbf{Parameters scaling} Barchart comparing the best final performances of the QNNs without and with trained parameters scaling on the \textbf{(a)} \texttt{sin} dataset, on the \textbf{(c)} \texttt{module} dataset and on the \textbf{(e)} \texttt{moon} dataset. Standard deviation over 10 runs. \textbf{(b)}, \textbf{(d)} Average MSE and \textbf{(f)} average CCE trend when scaling the trained parameters with $s=(1-p)^{1/k}$ as a function of $k$. }
        \label{fig:params scaling regression}
\end{figure*}

We now show that for deep QNNs the scaling factor $s$ is not the same as in the classical case. Interestingly, our experiments have shown that by \emph{not rescaling} the parameters one achieves (on average) better results compared to the ones obtained by scaling with the classical linear scaling factor. To illustrate this finding, we have included a comparison of test errors obtained with and without rescaled parameters in our experiments, as shown in Fig.~\ref{fig:params scaling regression}a,~c,~e. We find either negligible differences or no discernible change in the error rate across all quantum dropout strategies; the only exception is canonical-forward in the \texttt{module} regression task, which nevertheless shows a very little but not statistically significant improvement with classical rescaling.

This general behaviour may be related to the fact that any change on one qubit could affect all the others, since we are working with qubits that in principle share multipartite entanglement, as explained above. Further details on the entanglement of the employed ansatz can be found in Appendix~\ref{appendix: multip ent}. In addition, in a QNN we consider as neurons the rotations but, in practice, the units of computations are only the $N$ qubits on which we act multiple times and this is a strong difference from classical NNs. Taking these two facts into account, one intuitive explanation could be that if we scale a parameter, the scaling factor is propagated across the whole structure to some extent multiplicatively. In this picture, to obtain the same effect of the classical scaling one might apply a scaling factor which is the $k$-root of the classical one, where $k$ is some real number possibly dependent on the concentrable entanglement~\cite{beckey2021concentrable} in the QNN, which is directly related to the number of qubits N.
Since $p\leq 1$, if $k>>1 \implies 1/k \approx 0$ leads to:
\begin{equation}
\label{eq:params scaling}
    s=(1-p)^{1/k}\approx 1\,,
\end{equation}
corresponding to not scaling the parameters. We extended our analysis by exploring the use of a $k$-root attenuating factor for rescaling, as shown in Fig.~\ref{fig:params scaling regression}b,~d,~f. These results confirm our previous finding that rescaling does not provide any significant advantage in terms of generalization, because the general trend for all the quantum dropout strategies is to enhance the performance as $k$ grows, i.e. $s$ tends to $1$. The canonical-forward strategy in the regression tasks is the only exception to this trend: we observed a slight yet not significant improvement in performance at $k=2$ for the \texttt{sin} dataset and at $k=1$ for the \texttt{module} dataset.


In order to fully comprehend the mechanism behind quantum dropout and to unlock its potential as in the classical case, future works should dive deeper into the relationship between parameters scaling and entanglement within the QNN model.

\section{Conclusions}
\label{sec:conclusions}

Starting from the comparison between classical and quantum NNs, we have proposed a general approach to exploit the dropout technique to avoid overfitting in learning with QNNs. In this framework, depending on the dropped unitary operations, we define and apply different quantum dropout strategies. Our results generalize and include special cases that are already present in the literature, such as entangling dropout~\cite{kobayashi_overfitting_2022}. In particular, our results highlight the importance of dropping parametrized operations, leading to improved generalization performance with respect to the case without dropout as well as to entangling dropout. In addition, aware of the importance of having an overparametrized QNN, we also provide guidelines on how to determine the maximally allowed dropout probability.

In addition, we investigate for the first time the detailed behaviour of genuine QNN features upon quantum dropout application. Our analysis reveals that these techniques seem to better mitigate overfitting with moderate dropout probabilities, differently from the classical case where typical probability values are in the order 50-80\%. In this working regime, quantum dropout does not reduce expressibility nor entanglement, as the QNN is still highly overparametrized. These findings are in contrast with conjectures previously made in the literature about the effectiveness of entangling dropout~\cite{kobayashi_overfitting_2022}. 

Last but not least, we show that quantum dropout does not require parameter rescaling, differently from the classical dropout technique applied to NNs. Interestingly, the performance is often actually worse when the variational parameters are rescaled. 

In conclusion, our findings witness that, once the QNN model is fixed, slight random modifications in the structure during the training procedure allow generalization when making inferences with the QNN original model.
This work may pave the way for the efficient employment of Deep Quantum Machine Learning Models, thanks to a robust training methodology that encourages generalization.

Future works may investigate the theoretical generalization of quantum dropout more in detail, as well as its effectiveness by running experiments on real quantum hardware. Also, the scaling of performances with an increasing number of qubits and the relationship between parameters rescaling and entanglement produced by the QNN are worth being further investigated.






\section{Methods}
\label{sec:methods}
To assess the effectiveness of our approach, we conduct two regression tasks and one binary classification task. In every regression experiment, the corresponding dataset (with 20 data samples) was divided into 75\% train samples and 25\% test samples, whereas for classification we have 50 data samples split into $40\%$ train and $60\%$ test in order to emphasize the overfitting phenomenon.  In a data preprocessing phase, raw data were scaled in the range $[-1,1]$ to best suit the input and output of the QNNs; the scaler was fitted using training data only. 

To ensure a good and fast convergence, the models were trained with the Adam optimization algorithm \cite{kingma2014adam} and a thorough grid search was conducted to tune the training hyperparameters. For the regression tasks, we observed that different hyperparameters led to similar results and did not impact substantially the outcomes of the experiments. As a consequence, and in order to make a fair comparison among all the QNN models, the hyperparameters remained the same for these experiments. The learning rate was set to 0.01, the number of training epochs was set to 1000 and the QNN was made up of 10 layers for regression. On the other hand, for the classification problem, the number of epochs was set to 5000, whereas the number of QNN layers was increased to 20 due to both the learning task and the different structure of the single layer, as already mentioned in Sec.~\ref{subsec:classif}.

The drop rates space was extensively analysed in the case of \texttt{sin} dataset through a grid search with $p_L\in[0.1,0.2, \dots, 0.7]$ and $p_G\in[0.1,0.2, \dots, 0.9]$. This choice of hyperparameters was dictated by eq.~\eqref{eq:max_drop_params} for the regression QNN, for which one can safely drop up to 60\% of the parameters (see Appendix~\ref{appendix: overparam}). For the other two datasets, we conducted a more limited and coarse-grained analysis, since we noticed that close drop probabilities led to similar performance and dropping many gates hindered the trainability.

The MSE was selected as the loss function to train the networks and the error metric to evaluate the performances of the models in the regression tasks, as it is more sensitive to larger errors and a standard error metric in supervised learning. CCE was used as loss function for the classification task instead, while Accuracy score was employed as the main error metric to assess the goodness of the classification.
In order to prove the robustness of our approach against random parameters initializations, we performed 10 runs of the algorithms for every test case. 

The QNN models were implemented in Python 3.8 with PennyLane~\cite{pennylane}, a framework that supports local simulations of quantum circuits and integration with NN optimization tools. To improve simulation times, we employed the JAX~\cite{jax2018github} linear algebra framework as the simulation backend. JAX is a software library for
high-performance ML research which guarantees a fast and efficient way to execute quantum circuits through automatic differentiation and Just-in-time (JIT) circuit compilation. 
Due to the difficulty in simulating deep quantum circuits with a high degree of entanglement on local devices, JIT compilation has been particularly beneficial for the simulation of our experiments. A machine equipped with an AMD Ryzen 7™ 5800X 8-Core CPU at 3.80 GHz and with 64 GB of RAM was used for the experiments.

\bibliography{bibliography}

\begin{thebibliography}{56}%
\makeatletter
\providecommand \@ifxundefined [1]{%
 \@ifx{#1\undefined}
}%
\providecommand \@ifnum [1]{%
 \ifnum #1\expandafter \@firstoftwo
 \else \expandafter \@secondoftwo
 \fi
}%
\providecommand \@ifx [1]{%
 \ifx #1\expandafter \@firstoftwo
 \else \expandafter \@secondoftwo
 \fi
}%
\providecommand \natexlab [1]{#1}%
\providecommand \enquote  [1]{``#1''}%
\providecommand \bibnamefont  [1]{#1}%
\providecommand \bibfnamefont [1]{#1}%
\providecommand \citenamefont [1]{#1}%
\providecommand \href@noop [0]{\@secondoftwo}%
\providecommand \href [0]{\begingroup \@sanitize@url \@href}%
\providecommand \@href[1]{\@@startlink{#1}\@@href}%
\providecommand \@@href[1]{\endgroup#1\@@endlink}%
\providecommand \@sanitize@url [0]{\catcode `\\12\catcode `\$12\catcode
  `\&12\catcode `\#12\catcode `\^12\catcode `\_12\catcode `\%12\relax}%
\providecommand \@@startlink[1]{}%
\providecommand \@@endlink[0]{}%
\providecommand \url  [0]{\begingroup\@sanitize@url \@url }%
\providecommand \@url [1]{\endgroup\@href {#1}{\urlprefix }}%
\providecommand \urlprefix  [0]{URL }%
\providecommand \Eprint [0]{\href }%
\providecommand \doibase [0]{https://doi.org/}%
\providecommand \selectlanguage [0]{\@gobble}%
\providecommand \bibinfo  [0]{\@secondoftwo}%
\providecommand \bibfield  [0]{\@secondoftwo}%
\providecommand \translation [1]{[#1]}%
\providecommand \BibitemOpen [0]{}%
\providecommand \bibitemStop [0]{}%
\providecommand \bibitemNoStop [0]{.\EOS\space}%
\providecommand \EOS [0]{\spacefactor3000\relax}%
\providecommand \BibitemShut  [1]{\csname bibitem#1\endcsname}%
\let\auto@bib@innerbib\@empty
\bibitem [{\citenamefont {Condie}\ \emph {et~al.}(2013)\citenamefont {Condie},
  \citenamefont {Mineiro}, \citenamefont {Polyzotis},\ and\ \citenamefont
  {Weimer}}]{tysonMachineLearning2020}%
  \BibitemOpen
  \bibfield  {author} {\bibinfo {author} {\bibfnamefont {T.}~\bibnamefont
  {Condie}}, \bibinfo {author} {\bibfnamefont {P.}~\bibnamefont {Mineiro}},
  \bibinfo {author} {\bibfnamefont {N.}~\bibnamefont {Polyzotis}},\ and\
  \bibinfo {author} {\bibfnamefont {M.}~\bibnamefont {Weimer}},\ }\bibfield
  {title} {\bibinfo {title} {Machine learning on big data},\ }in\ \href
  {https://doi.org/10.1109/ICDE.2013.6544913} {\emph {\bibinfo {booktitle}
  {2013 IEEE 29th International Conference on Data Engineering (ICDE)}}}\
  (\bibinfo {year} {2013})\ pp.\ \bibinfo {pages} {1242--1244}\BibitemShut
  {NoStop}%
\bibitem [{\citenamefont {Dhall}\ \emph {et~al.}(2020)\citenamefont {Dhall},
  \citenamefont {Kaur},\ and\ \citenamefont {Juneja}}]{dhall2020machine}%
  \BibitemOpen
  \bibfield  {author} {\bibinfo {author} {\bibfnamefont {D.}~\bibnamefont
  {Dhall}}, \bibinfo {author} {\bibfnamefont {R.}~\bibnamefont {Kaur}},\ and\
  \bibinfo {author} {\bibfnamefont {M.}~\bibnamefont {Juneja}},\ }\bibfield
  {title} {\bibinfo {title} {Machine learning: a review of the algorithms and
  its applications},\ }\href@noop {} {\bibfield  {journal} {\bibinfo  {journal}
  {Proceedings of ICRIC 2019}\ ,\ \bibinfo {pages} {47}} (\bibinfo {year}
  {2020})}\BibitemShut {NoStop}%
\bibitem [{\citenamefont {Singh}\ \emph {et~al.}(2016)\citenamefont {Singh},
  \citenamefont {Thakur},\ and\ \citenamefont
  {Sharma}}]{amanpreetReviewSupervised2016}%
  \BibitemOpen
  \bibfield  {author} {\bibinfo {author} {\bibfnamefont {A.}~\bibnamefont
  {Singh}}, \bibinfo {author} {\bibfnamefont {N.}~\bibnamefont {Thakur}},\ and\
  \bibinfo {author} {\bibfnamefont {A.}~\bibnamefont {Sharma}},\ }\bibfield
  {title} {\bibinfo {title} {A review of supervised machine learning
  algorithms},\ }in\ \href@noop {} {\emph {\bibinfo {booktitle} {2016 3rd
  International Conference on Computing for Sustainable Global Development
  (INDIACom)}}}\ (\bibinfo {year} {2016})\ pp.\ \bibinfo {pages}
  {1310--1315}\BibitemShut {NoStop}%
\bibitem [{\citenamefont {Hawkins}(2004)}]{hawkins2004problem}%
  \BibitemOpen
  \bibfield  {author} {\bibinfo {author} {\bibfnamefont {D.~M.}\ \bibnamefont
  {Hawkins}},\ }\bibfield  {title} {\bibinfo {title} {The problem of
  overfitting},\ }\href@noop {} {\bibfield  {journal} {\bibinfo  {journal}
  {Journal of chemical information and computer sciences}\ }\textbf {\bibinfo
  {volume} {44}},\ \bibinfo {pages} {1} (\bibinfo {year} {2004})}\BibitemShut
  {NoStop}%
\bibitem [{\citenamefont {Ying}(2019)}]{ying2019overview}%
  \BibitemOpen
  \bibfield  {author} {\bibinfo {author} {\bibfnamefont {X.}~\bibnamefont
  {Ying}},\ }\bibfield  {title} {\bibinfo {title} {An overview of overfitting
  and its solutions},\ }in\ \href@noop {} {\emph {\bibinfo {booktitle} {Journal
  of physics: Conference series}}},\ Vol.\ \bibinfo {volume} {1168}\ (\bibinfo
  {organization} {IOP Publishing},\ \bibinfo {year} {2019})\ p.\ \bibinfo
  {pages} {022022}\BibitemShut {NoStop}%
\bibitem [{\citenamefont {Srivastava}\ \emph {et~al.}(2014)\citenamefont
  {Srivastava}, \citenamefont {Hinton}, \citenamefont {Krizhevsky},
  \citenamefont {Sutskever},\ and\ \citenamefont
  {Salakhutdinov}}]{srivastava_dropout_2014}%
  \BibitemOpen
  \bibfield  {author} {\bibinfo {author} {\bibfnamefont {N.}~\bibnamefont
  {Srivastava}}, \bibinfo {author} {\bibfnamefont {G.}~\bibnamefont {Hinton}},
  \bibinfo {author} {\bibfnamefont {A.}~\bibnamefont {Krizhevsky}}, \bibinfo
  {author} {\bibfnamefont {I.}~\bibnamefont {Sutskever}},\ and\ \bibinfo
  {author} {\bibfnamefont {R.}~\bibnamefont {Salakhutdinov}},\ }\bibfield
  {title} {\bibinfo {title} {Dropout: A simple way to prevent neural networks
  from overfitting},\ }\href {http://jmlr.org/papers/v15/srivastava14a.html}
  {\bibfield  {journal} {\bibinfo  {journal} {Journal of Machine Learning
  Research}\ }\textbf {\bibinfo {volume} {15}},\ \bibinfo {pages} {1929}
  (\bibinfo {year} {2014})}\BibitemShut {NoStop}%
\bibitem [{\citenamefont {Huang}\ \emph {et~al.}(2022)\citenamefont {Huang},
  \citenamefont {Broughton}, \citenamefont {Cotler}, \citenamefont {Chen},
  \citenamefont {Li}, \citenamefont {Mohseni}, \citenamefont {Neven},
  \citenamefont {Babbush}, \citenamefont {Kueng}, \citenamefont {Preskill}
  \emph {et~al.}}]{huang2022quantum}%
  \BibitemOpen
  \bibfield  {author} {\bibinfo {author} {\bibfnamefont {H.-Y.}\ \bibnamefont
  {Huang}}, \bibinfo {author} {\bibfnamefont {M.}~\bibnamefont {Broughton}},
  \bibinfo {author} {\bibfnamefont {J.}~\bibnamefont {Cotler}}, \bibinfo
  {author} {\bibfnamefont {S.}~\bibnamefont {Chen}}, \bibinfo {author}
  {\bibfnamefont {J.}~\bibnamefont {Li}}, \bibinfo {author} {\bibfnamefont
  {M.}~\bibnamefont {Mohseni}}, \bibinfo {author} {\bibfnamefont
  {H.}~\bibnamefont {Neven}}, \bibinfo {author} {\bibfnamefont
  {R.}~\bibnamefont {Babbush}}, \bibinfo {author} {\bibfnamefont
  {R.}~\bibnamefont {Kueng}}, \bibinfo {author} {\bibfnamefont
  {J.}~\bibnamefont {Preskill}}, \emph {et~al.},\ }\bibfield  {title} {\bibinfo
  {title} {Quantum advantage in learning from experiments},\ }\href@noop {}
  {\bibfield  {journal} {\bibinfo  {journal} {Science}\ }\textbf {\bibinfo
  {volume} {376}},\ \bibinfo {pages} {1182} (\bibinfo {year}
  {2022})}\BibitemShut {NoStop}%
\bibitem [{\citenamefont {Caro}\ \emph {et~al.}(2022)\citenamefont {Caro},
  \citenamefont {Huang}, \citenamefont {Cerezo}, \citenamefont {Sharma},
  \citenamefont {Sornborger}, \citenamefont {Cincio},\ and\ \citenamefont
  {Coles}}]{caro_generalization_2022}%
  \BibitemOpen
  \bibfield  {author} {\bibinfo {author} {\bibfnamefont {M.~C.}\ \bibnamefont
  {Caro}}, \bibinfo {author} {\bibfnamefont {H.-Y.}\ \bibnamefont {Huang}},
  \bibinfo {author} {\bibfnamefont {M.}~\bibnamefont {Cerezo}}, \bibinfo
  {author} {\bibfnamefont {K.}~\bibnamefont {Sharma}}, \bibinfo {author}
  {\bibfnamefont {A.}~\bibnamefont {Sornborger}}, \bibinfo {author}
  {\bibfnamefont {L.}~\bibnamefont {Cincio}},\ and\ \bibinfo {author}
  {\bibfnamefont {P.~J.}\ \bibnamefont {Coles}},\ }\bibfield  {title} {\bibinfo
  {title} {Generalization in quantum machine learning from few training data},\
  }\href {https://doi.org/10.1038/s41467-022-32550-3} {\bibfield  {journal}
  {\bibinfo  {journal} {Nature Communications}\ }\textbf {\bibinfo {volume}
  {13}},\ \bibinfo {pages} {4919} (\bibinfo {year} {2022})}\BibitemShut
  {NoStop}%
\bibitem [{\citenamefont {Biamonte}\ \emph {et~al.}(2017)\citenamefont
  {Biamonte}, \citenamefont {Wittek}, \citenamefont {Pancotti}, \citenamefont
  {Rebentrost}, \citenamefont {Wiebe},\ and\ \citenamefont
  {Lloyd}}]{biamonte2017quantum}%
  \BibitemOpen
  \bibfield  {author} {\bibinfo {author} {\bibfnamefont {J.}~\bibnamefont
  {Biamonte}}, \bibinfo {author} {\bibfnamefont {P.}~\bibnamefont {Wittek}},
  \bibinfo {author} {\bibfnamefont {N.}~\bibnamefont {Pancotti}}, \bibinfo
  {author} {\bibfnamefont {P.}~\bibnamefont {Rebentrost}}, \bibinfo {author}
  {\bibfnamefont {N.}~\bibnamefont {Wiebe}},\ and\ \bibinfo {author}
  {\bibfnamefont {S.}~\bibnamefont {Lloyd}},\ }\bibfield  {title} {\bibinfo
  {title} {Quantum machine learning},\ }\href@noop {} {\bibfield  {journal}
  {\bibinfo  {journal} {Nature}\ }\textbf {\bibinfo {volume} {549}},\ \bibinfo
  {pages} {195} (\bibinfo {year} {2017})}\BibitemShut {NoStop}%
\bibitem [{\citenamefont {Sajjan}\ \emph {et~al.}(2022)\citenamefont {Sajjan},
  \citenamefont {Li}, \citenamefont {Selvarajan}, \citenamefont {Sureshbabu},
  \citenamefont {Kale}, \citenamefont {Gupta}, \citenamefont {Singh},\ and\
  \citenamefont {Kais}}]{sajjan2022quantum}%
  \BibitemOpen
  \bibfield  {author} {\bibinfo {author} {\bibfnamefont {M.}~\bibnamefont
  {Sajjan}}, \bibinfo {author} {\bibfnamefont {J.}~\bibnamefont {Li}}, \bibinfo
  {author} {\bibfnamefont {R.}~\bibnamefont {Selvarajan}}, \bibinfo {author}
  {\bibfnamefont {S.~H.}\ \bibnamefont {Sureshbabu}}, \bibinfo {author}
  {\bibfnamefont {S.~S.}\ \bibnamefont {Kale}}, \bibinfo {author}
  {\bibfnamefont {R.}~\bibnamefont {Gupta}}, \bibinfo {author} {\bibfnamefont
  {V.}~\bibnamefont {Singh}},\ and\ \bibinfo {author} {\bibfnamefont
  {S.}~\bibnamefont {Kais}},\ }\bibfield  {title} {\bibinfo {title} {Quantum
  machine learning for chemistry and physics},\ }\href@noop {} {\bibfield
  {journal} {\bibinfo  {journal} {Chemical Society Reviews}\ } (\bibinfo {year}
  {2022})}\BibitemShut {NoStop}%
\bibitem [{\citenamefont {Linke}\ \emph {et~al.}(2017)\citenamefont {Linke},
  \citenamefont {Maslov}, \citenamefont {Roetteler}, \citenamefont {Debnath},
  \citenamefont {Figgatt}, \citenamefont {Landsman}, \citenamefont {Wright},\
  and\ \citenamefont {Monroe}}]{Linke2017}%
  \BibitemOpen
  \bibfield  {author} {\bibinfo {author} {\bibfnamefont {N.~M.}\ \bibnamefont
  {Linke}}, \bibinfo {author} {\bibfnamefont {D.}~\bibnamefont {Maslov}},
  \bibinfo {author} {\bibfnamefont {M.}~\bibnamefont {Roetteler}}, \bibinfo
  {author} {\bibfnamefont {S.}~\bibnamefont {Debnath}}, \bibinfo {author}
  {\bibfnamefont {C.}~\bibnamefont {Figgatt}}, \bibinfo {author} {\bibfnamefont
  {K.~A.}\ \bibnamefont {Landsman}}, \bibinfo {author} {\bibfnamefont
  {K.}~\bibnamefont {Wright}},\ and\ \bibinfo {author} {\bibfnamefont
  {C.}~\bibnamefont {Monroe}},\ }\bibfield  {title} {\bibinfo {title}
  {Experimental comparison of two quantum computing architectures},\
  }\href@noop {} {\bibfield  {journal} {\bibinfo  {journal} {Proceedings of the
  National Academy of Sciences}\ }\textbf {\bibinfo {volume} {114}},\ \bibinfo
  {pages} {3305} (\bibinfo {year} {2017})}\BibitemShut {NoStop}%
\bibitem [{\citenamefont {Preskill}(2018)}]{Preskill_2018}%
  \BibitemOpen
  \bibfield  {author} {\bibinfo {author} {\bibfnamefont {J.}~\bibnamefont
  {Preskill}},\ }\bibfield  {title} {\bibinfo {title} {Quantum computing in the
  nisq era and beyond},\ }\href@noop {} {\bibfield  {journal} {\bibinfo
  {journal} {Quantum}\ }\textbf {\bibinfo {volume} {2}},\ \bibinfo {pages} {79}
  (\bibinfo {year} {2018})}\BibitemShut {NoStop}%
\bibitem [{\citenamefont {Bharti}\ \emph {et~al.}(2022)\citenamefont {Bharti},
  \citenamefont {Cervera-Lierta}, \citenamefont {Kyaw}, \citenamefont {Haug},
  \citenamefont {Alperin-Lea}, \citenamefont {Anand}, \citenamefont {Degroote},
  \citenamefont {Heimonen}, \citenamefont {Kottmann}, \citenamefont {Menke},
  \citenamefont {Mok}, \citenamefont {Sim}, \citenamefont {Kwek},\ and\
  \citenamefont {Aspuru-Guzik}}]{Bharti_2022}%
  \BibitemOpen
  \bibfield  {author} {\bibinfo {author} {\bibfnamefont {K.}~\bibnamefont
  {Bharti}}, \bibinfo {author} {\bibfnamefont {A.}~\bibnamefont
  {Cervera-Lierta}}, \bibinfo {author} {\bibfnamefont {T.~H.}\ \bibnamefont
  {Kyaw}}, \bibinfo {author} {\bibfnamefont {T.}~\bibnamefont {Haug}}, \bibinfo
  {author} {\bibfnamefont {S.}~\bibnamefont {Alperin-Lea}}, \bibinfo {author}
  {\bibfnamefont {A.}~\bibnamefont {Anand}}, \bibinfo {author} {\bibfnamefont
  {M.}~\bibnamefont {Degroote}}, \bibinfo {author} {\bibfnamefont
  {H.}~\bibnamefont {Heimonen}}, \bibinfo {author} {\bibfnamefont {J.~S.}\
  \bibnamefont {Kottmann}}, \bibinfo {author} {\bibfnamefont {T.}~\bibnamefont
  {Menke}}, \bibinfo {author} {\bibfnamefont {W.-K.}\ \bibnamefont {Mok}},
  \bibinfo {author} {\bibfnamefont {S.}~\bibnamefont {Sim}}, \bibinfo {author}
  {\bibfnamefont {L.-C.}\ \bibnamefont {Kwek}},\ and\ \bibinfo {author}
  {\bibfnamefont {A.}~\bibnamefont {Aspuru-Guzik}},\ }\bibfield  {title}
  {\bibinfo {title} {Noisy intermediate-scale quantum algorithms},\ }\bibfield
  {journal} {\bibinfo  {journal} {Reviews of Modern Physics}\ }\textbf
  {\bibinfo {volume} {94}},\ \href
  {https://doi.org/10.1103/revmodphys.94.015004} {10.1103/revmodphys.94.015004}
  (\bibinfo {year} {2022})\BibitemShut {NoStop}%
\bibitem [{\citenamefont {Cerezo}\ \emph {et~al.}(2021)\citenamefont {Cerezo},
  \citenamefont {Arrasmith}, \citenamefont {Babbush}, \citenamefont {Benjamin},
  \citenamefont {Endo}, \citenamefont {Fujii}, \citenamefont {McClean},
  \citenamefont {Mitarai}, \citenamefont {Yuan}, \citenamefont {Cincio},\ and\
  \citenamefont {Coles}}]{Cerezo_2021VQA}%
  \BibitemOpen
  \bibfield  {author} {\bibinfo {author} {\bibfnamefont {M.}~\bibnamefont
  {Cerezo}}, \bibinfo {author} {\bibfnamefont {A.}~\bibnamefont {Arrasmith}},
  \bibinfo {author} {\bibfnamefont {R.}~\bibnamefont {Babbush}}, \bibinfo
  {author} {\bibfnamefont {S.~C.}\ \bibnamefont {Benjamin}}, \bibinfo {author}
  {\bibfnamefont {S.}~\bibnamefont {Endo}}, \bibinfo {author} {\bibfnamefont
  {K.}~\bibnamefont {Fujii}}, \bibinfo {author} {\bibfnamefont {J.~R.}\
  \bibnamefont {McClean}}, \bibinfo {author} {\bibfnamefont {K.}~\bibnamefont
  {Mitarai}}, \bibinfo {author} {\bibfnamefont {X.}~\bibnamefont {Yuan}},
  \bibinfo {author} {\bibfnamefont {L.}~\bibnamefont {Cincio}},\ and\ \bibinfo
  {author} {\bibfnamefont {P.~J.}\ \bibnamefont {Coles}},\ }\bibfield  {title}
  {\bibinfo {title} {Variational quantum algorithms},\ }\href
  {https://doi.org/10.1038/s42254-021-00348-9} {\bibfield  {journal} {\bibinfo
  {journal} {Nature Reviews Physics}\ }\textbf {\bibinfo {volume} {3}},\
  \bibinfo {pages} {625} (\bibinfo {year} {2021})}\BibitemShut {NoStop}%
\bibitem [{\citenamefont {Mangini}\ \emph {et~al.}(2021)\citenamefont
  {Mangini}, \citenamefont {Tacchino}, \citenamefont {Gerace}, \citenamefont
  {Bajoni},\ and\ \citenamefont {Macchiavello}}]{Mangini_2021}%
  \BibitemOpen
  \bibfield  {author} {\bibinfo {author} {\bibfnamefont {S.}~\bibnamefont
  {Mangini}}, \bibinfo {author} {\bibfnamefont {F.}~\bibnamefont {Tacchino}},
  \bibinfo {author} {\bibfnamefont {D.}~\bibnamefont {Gerace}}, \bibinfo
  {author} {\bibfnamefont {D.}~\bibnamefont {Bajoni}},\ and\ \bibinfo {author}
  {\bibfnamefont {C.}~\bibnamefont {Macchiavello}},\ }\bibfield  {title}
  {\bibinfo {title} {Quantum computing models for artificial neural networks},\
  }\href {https://doi.org/10.1209/0295-5075/134/10002} {\bibfield  {journal}
  {\bibinfo  {journal} {Europhysics Letters}\ }\textbf {\bibinfo {volume}
  {134}},\ \bibinfo {pages} {10002} (\bibinfo {year} {2021})}\BibitemShut
  {NoStop}%
\bibitem [{\citenamefont {Schuld}\ \emph {et~al.}(2021)\citenamefont {Schuld},
  \citenamefont {Sweke},\ and\ \citenamefont {Meyer}}]{schuld2021effect}%
  \BibitemOpen
  \bibfield  {author} {\bibinfo {author} {\bibfnamefont {M.}~\bibnamefont
  {Schuld}}, \bibinfo {author} {\bibfnamefont {R.}~\bibnamefont {Sweke}},\ and\
  \bibinfo {author} {\bibfnamefont {J.~J.}\ \bibnamefont {Meyer}},\ }\bibfield
  {title} {\bibinfo {title} {Effect of data encoding on the expressive power of
  variational quantum-machine-learning models},\ }\href
  {https://doi.org/10.1103/PhysRevA.103.032430} {\bibfield  {journal} {\bibinfo
   {journal} {Phys. Rev. A}\ }\textbf {\bibinfo {volume} {103}},\ \bibinfo
  {pages} {032430} (\bibinfo {year} {2021})}\BibitemShut {NoStop}%
\bibitem [{\citenamefont {Abbas}\ \emph {et~al.}(2021)\citenamefont {Abbas},
  \citenamefont {Sutter}, \citenamefont {Zoufal}, \citenamefont {Lucchi},
  \citenamefont {Figalli},\ and\ \citenamefont {Woerner}}]{Abbas_2021}%
  \BibitemOpen
  \bibfield  {author} {\bibinfo {author} {\bibfnamefont {A.}~\bibnamefont
  {Abbas}}, \bibinfo {author} {\bibfnamefont {D.}~\bibnamefont {Sutter}},
  \bibinfo {author} {\bibfnamefont {C.}~\bibnamefont {Zoufal}}, \bibinfo
  {author} {\bibfnamefont {A.}~\bibnamefont {Lucchi}}, \bibinfo {author}
  {\bibfnamefont {A.}~\bibnamefont {Figalli}},\ and\ \bibinfo {author}
  {\bibfnamefont {S.}~\bibnamefont {Woerner}},\ }\bibfield  {title} {\bibinfo
  {title} {The power of quantum neural networks},\ }\href
  {https://doi.org/10.1038/s43588-021-00084-1} {\bibfield  {journal} {\bibinfo
  {journal} {Nat Comput Sci}\ }\textbf {\bibinfo {volume} {1}},\ \bibinfo
  {pages} {403} (\bibinfo {year} {2021})}\BibitemShut {NoStop}%
\bibitem [{\citenamefont {Nguyen}\ \emph {et~al.}(2022)\citenamefont {Nguyen},
  \citenamefont {Schatzki}, \citenamefont {Braccia}, \citenamefont {Ragone},
  \citenamefont {Coles}, \citenamefont {Sauvage}, \citenamefont {Larocca},\
  and\ \citenamefont {Cerezo}}]{Nguyen2022equivariant}%
  \BibitemOpen
  \bibfield  {author} {\bibinfo {author} {\bibfnamefont {Q.~T.}\ \bibnamefont
  {Nguyen}}, \bibinfo {author} {\bibfnamefont {L.}~\bibnamefont {Schatzki}},
  \bibinfo {author} {\bibfnamefont {P.}~\bibnamefont {Braccia}}, \bibinfo
  {author} {\bibfnamefont {M.}~\bibnamefont {Ragone}}, \bibinfo {author}
  {\bibfnamefont {P.~J.}\ \bibnamefont {Coles}}, \bibinfo {author}
  {\bibfnamefont {F.}~\bibnamefont {Sauvage}}, \bibinfo {author} {\bibfnamefont
  {M.}~\bibnamefont {Larocca}},\ and\ \bibinfo {author} {\bibfnamefont
  {M.}~\bibnamefont {Cerezo}},\ }\href
  {https://doi.org/10.48550/ARXIV.2210.08566} {\bibinfo {title} {Theory for
  equivariant quantum neural networks}} (\bibinfo {year} {2022})\BibitemShut
  {NoStop}%
\bibitem [{\citenamefont {Peruzzo}\ \emph {et~al.}(2014)\citenamefont
  {Peruzzo}, \citenamefont {McClean}, \citenamefont {Shadbolt}, \citenamefont
  {Yung}, \citenamefont {Zhou}, \citenamefont {Love}, \citenamefont
  {Aspuru-Guzik},\ and\ \citenamefont {O'Brien}}]{Peruzzo_2014}%
  \BibitemOpen
  \bibfield  {author} {\bibinfo {author} {\bibfnamefont {A.}~\bibnamefont
  {Peruzzo}}, \bibinfo {author} {\bibfnamefont {J.}~\bibnamefont {McClean}},
  \bibinfo {author} {\bibfnamefont {P.}~\bibnamefont {Shadbolt}}, \bibinfo
  {author} {\bibfnamefont {M.-H.}\ \bibnamefont {Yung}}, \bibinfo {author}
  {\bibfnamefont {X.-Q.}\ \bibnamefont {Zhou}}, \bibinfo {author}
  {\bibfnamefont {P.~J.}\ \bibnamefont {Love}}, \bibinfo {author}
  {\bibfnamefont {A.}~\bibnamefont {Aspuru-Guzik}},\ and\ \bibinfo {author}
  {\bibfnamefont {J.~L.}\ \bibnamefont {O'Brien}},\ }\bibfield  {title}
  {\bibinfo {title} {A variational eigenvalue solver on a photonic quantum
  processor},\ }\href@noop {} {\bibfield  {journal} {\bibinfo  {journal} {Nat.
  Commun.}\ }\textbf {\bibinfo {volume} {5}} (\bibinfo {year}
  {2014})}\BibitemShut {NoStop}%
\bibitem [{\citenamefont {Benedetti}\ \emph {et~al.}(2020)\citenamefont
  {Benedetti}, \citenamefont {Lloyd}, \citenamefont {Sack},\ and\ \citenamefont
  {Fiorentini}}]{Benedetti2019}%
  \BibitemOpen
  \bibfield  {author} {\bibinfo {author} {\bibfnamefont {M.}~\bibnamefont
  {Benedetti}}, \bibinfo {author} {\bibfnamefont {E.}~\bibnamefont {Lloyd}},
  \bibinfo {author} {\bibfnamefont {S.}~\bibnamefont {Sack}},\ and\ \bibinfo
  {author} {\bibfnamefont {M.}~\bibnamefont {Fiorentini}},\ }\bibfield  {title}
  {\bibinfo {title} {Parameterized quantum circuits as machine learning
  models},\ }\href@noop {} {\bibfield  {journal} {\bibinfo  {journal} {Quantum
  Sci. Technol.}\ }\textbf {\bibinfo {volume} {5}},\ \bibinfo {pages} {019601}
  (\bibinfo {year} {2020})}\BibitemShut {NoStop}%
\bibitem [{\citenamefont {Tacchino}\ \emph {et~al.}(2021)\citenamefont
  {Tacchino}, \citenamefont {Mangini}, \citenamefont {Barkoutsos},
  \citenamefont {Macchiavello}, \citenamefont {Gerace}, \citenamefont
  {Tavernelli},\ and\ \citenamefont {Bajoni}}]{Tacchino2021variational}%
  \BibitemOpen
  \bibfield  {author} {\bibinfo {author} {\bibfnamefont {F.}~\bibnamefont
  {Tacchino}}, \bibinfo {author} {\bibfnamefont {S.}~\bibnamefont {Mangini}},
  \bibinfo {author} {\bibfnamefont {P.~K.}\ \bibnamefont {Barkoutsos}},
  \bibinfo {author} {\bibfnamefont {C.}~\bibnamefont {Macchiavello}}, \bibinfo
  {author} {\bibfnamefont {D.}~\bibnamefont {Gerace}}, \bibinfo {author}
  {\bibfnamefont {I.}~\bibnamefont {Tavernelli}},\ and\ \bibinfo {author}
  {\bibfnamefont {D.}~\bibnamefont {Bajoni}},\ }\bibfield  {title} {\bibinfo
  {title} {Variational learning for quantum artificial neural networks},\
  }\href@noop {} {\bibfield  {journal} {\bibinfo  {journal} {IEEE Transactions
  on Quantum Engineering}\ }\textbf {\bibinfo {volume} {2}},\ \bibinfo {pages}
  {1} (\bibinfo {year} {2021})}\BibitemShut {NoStop}%
\bibitem [{\citenamefont {Sim}\ \emph {et~al.}(2019)\citenamefont {Sim},
  \citenamefont {Johnson},\ and\ \citenamefont {Aspuru-Guzik}}]{Sim2019}%
  \BibitemOpen
  \bibfield  {author} {\bibinfo {author} {\bibfnamefont {S.}~\bibnamefont
  {Sim}}, \bibinfo {author} {\bibfnamefont {P.~D.}\ \bibnamefont {Johnson}},\
  and\ \bibinfo {author} {\bibfnamefont {A.}~\bibnamefont {Aspuru-Guzik}},\
  }\bibfield  {title} {\bibinfo {title} {Expressibility and entangling
  capability of parameterized quantum circuits for hybrid quantum-classical
  algorithms},\ }\href@noop {} {\bibfield  {journal} {\bibinfo  {journal} {Adv.
  Quantum Technol.}\ }\textbf {\bibinfo {volume} {2}},\ \bibinfo {pages}
  {1900070} (\bibinfo {year} {2019})}\BibitemShut {NoStop}%
\bibitem [{\citenamefont {Hubregtsen}\ \emph {et~al.}(2021)\citenamefont
  {Hubregtsen}, \citenamefont {Pichlmeier}, \citenamefont {Stecher},\ and\
  \citenamefont {Bertels}}]{Hubregtsen2021}%
  \BibitemOpen
  \bibfield  {author} {\bibinfo {author} {\bibfnamefont {T.}~\bibnamefont
  {Hubregtsen}}, \bibinfo {author} {\bibfnamefont {J.}~\bibnamefont
  {Pichlmeier}}, \bibinfo {author} {\bibfnamefont {P.}~\bibnamefont
  {Stecher}},\ and\ \bibinfo {author} {\bibfnamefont {K.}~\bibnamefont
  {Bertels}},\ }\bibfield  {title} {\bibinfo {title} {Evaluation of
  parameterized quantum circuits: on the relation between classification
  accuracy, expressibility, and entangling capability},\ }\href@noop {}
  {\bibfield  {journal} {\bibinfo  {journal} {Quantum Mach. Intell.}\ }\textbf
  {\bibinfo {volume} {3}} (\bibinfo {year} {2021})}\BibitemShut {NoStop}%
\bibitem [{\citenamefont {Ceschini}\ \emph {et~al.}(2022)\citenamefont
  {Ceschini}, \citenamefont {Rosato},\ and\ \citenamefont
  {Panella}}]{ceschini2022hybrid}%
  \BibitemOpen
  \bibfield  {author} {\bibinfo {author} {\bibfnamefont {A.}~\bibnamefont
  {Ceschini}}, \bibinfo {author} {\bibfnamefont {A.}~\bibnamefont {Rosato}},\
  and\ \bibinfo {author} {\bibfnamefont {M.}~\bibnamefont {Panella}},\
  }\bibfield  {title} {\bibinfo {title} {Hybrid quantum-classical recurrent
  neural networks for time series prediction},\ }in\ \href@noop {} {\emph
  {\bibinfo {booktitle} {2022 International Joint Conference on Neural Networks
  (IJCNN)}}}\ (\bibinfo {organization} {IEEE},\ \bibinfo {year} {2022})\ pp.\
  \bibinfo {pages} {1--8}\BibitemShut {NoStop}%
\bibitem [{\citenamefont {Scala}\ \emph {et~al.}(2022)\citenamefont {Scala},
  \citenamefont {Mangini}, \citenamefont {Macchiavello}, \citenamefont
  {Bajoni},\ and\ \citenamefont {Gerace}}]{scala2022quantum}%
  \BibitemOpen
  \bibfield  {author} {\bibinfo {author} {\bibfnamefont {F.}~\bibnamefont
  {Scala}}, \bibinfo {author} {\bibfnamefont {S.}~\bibnamefont {Mangini}},
  \bibinfo {author} {\bibfnamefont {C.}~\bibnamefont {Macchiavello}}, \bibinfo
  {author} {\bibfnamefont {D.}~\bibnamefont {Bajoni}},\ and\ \bibinfo {author}
  {\bibfnamefont {D.}~\bibnamefont {Gerace}},\ }\bibfield  {title} {\bibinfo
  {title} {Quantum variational learning for entanglement witnessing},\ }in\
  \href {https://doi.org/10.1109/IJCNN55064.2022.9892080} {\emph {\bibinfo
  {booktitle} {2022 International Joint Conference on Neural Networks
  (IJCNN)}}}\ (\bibinfo {year} {2022})\ pp.\ \bibinfo {pages}
  {1--8}\BibitemShut {NoStop}%
\bibitem [{\citenamefont {Leone}\ \emph {et~al.}(2022)\citenamefont {Leone},
  \citenamefont {Oliviero}, \citenamefont {Cincio},\ and\ \citenamefont
  {Cerezo}}]{leone2022practical}%
  \BibitemOpen
  \bibfield  {author} {\bibinfo {author} {\bibfnamefont {L.}~\bibnamefont
  {Leone}}, \bibinfo {author} {\bibfnamefont {S.~F.~E.}\ \bibnamefont
  {Oliviero}}, \bibinfo {author} {\bibfnamefont {L.}~\bibnamefont {Cincio}},\
  and\ \bibinfo {author} {\bibfnamefont {M.}~\bibnamefont {Cerezo}},\ }\href
  {https://doi.org/10.48550/ARXIV.2211.01477} {\bibinfo {title} {On the
  practical usefulness of the hardware efficient ansatz}} (\bibinfo {year}
  {2022})\BibitemShut {NoStop}%
\bibitem [{\citenamefont {Ballarin}\ \emph {et~al.}(2023)\citenamefont
  {Ballarin}, \citenamefont {Mangini}, \citenamefont {Montangero},
  \citenamefont {Macchiavello},\ and\ \citenamefont {Mengoni}}]{Ballarin_2023}%
  \BibitemOpen
  \bibfield  {author} {\bibinfo {author} {\bibfnamefont {M.}~\bibnamefont
  {Ballarin}}, \bibinfo {author} {\bibfnamefont {S.}~\bibnamefont {Mangini}},
  \bibinfo {author} {\bibfnamefont {S.}~\bibnamefont {Montangero}}, \bibinfo
  {author} {\bibfnamefont {C.}~\bibnamefont {Macchiavello}},\ and\ \bibinfo
  {author} {\bibfnamefont {R.}~\bibnamefont {Mengoni}},\ }\bibfield  {title}
  {\bibinfo {title} {Entanglement entropy production in quantum neural
  networks},\ }\href {https://doi.org/10.22331/q-2023-05-31-1023} {\bibfield
  {journal} {\bibinfo  {journal} {Quantum}\ }\textbf {\bibinfo {volume} {7}},\
  \bibinfo {pages} {1023} (\bibinfo {year} {2023})}\BibitemShut {NoStop}%
\bibitem [{\citenamefont {Banchi}\ \emph {et~al.}(2021)\citenamefont {Banchi},
  \citenamefont {Pereira},\ and\ \citenamefont
  {Pirandola}}]{banchi2021generalization}%
  \BibitemOpen
  \bibfield  {author} {\bibinfo {author} {\bibfnamefont {L.}~\bibnamefont
  {Banchi}}, \bibinfo {author} {\bibfnamefont {J.}~\bibnamefont {Pereira}},\
  and\ \bibinfo {author} {\bibfnamefont {S.}~\bibnamefont {Pirandola}},\
  }\bibfield  {title} {\bibinfo {title} {Generalization in quantum machine
  learning: A quantum information standpoint},\ }\href
  {https://doi.org/10.1103/PRXQuantum.2.040321} {\bibfield  {journal} {\bibinfo
   {journal} {PRX Quantum}\ }\textbf {\bibinfo {volume} {2}},\ \bibinfo {pages}
  {040321} (\bibinfo {year} {2021})}\BibitemShut {NoStop}%
\bibitem [{\citenamefont {Caro}\ \emph {et~al.}(2021)\citenamefont {Caro},
  \citenamefont {Gil-Fuster}, \citenamefont {Meyer}, \citenamefont {Eisert},\
  and\ \citenamefont {Sweke}}]{caro_encoding-dependent_2021}%
  \BibitemOpen
  \bibfield  {author} {\bibinfo {author} {\bibfnamefont {M.~C.}\ \bibnamefont
  {Caro}}, \bibinfo {author} {\bibfnamefont {E.}~\bibnamefont {Gil-Fuster}},
  \bibinfo {author} {\bibfnamefont {J.~J.}\ \bibnamefont {Meyer}}, \bibinfo
  {author} {\bibfnamefont {J.}~\bibnamefont {Eisert}},\ and\ \bibinfo {author}
  {\bibfnamefont {R.}~\bibnamefont {Sweke}},\ }\bibfield  {title} {\bibinfo
  {title} {Encoding-dependent generalization bounds for parametrized quantum
  circuits},\ }\href {https://doi.org/10.22331/q-2021-11-17-582} {\bibfield
  {journal} {\bibinfo  {journal} {Quantum}\ }\textbf {\bibinfo {volume} {5}},\
  \bibinfo {pages} {582} (\bibinfo {year} {2021})}\BibitemShut {NoStop}%
\bibitem [{\citenamefont {P{\'e}rez-Salinas}\ \emph {et~al.}(2020)\citenamefont
  {P{\'e}rez-Salinas}, \citenamefont {Cervera-Lierta}, \citenamefont
  {Gil-Fuster},\ and\ \citenamefont {Latorre}}]{perez2020data}%
  \BibitemOpen
  \bibfield  {author} {\bibinfo {author} {\bibfnamefont {A.}~\bibnamefont
  {P{\'e}rez-Salinas}}, \bibinfo {author} {\bibfnamefont {A.}~\bibnamefont
  {Cervera-Lierta}}, \bibinfo {author} {\bibfnamefont {E.}~\bibnamefont
  {Gil-Fuster}},\ and\ \bibinfo {author} {\bibfnamefont {J.~I.}\ \bibnamefont
  {Latorre}},\ }\bibfield  {title} {\bibinfo {title} {Data re-uploading for a
  universal quantum classifier},\ }\href@noop {} {\bibfield  {journal}
  {\bibinfo  {journal} {Quantum}\ }\textbf {\bibinfo {volume} {4}},\ \bibinfo
  {pages} {226} (\bibinfo {year} {2020})}\BibitemShut {NoStop}%
\bibitem [{\citenamefont {Haug}\ \emph {et~al.}(2021)\citenamefont {Haug},
  \citenamefont {Bharti},\ and\ \citenamefont {Kim}}]{Haug_2021}%
  \BibitemOpen
  \bibfield  {author} {\bibinfo {author} {\bibfnamefont {T.}~\bibnamefont
  {Haug}}, \bibinfo {author} {\bibfnamefont {K.}~\bibnamefont {Bharti}},\ and\
  \bibinfo {author} {\bibfnamefont {M.}~\bibnamefont {Kim}},\ }\bibfield
  {title} {\bibinfo {title} {Capacity and quantum geometry of parametrized
  quantum circuits},\ }\bibfield  {journal} {\bibinfo  {journal} {{PRX}
  Quantum}\ }\textbf {\bibinfo {volume} {2}},\ \href
  {https://doi.org/10.1103/prxquantum.2.040309} {10.1103/prxquantum.2.040309}
  (\bibinfo {year} {2021})\BibitemShut {NoStop}%
\bibitem [{\citenamefont {Larocca}\ \emph {et~al.}(2023)\citenamefont
  {Larocca}, \citenamefont {Ju}, \citenamefont {Garc{\'{\i}}a-Mart{\'{\i}}n},
  \citenamefont {Coles},\ and\ \citenamefont {Cerezo}}]{Larocca_2023}%
  \BibitemOpen
  \bibfield  {author} {\bibinfo {author} {\bibfnamefont {M.}~\bibnamefont
  {Larocca}}, \bibinfo {author} {\bibfnamefont {N.}~\bibnamefont {Ju}},
  \bibinfo {author} {\bibfnamefont {D.}~\bibnamefont
  {Garc{\'{\i}}a-Mart{\'{\i}}n}}, \bibinfo {author} {\bibfnamefont {P.~J.}\
  \bibnamefont {Coles}},\ and\ \bibinfo {author} {\bibfnamefont
  {M.}~\bibnamefont {Cerezo}},\ }\bibfield  {title} {\bibinfo {title} {Theory
  of overparametrization in quantum neural networks},\ }\href
  {https://doi.org/10.1038/s43588-023-00467-6} {\bibfield  {journal} {\bibinfo
  {journal} {Nat Comput Sci}\ }\textbf {\bibinfo {volume} {3}},\ \bibinfo
  {pages} {542} (\bibinfo {year} {2023})}\BibitemShut {NoStop}%
\bibitem [{\citenamefont {Peters}\ and\ \citenamefont
  {Schuld}(2022)}]{peters_generalization_2022}%
  \BibitemOpen
  \bibfield  {author} {\bibinfo {author} {\bibfnamefont {E.}~\bibnamefont
  {Peters}}\ and\ \bibinfo {author} {\bibfnamefont {M.}~\bibnamefont
  {Schuld}},\ }\href {http://arxiv.org/abs/2209.05523} {\bibinfo {title}
  {Generalization despite overfitting in quantum machine learning models}}
  (\bibinfo {year} {2022}),\ \bibinfo {note} {arXiv:2209.05523
  [quant-ph]}\BibitemShut {NoStop}%
\bibitem [{\citenamefont {García-Martín}\ \emph {et~al.}(2023)\citenamefont
  {García-Martín}, \citenamefont {Larocca},\ and\ \citenamefont
  {Cerezo}}]{garcia2023effects}%
  \BibitemOpen
  \bibfield  {author} {\bibinfo {author} {\bibfnamefont {D.}~\bibnamefont
  {García-Martín}}, \bibinfo {author} {\bibfnamefont {M.}~\bibnamefont
  {Larocca}},\ and\ \bibinfo {author} {\bibfnamefont {M.}~\bibnamefont
  {Cerezo}},\ }\href {https://doi.org/10.48550/ARXIV.2302.05059} {\bibinfo
  {title} {Effects of noise on the overparametrization of quantum neural
  networks}} (\bibinfo {year} {2023})\BibitemShut {NoStop}%
\bibitem [{\citenamefont {Arrasmith}\ \emph {et~al.}(2021)\citenamefont
  {Arrasmith}, \citenamefont {Cerezo}, \citenamefont {Czarnik}, \citenamefont
  {Cincio},\ and\ \citenamefont {Coles}}]{arrasmith2020effect}%
  \BibitemOpen
  \bibfield  {author} {\bibinfo {author} {\bibfnamefont {A.}~\bibnamefont
  {Arrasmith}}, \bibinfo {author} {\bibfnamefont {M.}~\bibnamefont {Cerezo}},
  \bibinfo {author} {\bibfnamefont {P.}~\bibnamefont {Czarnik}}, \bibinfo
  {author} {\bibfnamefont {L.}~\bibnamefont {Cincio}},\ and\ \bibinfo {author}
  {\bibfnamefont {P.~J.}\ \bibnamefont {Coles}},\ }\bibfield  {title} {\bibinfo
  {title} {Effect of barren plateaus on gradient-free optimization},\
  }\href@noop {} {\bibfield  {journal} {\bibinfo  {journal} {{Quantum}}\
  }\textbf {\bibinfo {volume} {5}},\ \bibinfo {pages} {558} (\bibinfo {year}
  {2021})}\BibitemShut {NoStop}%
\bibitem [{\citenamefont {Patti}\ \emph {et~al.}(2021)\citenamefont {Patti},
  \citenamefont {Najafi}, \citenamefont {Gao},\ and\ \citenamefont
  {Yelin}}]{Patti_2021}%
  \BibitemOpen
  \bibfield  {author} {\bibinfo {author} {\bibfnamefont {T.~L.}\ \bibnamefont
  {Patti}}, \bibinfo {author} {\bibfnamefont {K.}~\bibnamefont {Najafi}},
  \bibinfo {author} {\bibfnamefont {X.}~\bibnamefont {Gao}},\ and\ \bibinfo
  {author} {\bibfnamefont {S.~F.}\ \bibnamefont {Yelin}},\ }\bibfield  {title}
  {\bibinfo {title} {Entanglement devised barren plateau mitigation},\
  }\href@noop {} {\bibfield  {journal} {\bibinfo  {journal} {Physical Review
  Research}\ }\textbf {\bibinfo {volume} {3}} (\bibinfo {year}
  {2021})}\BibitemShut {NoStop}%
\bibitem [{\citenamefont {Ortiz~Marrero}\ \emph {et~al.}(2021)\citenamefont
  {Ortiz~Marrero}, \citenamefont {Kieferov\'a},\ and\ \citenamefont
  {Wiebe}}]{marrero2021entanglement}%
  \BibitemOpen
  \bibfield  {author} {\bibinfo {author} {\bibfnamefont {C.}~\bibnamefont
  {Ortiz~Marrero}}, \bibinfo {author} {\bibfnamefont {M.}~\bibnamefont
  {Kieferov\'a}},\ and\ \bibinfo {author} {\bibfnamefont {N.}~\bibnamefont
  {Wiebe}},\ }\bibfield  {title} {\bibinfo {title} {Entanglement-induced barren
  plateaus},\ }\href@noop {} {\bibfield  {journal} {\bibinfo  {journal} {PRX
  Quantum}\ }\textbf {\bibinfo {volume} {2}},\ \bibinfo {pages} {040316}
  (\bibinfo {year} {2021})}\BibitemShut {NoStop}%
\bibitem [{\citenamefont {Holmes}\ \emph {et~al.}(2022)\citenamefont {Holmes},
  \citenamefont {Sharma}, \citenamefont {Cerezo},\ and\ \citenamefont
  {Coles}}]{holmes2022connecting}%
  \BibitemOpen
  \bibfield  {author} {\bibinfo {author} {\bibfnamefont {Z.}~\bibnamefont
  {Holmes}}, \bibinfo {author} {\bibfnamefont {K.}~\bibnamefont {Sharma}},
  \bibinfo {author} {\bibfnamefont {M.}~\bibnamefont {Cerezo}},\ and\ \bibinfo
  {author} {\bibfnamefont {P.~J.}\ \bibnamefont {Coles}},\ }\bibfield  {title}
  {\bibinfo {title} {Connecting ansatz expressibility to gradient magnitudes
  and barren plateaus},\ }\href {https://doi.org/10.1103/PRXQuantum.3.010313}
  {\bibfield  {journal} {\bibinfo  {journal} {PRX Quantum}\ }\textbf {\bibinfo
  {volume} {3}},\ \bibinfo {pages} {010313} (\bibinfo {year}
  {2022})}\BibitemShut {NoStop}%
\bibitem [{\citenamefont {Arrasmith}\ \emph {et~al.}(2022)\citenamefont
  {Arrasmith}, \citenamefont {Holmes}, \citenamefont {Cerezo},\ and\
  \citenamefont {Coles}}]{Arrasmith_2022equivalence}%
  \BibitemOpen
  \bibfield  {author} {\bibinfo {author} {\bibfnamefont {A.}~\bibnamefont
  {Arrasmith}}, \bibinfo {author} {\bibfnamefont {Z.}~\bibnamefont {Holmes}},
  \bibinfo {author} {\bibfnamefont {M.}~\bibnamefont {Cerezo}},\ and\ \bibinfo
  {author} {\bibfnamefont {P.~J.}\ \bibnamefont {Coles}},\ }\bibfield  {title}
  {\bibinfo {title} {Equivalence of quantum barren plateaus to cost
  concentration and narrow gorges},\ }\href
  {https://doi.org/10.1088/2058-9565/ac7d06} {\bibfield  {journal} {\bibinfo
  {journal} {Quantum Science and Technology}\ }\textbf {\bibinfo {volume}
  {7}},\ \bibinfo {pages} {045015} (\bibinfo {year} {2022})}\BibitemShut
  {NoStop}%
\bibitem [{\citenamefont {Thanasilp}\ \emph {et~al.}(2022)\citenamefont
  {Thanasilp}, \citenamefont {Wang}, \citenamefont {Cerezo},\ and\
  \citenamefont {Holmes}}]{thanasilp_exponential_2022}%
  \BibitemOpen
  \bibfield  {author} {\bibinfo {author} {\bibfnamefont {S.}~\bibnamefont
  {Thanasilp}}, \bibinfo {author} {\bibfnamefont {S.}~\bibnamefont {Wang}},
  \bibinfo {author} {\bibfnamefont {M.}~\bibnamefont {Cerezo}},\ and\ \bibinfo
  {author} {\bibfnamefont {Z.}~\bibnamefont {Holmes}},\ }\href
  {http://arxiv.org/abs/2208.11060} {\bibinfo {title} {Exponential
  concentration and untrainability in quantum kernel methods}} (\bibinfo {year}
  {2022}),\ \bibinfo {note} {arXiv:2208.11060 [quant-ph, stat]}\BibitemShut
  {NoStop}%
\bibitem [{\citenamefont {Gil-Fuster}\ \emph {et~al.}(2023)\citenamefont
  {Gil-Fuster}, \citenamefont {Eisert},\ and\ \citenamefont
  {Bravo-Prieto}}]{gilfuster2023understanding}%
  \BibitemOpen
  \bibfield  {author} {\bibinfo {author} {\bibfnamefont {E.}~\bibnamefont
  {Gil-Fuster}}, \bibinfo {author} {\bibfnamefont {J.}~\bibnamefont {Eisert}},\
  and\ \bibinfo {author} {\bibfnamefont {C.}~\bibnamefont {Bravo-Prieto}},\
  }\href@noop {} {\bibinfo {title} {Understanding quantum machine learning also
  requires rethinking generalization}} (\bibinfo {year} {2023}),\ \Eprint
  {https://arxiv.org/abs/2306.13461} {arXiv:2306.13461 [quant-ph]} \BibitemShut
  {NoStop}%
\bibitem [{\citenamefont {Verdon}\ \emph {et~al.}(2018)\citenamefont {Verdon},
  \citenamefont {Pye},\ and\ \citenamefont {Broughton}}]{verdon2018universal}%
  \BibitemOpen
  \bibfield  {author} {\bibinfo {author} {\bibfnamefont {G.}~\bibnamefont
  {Verdon}}, \bibinfo {author} {\bibfnamefont {J.}~\bibnamefont {Pye}},\ and\
  \bibinfo {author} {\bibfnamefont {M.}~\bibnamefont {Broughton}},\ }\href@noop
  {} {\bibinfo {title} {A universal training algorithm for quantum deep
  learning}} (\bibinfo {year} {2018}),\ \Eprint
  {https://arxiv.org/abs/1806.09729} {arXiv:1806.09729 [quant-ph]} \BibitemShut
  {NoStop}%
\bibitem [{\citenamefont {Schuld}\ \emph {et~al.}(2020)\citenamefont {Schuld},
  \citenamefont {Bocharov}, \citenamefont {Svore},\ and\ \citenamefont
  {Wiebe}}]{schuld_circuit-centric_2020}%
  \BibitemOpen
  \bibfield  {author} {\bibinfo {author} {\bibfnamefont {M.}~\bibnamefont
  {Schuld}}, \bibinfo {author} {\bibfnamefont {A.}~\bibnamefont {Bocharov}},
  \bibinfo {author} {\bibfnamefont {K.~M.}\ \bibnamefont {Svore}},\ and\
  \bibinfo {author} {\bibfnamefont {N.}~\bibnamefont {Wiebe}},\ }\bibfield
  {title} {\bibinfo {title} {Circuit-centric quantum classifiers},\ }\href
  {https://doi.org/10.1103/PhysRevA.101.032308} {\bibfield  {journal} {\bibinfo
   {journal} {Physical Review A}\ }\textbf {\bibinfo {volume} {101}},\ \bibinfo
  {pages} {032308} (\bibinfo {year} {2020})}\BibitemShut {NoStop}%
\bibitem [{\citenamefont {Kobayashi}\ \emph {et~al.}(2022)\citenamefont
  {Kobayashi}, \citenamefont {Nakaji},\ and\ \citenamefont
  {Yamamoto}}]{kobayashi_overfitting_2022}%
  \BibitemOpen
  \bibfield  {author} {\bibinfo {author} {\bibfnamefont {M.}~\bibnamefont
  {Kobayashi}}, \bibinfo {author} {\bibfnamefont {K.}~\bibnamefont {Nakaji}},\
  and\ \bibinfo {author} {\bibfnamefont {N.}~\bibnamefont {Yamamoto}},\
  }\bibfield  {title} {\bibinfo {title} {Overfitting in quantum machine
  learning and entangling dropout},\ }\bibfield  {journal} {\bibinfo  {journal}
  {Quantum Mach. Intell.}\ }\textbf {\bibinfo {volume} {4}},\ \href
  {https://doi.org/10.1007/s42484-022-00087-9} {10.1007/s42484-022-00087-9}
  (\bibinfo {year} {2022})\BibitemShut {NoStop}%
\bibitem [{\citenamefont {Wang}\ \emph {et~al.}(2023)\citenamefont {Wang},
  \citenamefont {Zheng}, \citenamefont {Wu},\ and\ \citenamefont
  {Zhang}}]{wang_2023_qdrop}%
  \BibitemOpen
  \bibfield  {author} {\bibinfo {author} {\bibfnamefont {Z.}~\bibnamefont
  {Wang}}, \bibinfo {author} {\bibfnamefont {P.-L.}\ \bibnamefont {Zheng}},
  \bibinfo {author} {\bibfnamefont {B.}~\bibnamefont {Wu}},\ and\ \bibinfo
  {author} {\bibfnamefont {Y.}~\bibnamefont {Zhang}},\ }\bibfield  {title}
  {\bibinfo {title} {Quantum dropout: On and over the hardness of quantum
  approximate optimization algorithm},\ }\href
  {https://doi.org/10.1103/PhysRevResearch.5.023171} {\bibfield  {journal}
  {\bibinfo  {journal} {Phys. Rev. Res.}\ }\textbf {\bibinfo {volume} {5}},\
  \bibinfo {pages} {023171} (\bibinfo {year} {2023})}\BibitemShut {NoStop}%
\bibitem [{\citenamefont {Chen}\ \emph {et~al.}(2022)\citenamefont {Chen},
  \citenamefont {Wei}, \citenamefont {Zhang}, \citenamefont {Yu},\ and\
  \citenamefont {Yoo}}]{Chen_2022_QCNN}%
  \BibitemOpen
  \bibfield  {author} {\bibinfo {author} {\bibfnamefont {S.~Y.-C.}\
  \bibnamefont {Chen}}, \bibinfo {author} {\bibfnamefont {T.-C.}\ \bibnamefont
  {Wei}}, \bibinfo {author} {\bibfnamefont {C.}~\bibnamefont {Zhang}}, \bibinfo
  {author} {\bibfnamefont {H.}~\bibnamefont {Yu}},\ and\ \bibinfo {author}
  {\bibfnamefont {S.}~\bibnamefont {Yoo}},\ }\bibfield  {title} {\bibinfo
  {title} {Quantum convolutional neural networks for high energy physics data
  analysis},\ }\href {https://doi.org/10.1103/PhysRevResearch.4.013231}
  {\bibfield  {journal} {\bibinfo  {journal} {Phys. Rev. Res.}\ }\textbf
  {\bibinfo {volume} {4}},\ \bibinfo {pages} {013231} (\bibinfo {year}
  {2022})}\BibitemShut {NoStop}%
\bibitem [{\citenamefont {Beckey}\ \emph {et~al.}(2021)\citenamefont {Beckey},
  \citenamefont {Gigena}, \citenamefont {Coles},\ and\ \citenamefont
  {Cerezo}}]{beckey2021concentrable}%
  \BibitemOpen
  \bibfield  {author} {\bibinfo {author} {\bibfnamefont {J.~L.}\ \bibnamefont
  {Beckey}}, \bibinfo {author} {\bibfnamefont {N.}~\bibnamefont {Gigena}},
  \bibinfo {author} {\bibfnamefont {P.~J.}\ \bibnamefont {Coles}},\ and\
  \bibinfo {author} {\bibfnamefont {M.}~\bibnamefont {Cerezo}},\ }\bibfield
  {title} {\bibinfo {title} {Computable and operationally meaningful
  multipartite entanglement measures},\ }\href
  {https://doi.org/10.1103/PhysRevLett.127.140501} {\bibfield  {journal}
  {\bibinfo  {journal} {Phys. Rev. Lett.}\ }\textbf {\bibinfo {volume} {127}},\
  \bibinfo {pages} {140501} (\bibinfo {year} {2021})}\BibitemShut {NoStop}%
\bibitem [{\citenamefont {Gühne}\ and\ \citenamefont
  {Tóth}(2009)}]{Guhne_2009ent}%
  \BibitemOpen
  \bibfield  {author} {\bibinfo {author} {\bibfnamefont {O.}~\bibnamefont
  {Gühne}}\ and\ \bibinfo {author} {\bibfnamefont {G.}~\bibnamefont {Tóth}},\
  }\bibfield  {title} {\bibinfo {title} {Entanglement detection},\ }\href@noop
  {} {\bibfield  {journal} {\bibinfo  {journal} {Physics Reports}\ }\textbf
  {\bibinfo {volume} {474}},\ \bibinfo {pages} {1} (\bibinfo {year}
  {2009})}\BibitemShut {NoStop}%
\bibitem [{\citenamefont {Blalock}\ \emph {et~al.}(2020)\citenamefont
  {Blalock}, \citenamefont {Ortiz}, \citenamefont {Frankle},\ and\
  \citenamefont {Guttag}}]{pruning}%
  \BibitemOpen
  \bibfield  {author} {\bibinfo {author} {\bibfnamefont {D.}~\bibnamefont
  {Blalock}}, \bibinfo {author} {\bibfnamefont {J.~J.~G.}\ \bibnamefont
  {Ortiz}}, \bibinfo {author} {\bibfnamefont {J.}~\bibnamefont {Frankle}},\
  and\ \bibinfo {author} {\bibfnamefont {J.}~\bibnamefont {Guttag}},\
  }\href@noop {} {\bibinfo {title} {What is the state of neural network
  pruning?}} (\bibinfo {year} {2020}),\ \Eprint
  {https://arxiv.org/abs/2003.03033} {arXiv:2003.03033 [cs.LG]} \BibitemShut
  {NoStop}%
\bibitem [{\citenamefont {Carslaw}(1921)}]{introductionFourier}%
  \BibitemOpen
  \bibfield  {author} {\bibinfo {author} {\bibfnamefont {H.}~\bibnamefont
  {Carslaw}},\ }\href {https://books.google.it/books?id=JNVAAAAAIAAJ} {\emph
  {\bibinfo {title} {Introduction to the Theory of Fourier's Series and
  Integrals}}},\ \bibinfo {series} {Introduction to the Theory of Fourier's
  Series and Integrals}\ No.\ \bibinfo {number} {v. 1}\ (\bibinfo  {publisher}
  {Macmillan and Company, limited},\ \bibinfo {year} {1921})\BibitemShut
  {NoStop}%
\bibitem [{\citenamefont {Kingma}\ and\ \citenamefont
  {Ba}(2014)}]{kingma2014adam}%
  \BibitemOpen
  \bibfield  {author} {\bibinfo {author} {\bibfnamefont {D.~P.}\ \bibnamefont
  {Kingma}}\ and\ \bibinfo {author} {\bibfnamefont {J.}~\bibnamefont {Ba}},\
  }\bibfield  {title} {\bibinfo {title} {Adam: A method for stochastic
  optimization},\ }\href@noop {} {\bibfield  {journal} {\bibinfo  {journal}
  {arXiv preprint arXiv:1412.6980}\ } (\bibinfo {year} {2014})}\BibitemShut
  {NoStop}%
\bibitem [{\citenamefont {Bergholm}\ \emph {et~al.}(2018)\citenamefont
  {Bergholm}, \citenamefont {Izaac}, \citenamefont {Schuld}, \citenamefont
  {Gogolin}, \citenamefont {Ahmed}, \citenamefont {Ajith}, \citenamefont
  {Alam}, \citenamefont {Alonso-Linaje}, \citenamefont {AkashNarayanan},
  \citenamefont {Asadi}, \citenamefont {Arrazola}, \citenamefont {Azad},
  \citenamefont {Banning}, \citenamefont {Blank}, \citenamefont {Bromley},
  \citenamefont {Cordier}, \citenamefont {Ceroni}, \citenamefont {Delgado},
  \citenamefont {Di~Matteo}, \citenamefont {Dusko}, \citenamefont {Garg},
  \citenamefont {Guala}, \citenamefont {Hayes}, \citenamefont {Hill},
  \citenamefont {Ijaz}, \citenamefont {Isacsson}, \citenamefont {Ittah},
  \citenamefont {Jahangiri}, \citenamefont {Jain}, \citenamefont {Jiang},
  \citenamefont {Khandelwal}, \citenamefont {Kottmann}, \citenamefont {Lang},
  \citenamefont {Lee}, \citenamefont {Loke}, \citenamefont {Lowe},
  \citenamefont {McKiernan}, \citenamefont {Meyer}, \citenamefont
  {Montañez-Barrera}, \citenamefont {Moyard}, \citenamefont {Niu},
  \citenamefont {O'Riordan}, \citenamefont {Oud}, \citenamefont {Panigrahi},
  \citenamefont {Park}, \citenamefont {Polatajko}, \citenamefont {Quesada},
  \citenamefont {Roberts}, \citenamefont {Sá}, \citenamefont {Schoch},
  \citenamefont {Shi}, \citenamefont {Shu}, \citenamefont {Sim}, \citenamefont
  {Singh}, \citenamefont {Strandberg}, \citenamefont {Soni}, \citenamefont
  {Száva}, \citenamefont {Thabet}, \citenamefont {Vargas-Hernández},
  \citenamefont {Vincent}, \citenamefont {Vitucci}, \citenamefont {Weber},
  \citenamefont {Wierichs}, \citenamefont {Wiersema}, \citenamefont {Willmann},
  \citenamefont {Wong}, \citenamefont {Zhang},\ and\ \citenamefont
  {Killoran}}]{pennylane}%
  \BibitemOpen
  \bibfield  {author} {\bibinfo {author} {\bibfnamefont {V.}~\bibnamefont
  {Bergholm}}, \bibinfo {author} {\bibfnamefont {J.}~\bibnamefont {Izaac}},
  \bibinfo {author} {\bibfnamefont {M.}~\bibnamefont {Schuld}}, \bibinfo
  {author} {\bibfnamefont {C.}~\bibnamefont {Gogolin}}, \bibinfo {author}
  {\bibfnamefont {S.}~\bibnamefont {Ahmed}}, \bibinfo {author} {\bibfnamefont
  {V.}~\bibnamefont {Ajith}}, \bibinfo {author} {\bibfnamefont {M.~S.}\
  \bibnamefont {Alam}}, \bibinfo {author} {\bibfnamefont {G.}~\bibnamefont
  {Alonso-Linaje}}, \bibinfo {author} {\bibfnamefont {B.}~\bibnamefont
  {AkashNarayanan}}, \bibinfo {author} {\bibfnamefont {A.}~\bibnamefont
  {Asadi}}, \bibinfo {author} {\bibfnamefont {J.~M.}\ \bibnamefont {Arrazola}},
  \bibinfo {author} {\bibfnamefont {U.}~\bibnamefont {Azad}}, \bibinfo {author}
  {\bibfnamefont {S.}~\bibnamefont {Banning}}, \bibinfo {author} {\bibfnamefont
  {C.}~\bibnamefont {Blank}}, \bibinfo {author} {\bibfnamefont {T.~R.}\
  \bibnamefont {Bromley}}, \bibinfo {author} {\bibfnamefont {B.~A.}\
  \bibnamefont {Cordier}}, \bibinfo {author} {\bibfnamefont {J.}~\bibnamefont
  {Ceroni}}, \bibinfo {author} {\bibfnamefont {A.}~\bibnamefont {Delgado}},
  \bibinfo {author} {\bibfnamefont {O.}~\bibnamefont {Di~Matteo}}, \bibinfo
  {author} {\bibfnamefont {A.}~\bibnamefont {Dusko}}, \bibinfo {author}
  {\bibfnamefont {T.}~\bibnamefont {Garg}}, \bibinfo {author} {\bibfnamefont
  {D.}~\bibnamefont {Guala}}, \bibinfo {author} {\bibfnamefont
  {A.}~\bibnamefont {Hayes}}, \bibinfo {author} {\bibfnamefont
  {R.}~\bibnamefont {Hill}}, \bibinfo {author} {\bibfnamefont {A.}~\bibnamefont
  {Ijaz}}, \bibinfo {author} {\bibfnamefont {T.}~\bibnamefont {Isacsson}},
  \bibinfo {author} {\bibfnamefont {D.}~\bibnamefont {Ittah}}, \bibinfo
  {author} {\bibfnamefont {S.}~\bibnamefont {Jahangiri}}, \bibinfo {author}
  {\bibfnamefont {P.}~\bibnamefont {Jain}}, \bibinfo {author} {\bibfnamefont
  {E.}~\bibnamefont {Jiang}}, \bibinfo {author} {\bibfnamefont
  {A.}~\bibnamefont {Khandelwal}}, \bibinfo {author} {\bibfnamefont
  {K.}~\bibnamefont {Kottmann}}, \bibinfo {author} {\bibfnamefont {R.~A.}\
  \bibnamefont {Lang}}, \bibinfo {author} {\bibfnamefont {C.}~\bibnamefont
  {Lee}}, \bibinfo {author} {\bibfnamefont {T.}~\bibnamefont {Loke}}, \bibinfo
  {author} {\bibfnamefont {A.}~\bibnamefont {Lowe}}, \bibinfo {author}
  {\bibfnamefont {K.}~\bibnamefont {McKiernan}}, \bibinfo {author}
  {\bibfnamefont {J.~J.}\ \bibnamefont {Meyer}}, \bibinfo {author}
  {\bibfnamefont {J.~A.}\ \bibnamefont {Montañez-Barrera}}, \bibinfo {author}
  {\bibfnamefont {R.}~\bibnamefont {Moyard}}, \bibinfo {author} {\bibfnamefont
  {Z.}~\bibnamefont {Niu}}, \bibinfo {author} {\bibfnamefont {L.~J.}\
  \bibnamefont {O'Riordan}}, \bibinfo {author} {\bibfnamefont {S.}~\bibnamefont
  {Oud}}, \bibinfo {author} {\bibfnamefont {A.}~\bibnamefont {Panigrahi}},
  \bibinfo {author} {\bibfnamefont {C.-Y.}\ \bibnamefont {Park}}, \bibinfo
  {author} {\bibfnamefont {D.}~\bibnamefont {Polatajko}}, \bibinfo {author}
  {\bibfnamefont {N.}~\bibnamefont {Quesada}}, \bibinfo {author} {\bibfnamefont
  {C.}~\bibnamefont {Roberts}}, \bibinfo {author} {\bibfnamefont
  {N.}~\bibnamefont {Sá}}, \bibinfo {author} {\bibfnamefont {I.}~\bibnamefont
  {Schoch}}, \bibinfo {author} {\bibfnamefont {B.}~\bibnamefont {Shi}},
  \bibinfo {author} {\bibfnamefont {S.}~\bibnamefont {Shu}}, \bibinfo {author}
  {\bibfnamefont {S.}~\bibnamefont {Sim}}, \bibinfo {author} {\bibfnamefont
  {A.}~\bibnamefont {Singh}}, \bibinfo {author} {\bibfnamefont
  {I.}~\bibnamefont {Strandberg}}, \bibinfo {author} {\bibfnamefont
  {J.}~\bibnamefont {Soni}}, \bibinfo {author} {\bibfnamefont {A.}~\bibnamefont
  {Száva}}, \bibinfo {author} {\bibfnamefont {S.}~\bibnamefont {Thabet}},
  \bibinfo {author} {\bibfnamefont {R.~A.}\ \bibnamefont {Vargas-Hernández}},
  \bibinfo {author} {\bibfnamefont {T.}~\bibnamefont {Vincent}}, \bibinfo
  {author} {\bibfnamefont {N.}~\bibnamefont {Vitucci}}, \bibinfo {author}
  {\bibfnamefont {M.}~\bibnamefont {Weber}}, \bibinfo {author} {\bibfnamefont
  {D.}~\bibnamefont {Wierichs}}, \bibinfo {author} {\bibfnamefont
  {R.}~\bibnamefont {Wiersema}}, \bibinfo {author} {\bibfnamefont
  {M.}~\bibnamefont {Willmann}}, \bibinfo {author} {\bibfnamefont
  {V.}~\bibnamefont {Wong}}, \bibinfo {author} {\bibfnamefont {S.}~\bibnamefont
  {Zhang}},\ and\ \bibinfo {author} {\bibfnamefont {N.}~\bibnamefont
  {Killoran}},\ }\href {https://doi.org/10.48550/ARXIV.1811.04968} {\bibinfo
  {title} {Pennylane: Automatic differentiation of hybrid quantum-classical
  computations}} (\bibinfo {year} {2018})\BibitemShut {NoStop}%
\bibitem [{\citenamefont {Bradbury}\ \emph {et~al.}(2018)\citenamefont
  {Bradbury}, \citenamefont {Frostig}, \citenamefont {Hawkins}, \citenamefont
  {Johnson}, \citenamefont {Leary}, \citenamefont {Maclaurin}, \citenamefont
  {Necula}, \citenamefont {Paszke}, \citenamefont {Vander{P}las}, \citenamefont
  {Wanderman-{M}ilne},\ and\ \citenamefont {Zhang}}]{jax2018github}%
  \BibitemOpen
  \bibfield  {author} {\bibinfo {author} {\bibfnamefont {J.}~\bibnamefont
  {Bradbury}}, \bibinfo {author} {\bibfnamefont {R.}~\bibnamefont {Frostig}},
  \bibinfo {author} {\bibfnamefont {P.}~\bibnamefont {Hawkins}}, \bibinfo
  {author} {\bibfnamefont {M.~J.}\ \bibnamefont {Johnson}}, \bibinfo {author}
  {\bibfnamefont {C.}~\bibnamefont {Leary}}, \bibinfo {author} {\bibfnamefont
  {D.}~\bibnamefont {Maclaurin}}, \bibinfo {author} {\bibfnamefont
  {G.}~\bibnamefont {Necula}}, \bibinfo {author} {\bibfnamefont
  {A.}~\bibnamefont {Paszke}}, \bibinfo {author} {\bibfnamefont
  {J.}~\bibnamefont {Vander{P}las}}, \bibinfo {author} {\bibfnamefont
  {S.}~\bibnamefont {Wanderman-{M}ilne}},\ and\ \bibinfo {author}
  {\bibfnamefont {Q.}~\bibnamefont {Zhang}},\ }\href
  {http://github.com/google/jax} {\bibinfo {title} {{JAX}: composable
  transformations of {P}ython+{N}um{P}y programs}} (\bibinfo {year}
  {2018})\BibitemShut {NoStop}%
\bibitem [{\citenamefont {Yamamoto}(2019)}]{Yamamoto2019}%
  \BibitemOpen
  \bibfield  {author} {\bibinfo {author} {\bibfnamefont {N.}~\bibnamefont
  {Yamamoto}},\ }\href {https://doi.org/10.48550/ARXIV.1909.05074} {\bibinfo
  {title} {On the natural gradient for variational quantum eigensolver}}
  (\bibinfo {year} {2019})\BibitemShut {NoStop}%
\bibitem [{\citenamefont {Stokes}\ \emph {et~al.}(2020)\citenamefont {Stokes},
  \citenamefont {Izaac}, \citenamefont {Killoran},\ and\ \citenamefont
  {Carleo}}]{Stokes_2020}%
  \BibitemOpen
  \bibfield  {author} {\bibinfo {author} {\bibfnamefont {J.}~\bibnamefont
  {Stokes}}, \bibinfo {author} {\bibfnamefont {J.}~\bibnamefont {Izaac}},
  \bibinfo {author} {\bibfnamefont {N.}~\bibnamefont {Killoran}},\ and\
  \bibinfo {author} {\bibfnamefont {G.}~\bibnamefont {Carleo}},\ }\bibfield
  {title} {\bibinfo {title} {Quantum natural gradient},\ }\href
  {https://doi.org/10.22331/q-2020-05-25-269} {\bibfield  {journal} {\bibinfo
  {journal} {Quantum}\ }\textbf {\bibinfo {volume} {4}},\ \bibinfo {pages}
  {269} (\bibinfo {year} {2020})}\BibitemShut {NoStop}%
\bibitem [{\citenamefont {Schatzki}\ \emph {et~al.}(2022)\citenamefont
  {Schatzki}, \citenamefont {Liu}, \citenamefont {Cerezo},\ and\ \citenamefont
  {Chitambar}}]{Schatzki2022hierarchy}%
  \BibitemOpen
  \bibfield  {author} {\bibinfo {author} {\bibfnamefont {L.}~\bibnamefont
  {Schatzki}}, \bibinfo {author} {\bibfnamefont {G.}~\bibnamefont {Liu}},
  \bibinfo {author} {\bibfnamefont {M.}~\bibnamefont {Cerezo}},\ and\ \bibinfo
  {author} {\bibfnamefont {E.}~\bibnamefont {Chitambar}},\ }\href
  {https://doi.org/10.48550/ARXIV.2209.07607} {\bibinfo {title} {A hierarchy of
  multipartite correlations based on concentratable entanglement}} (\bibinfo
  {year} {2022})\BibitemShut {NoStop}%
\end{thebibliography}%

\clearpage

\appendix

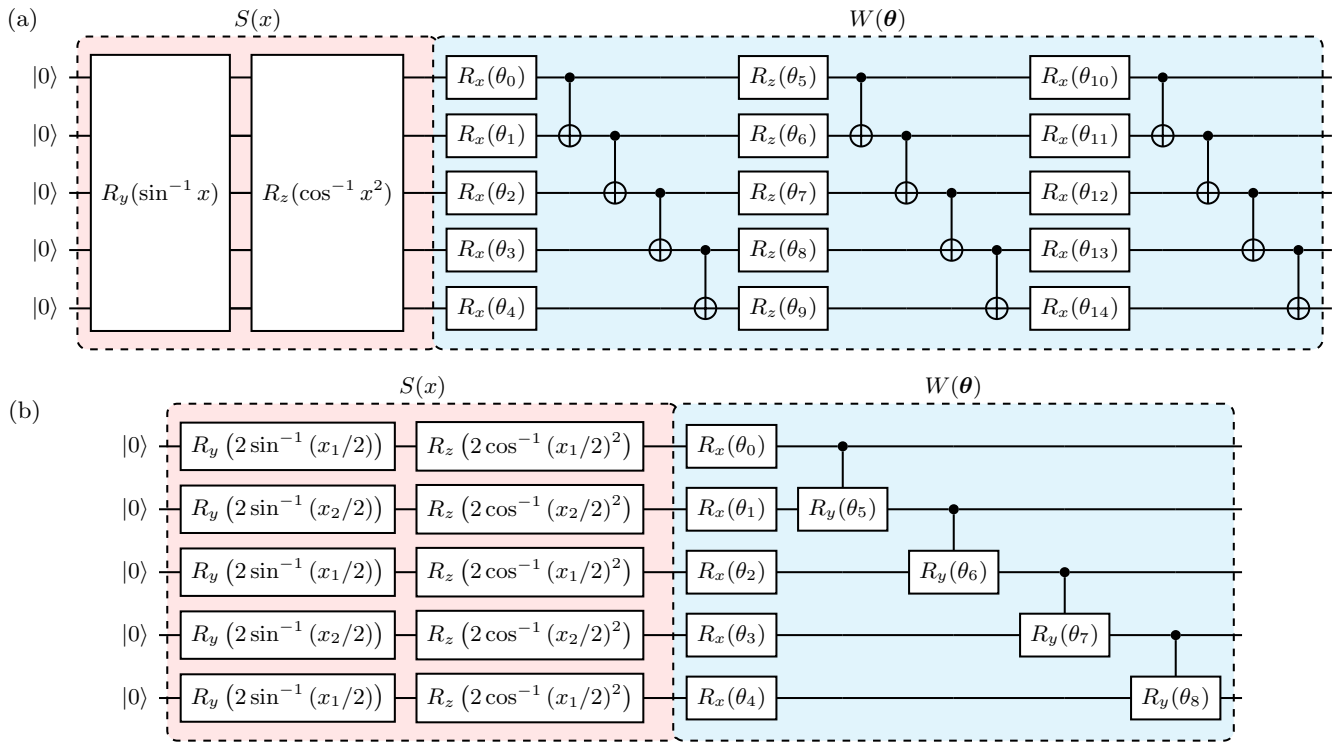
\begin{figure*}
\centering
    \begin{tikzpicture}
    \node[scale=.95]{
    \begin{quantikz}[row sep=0.2cm, column sep=0.3cm]
            \lstick{$\ket{0}$} & \gate[5]{\text{$R_y(\sin^{-1}x)$}}\gategroup[5,steps=3,style={dashed,rounded corners,fill=red!10, inner xsep=2pt}, background]{{\sc $S(x)$}}&\gate[5]{\text{$R_z(\cos^{-1}x^2)$}}&\qw & \gate[1]{\text{$R_x(\theta_0)$}}\gategroup[5,steps=15,style={dashed,rounded corners,fill=cyan!10, inner xsep=2pt}, background]{{\sc $W(\boldsymbol{\theta})$}} & \ctrl{1}&\qw &\qw &\qw & \gate[1]{\text{$R_z(\theta_5)$}} & \ctrl{1}&\qw&\qw &\qw & \gate[1]{\text{$R_x(\theta_{10})$}}& \ctrl{1}&\qw&\qw&\qw&\qw\\
            \lstick{$\ket{0}$} &&\qw&\qw & \gate[1]{\text{$R_x(\theta_1)$}}&  \targ{}  & \ctrl{1} &\qw &\qw& \gate[1]{\text{$R_z(\theta_6)$}}&  \targ{}  & \ctrl{1} &\qw &\qw & \gate[1]{\text{$R_x(\theta_{11})$}} & \targ{}  & \ctrl{1} &\qw&\qw&\qw\\
            \lstick{$\ket{0}$} &&\qw&\qw & \gate[1]{\text{$R_x(\theta_2)$}}& \qw &  \targ{}  & \ctrl{1}  &\qw & \gate[1]{\text{$R_z(\theta_7)$}}& \qw&  \targ{}  & \ctrl{1} &\qw & \gate[1]{\text{$R_x(\theta_{12})$}} &\qw & \targ{}  & \ctrl{1} &\qw&\qw\\
            \lstick{$\ket{0}$} &&\qw&\qw & \gate[1]{\text{$R_x(\theta_3)$}}&\qw&\qw &   \targ{}  & \ctrl{1} & \gate[1]{\text{$R_z(\theta_8)$}}&\qw&\qw&  \targ{}  & \ctrl{1}  & \gate[1]{\text{$R_x(\theta_{13})$}} &\qw&\qw& \targ{}  & \ctrl{1} &\qw\\
            \lstick{$\ket{0}$} &&\qw&\qw & \gate[1]{\text{$R_x(\theta_4)$}}&\qw&\qw&\qw&  \targ{}   & \gate[1]{\text{$R_z(\theta_9)$}}&\qw&\qw&\qw&  \targ{}   & \gate[1]{\text{$R_x(\theta_{14})$}} &\qw&\qw&\qw& \targ{}  &\qw
        \end{quantikz}};
        \end{tikzpicture}
        \shiftleft{17.75cm}{\raisebox{4.4cm}[0cm][0cm]{(a)}}
        \begin{tikzpicture}
    \node[scale=.95]{
    \begin{quantikz}[row sep=0.2cm, column sep=0.3cm]
            \lstick{$\ket{0}$} & \gate[1]{\text{$R_y\left(2\sin^{-1}\left(x_1 / 2\right)\right)$}}\gategroup[5,steps=3,style={dashed,rounded corners,fill=red!10, inner xsep=2pt}, background]{{\sc $S(x)$}}&\gate[1]{\text{$R_z\left(2\cos^{-1}\left(x_1 / 2\right)^2\right)$}}&\qw & \gate[1]{\text{$R_x(\theta_0)$}}\gategroup[5,steps=5,style={dashed,rounded corners,fill=cyan!10, inner xsep=2pt}, background]{{\sc $W(\boldsymbol{\theta})$}} & \ctrl{1}&\qw &\qw &\qw &\qw\\
            \lstick{$\ket{0}$} &\gate[1]{\text{$R_y\left(2\sin^{-1}\left(x_2 / 2\right)\right)$}}&\gate[1]{\text{$R_z\left(2\cos^{-1}\left(x_2 / 2\right)^2\right)$}}&\qw & \gate[1]{\text{$R_x(\theta_1)$}}&  \gate[1]{\text{$R_y(\theta_5)$}}  & \ctrl{1} &\qw &\qw &\qw\\
            \lstick{$\ket{0}$} &\gate[1]{\text{$R_y\left(2\sin^{-1}\left(x_1 / 2\right)\right)$}}&\gate[1]{\text{$R_z\left(2\cos^{-1}\left(x_1 / 2\right)^2\right)$}}&\qw & \gate[1]{\text{$R_x(\theta_2)$}}& \qw &  \gate[1]{\text{$R_y(\theta_6)$}}  & \ctrl{1}  &\qw & \qw\\
            \lstick{$\ket{0}$} &\gate[1]{\text{$R_y\left(2\sin^{-1}\left(x_2 / 2\right)\right)$}}&\gate[1]{\text{$R_z\left(2\cos^{-1}\left(x_2 / 2\right)^2\right)$}}&\qw & \gate[1]{\text{$R_x(\theta_3)$}}&\qw&\qw &   \gate[1]{\text{$R_y(\theta_7)$}}  & \ctrl{1} & \qw\\
            \lstick{$\ket{0}$} &\gate[1]{\text{$R_y\left(2\sin^{-1}\left(x_1 / 2\right)\right)$}}&\gate[1]{\text{$R_z\left(2\cos^{-1}\left(x_1 / 2\right)^2\right)$}}&\qw & \gate[1]{\text{$R_x(\theta_4)$}}&\qw&\qw&\qw&  \gate[1]{\text{$R_y(\theta_8)$}}   & \qw
        \end{quantikz}};
        \end{tikzpicture}
        \shiftleft{16.55cm}{\raisebox{4.4cm}[0cm][0cm]{(b)}}
\caption{One layer of the QNN models employed in this work for regression \textbf{(a)}, taken from~\cite{kobayashi_overfitting_2022}, and classification \textbf{(b)}. $S(x)$ encodes the data while $W(\boldsymbol{\theta})$ is the variational part of the circuit. This structure can be repeated multiple times in a data-reuploading fashion.}
\label{fig:QNN circuit}
\end{figure*}
\section{QNNs circuits}
\label{appendix:QNN circs}
Both QNN architectures employed in this work are built exploiting the data re-uploading scheme, explained in Sec.~\ref{sec:intro} of the main text, in which we alternate embedding circuits for the classical data and parametrized blocks.

\subsection{Regression QNN}
The first QNN is the one described in~\cite{kobayashi_overfitting_2022} and one of its layers is sketched in Fig.~\ref{fig:QNN circuit}a. The embedding consists of two layers of rotations where the classical features are transformed into angles by means of inverse trigonometric functions. These functions are applied to the classical input to provide the QNN with additional nonlinearity. The first layer is composed of $R_y$ rotations with $\sin^{-1}x$ function, while the second has $R_z$ gates with $\cos^{-1}x^2$. If we have one-dimensional features all the rotations are of course embedding the same feature, if the features are more than one then the embedding is alternate. A similar strategy is explicitly shown in the embedding of Fig.~\ref{fig:QNN circuit}b.

The parametrized block is composed of 3 sublayers of $R_x$, $R_z$, $R_x$ rotations followed by CNOTs linearly connecting all the qubits.

The full QNN is made of 10 such defined layers, implying 150 trainable parameters, and is employed in the regression tasks.

\subsection{Classification QNN}
The second QNN structure is similar to the first one, as illustrated in Fig.~\ref{fig:QNN circuit}b. The embedding only differs in the fact that we are dividing the value of the feature by a factor 2 in order to avoid non-admissible values for $\arcsin$ and $\arccos$ when rescaling test data. The parametrized block has only one sublayer with $R_x$ rotations followed by C$R_y$ linearly connecting all the qubits. The parametrized two-qubit gates are the main difference between the two architectures.

The full QNN is made of 20 such defined layers, implying 180 trainable parameters, and is utilised for the classification task.

\section{Overparametrization}
\label{appendix: overparam}
In this section, we provide useful definitions linked to the overparametrization of QNNs and we then verify that we are working in the overparametrized regime with both the architectures presented above.

In~\cite{Haug_2021}, the authors measure the capabilities of a PQC in terms of the number of independent parameters needed to describe a state of $N$ qubits, i.e. the \emph{parameter dimension} 

\begin{equation}
\label{eq:param dimension}
   D \leq D_{max}= 2^{N+1}-2\,,
\end{equation}
where the upper bound $D_{max}$ represents the number of degrees of freedom of the quantum state.
From this quantity, one can then calculate the \emph{redundancy}, which is the fraction of dependent parameters of the PQC that do not contribute to changing the quantum state:
\begin{equation}
\label{eq:redundancy}
    R=\frac{M-D}{M}\,,
\end{equation}
where $M$ is the total number of parameters in the QNN. Of course, $R\neq 0$ only if $M>D$. In this context, a model is said to be overparameterized when it has more parameters than degrees of
freedom. Another measure introduced in the same work is the \emph{effective quantum dimension}\footnote{Not to be confused with the use effective dimension, defined from classical Fisher information matrix, employed in~\cite{Abbas_2021}.} 
\begin{equation}
\label{eq: eff quantum dim}
    G(\boldsymbol{\theta})= \mathrm{rank}(\mathcal{F}(\boldsymbol{\theta}))\,,
\end{equation}
where $\mathcal{F}(\boldsymbol{\theta})$ is the Quantum Fisher Information Matrix (QFIM)~\cite{Yamamoto2019, Stokes_2020} that we will define here below. 

It can be that a certain QNN has $R>0$ ($D \leq M$) even when $M<D_{max}$, depending on the choice of the embedding circuit and of the variational ansatz. A simple example can be found in the QNN of~\cite{kobayashi_overfitting_2022} (showed in Fig.~\ref{fig:QNN circuit}a and employed for the regressions task of this work), if one takes an input equal to $\pm 1$ the first sub-layer of $R_x$ rotations acts trivially on the prepared state and, consequently, the corresponding parameters are useless, implying $D<M$ even if $M<2^{N+1}-2$. The proof is simple given, considering that
\begin{equation}
    \sin^{-1}(\pm1)=\pm \frac{\pi}{2} \ \text{and} \ \cos^{-1}(\pm1)=0\,,
\end{equation}
which means that the only operation we are applying is $R_y(\pm \pi/2)$ and hence, each qubit state is mapped on the x-axis of the Bloch sphere and therefore is invariant under $R_x$ operations. This in the end implies that we have $N$ useless parameters and then the parameter dimension is equal to $D=M-N<M$. 

\begin{figure}
    \centering
    \includegraphics[trim={0.cm 0.2cm 0cm 1cm},clip,width=0.49\textwidth]{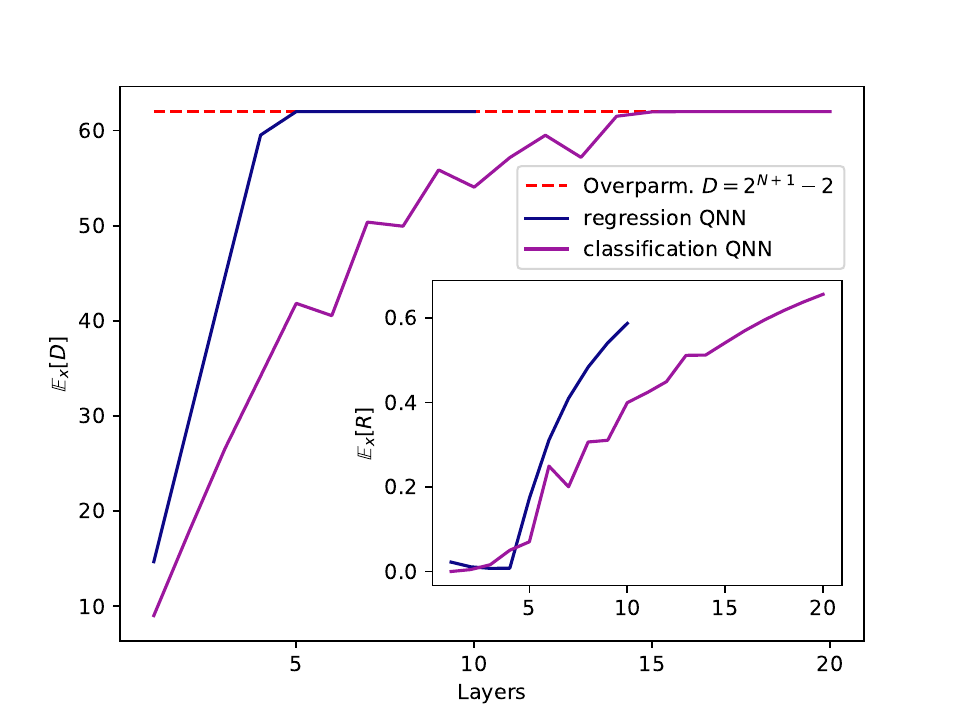}
    \caption{\textbf{Overparametrization} The average parameter dimension is constant and equal to $2^{N+1}-2$ after 5 layers. Inset: after the same threshold the average redundancy increases monotonically towards 1.}
    \label{fig:overparam}
\end{figure}

Another work~\cite{Larocca_2023} discusses the theory of overparametrization, labelling a QNN as overparametrized when, for at least one parameter vector $\boldsymbol{\theta}$, the maximum achievable rank of the QFIM is saturated for \emph{all} the $x$ in the training set.

\begin{definition}[Overparametrization]
\label{def:overparam}
A QNN is said to be overparametrized if the number of parameters $M$ is such that the QFIMs, for all the $x$ in the training set, simultaneously saturate their achievable rank $A_x$ for at least one parameter vector $\boldsymbol{\theta}$. That is, if increasing the number of parameters past some minimal (critical) value $M_c$ does not further increase the rank of any QFIM:
    \begin{equation}
    \label{eq:overparametrization}
        \max_{M\ge M_c, \boldsymbol{\theta}} \emph{rank}[\mathcal{F}_x(\boldsymbol{\theta})]=A_x\,,
    \end{equation}
\end{definition}

The Quantum Fisher Information Matrix for the training sample $x$ is the $M\times M$ square matrix:

\begin{align}
\label{eq:QFIM}
    [\mathcal{F}_x(\boldsymbol{\theta})]_{ij}=4\text{Re}[&\langle \partial_i\psi(x,\boldsymbol{\theta})\ket{\partial_j\psi(x,\boldsymbol{\theta})}\\
    &-\langle\partial_i\psi(x,\boldsymbol{\theta})\ket{\psi(x,\boldsymbol{\theta})} \langle\psi(x,\boldsymbol{\theta})\ket{\partial_j\psi(x,\boldsymbol{\theta})}] 
    \nonumber
\end{align}
where $\ket{\partial_i\psi(x,\boldsymbol{\theta})}=\partial\ket{\psi(x,\boldsymbol{\theta})}/\partial \theta_i$. Note that this definition of overparametrization only depends on features of the QNN while it is independent of the particular loss function employed and of the given problem. One can find more details about overparametrization in QML in~\cite{Larocca_2023}.

In addition, the following relations hold~\cite{Haug_2021}:
\begin{equation}
\label{eq:param dim bounds}
    G(\boldsymbol{\theta}) \leq D \leq M  \,, \quad G(\boldsymbol{\theta}_{rand}) \approx D\,,
\end{equation}
and considering eqs.~\eqref{eq:redundancy},\eqref{eq: eff quantum dim} together with the Def.~\ref{def:overparam}, it follows that a QNN can be considered overparametrized when 
\begin{equation}
    \label{eq:overparametriztion}
    \mathbb{E}_x[R]>0 \implies \mathbb{E}_x[\mathrm{rank}[\mathcal{F}_x(\boldsymbol{\theta}_{rand})]  ] < M \,,
\end{equation}
where we average over all the training data (labelled with $x$).

In Fig.~\ref{fig:overparam} we show overparametrized parameter dimension threshold for $N=5$ qubits ($D_{max}=62$) and the trends of the average parameter dimension, estimated as $\mathbb{E}_x[G(\boldsymbol{\theta}_{rand})] \approx\mathbb{E}_x[D]$ for both the QNN architecture utilised in this work. The inset reports the average redundancy $\mathbb{E}_x[R]$. After 4 layers ($M=60$) the regression QNN starts to be overparametrized, this is highlighted by the presence of a plateau in the parameter dimension and by increasing redundancy. As a matter of fact, if the parameter dimension grows linearly as the number of layers this implies that (on average) all the parameters are needed to describe a quantum state of $N=5$ qubits and this is reflected by approximately 0 redundancy (it is not exactly 0 because of the issue with the encoding of some data explained above). On the opposite, a plateau in the parameter dimension witnesses the fact that the QNN is fully capable of representing every possible quantum state on $N$ qubits and as a consequence adding more trainable parameters increases the redundancy monotonically towards 1, meaning that we are in the overparametrized regime. To exit, on average, the overparametrized regime with the regression QNN with 10 layers ($M=150$) one should remove at least $M - D_{max}= 88$ parametrized gates that are approximately $59\%$ of the total number of parametrized gates $M$. The behaviour of the classification QNN is slightly different, in fact as we increase the number of layers the redundancy grows, even though the QNN is not capable of representing all the states in the Hilbert space yet, i.e. it is not in the overparametrized regime. It begins to be overparametrized from 16 layers ($M=144$) on implying an average redundancy of $\mathbb{E}_x[R]\approx0.57$ equivalent to $M^{red}=82$ redundant parameters on average. We can estimate what is the maximal number of parametrized gates that one can remove without escaping overparametrization when we have 20 layers  ($M=180$): $M-D_{max}=118$ which is about 66\% of the total number of parametrized gates $M$. Nevertheless, it is shown in Appendix~\ref{appendix:expressib}~and~\ref{appendix: multip ent} that a dropout probability of 20\% is already sufficient to affect the expressibility and the entanglement produced by this QNN. This may be related to the redundancy level already present in the QNN before entering the overparametrized regime or the fact that this QNN has parametrized entangling gates.

Even if in this work we refer to the purely quantum concept of overparametrization that we just explained, before concluding the section it is worth mentioning that another definition of overparametrization can be used in QML. Viewing a QNN as a Fourier series, one can say that it is overparametrized when the degree of the Fourier series is much larger than the degree of the target function, in which case the model has many more frequencies available to perform fitting than there are present in the underlying signal to fit~\cite{peters_generalization_2022}.

\section{Regression of modulus function plots}
\label{appendix:experiment2}


\begin{figure*}
     \centering
     \begin{subfigure}[b]{0.49\textwidth}
         \centering
         \includegraphics[trim={0.cm 0cm 1.2cm 1cm},clip,width=\textwidth]{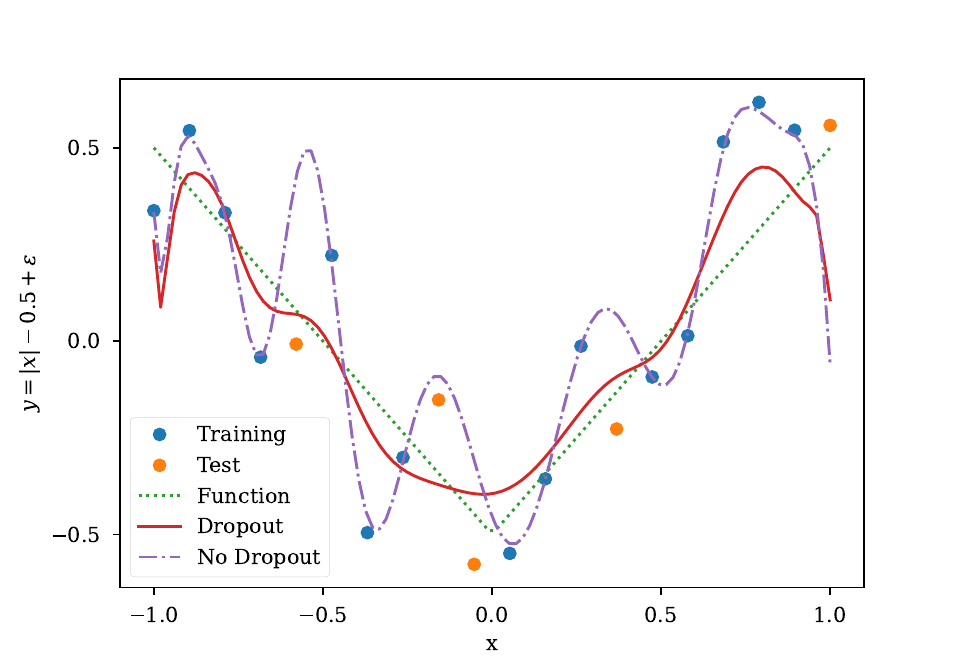}
         \caption{}
     \end{subfigure}
     \begin{subfigure}[b]{0.49\textwidth}
         \centering
         \includegraphics[trim={0.cm 0cm 1.2cm 1cm},clip,width=\textwidth]{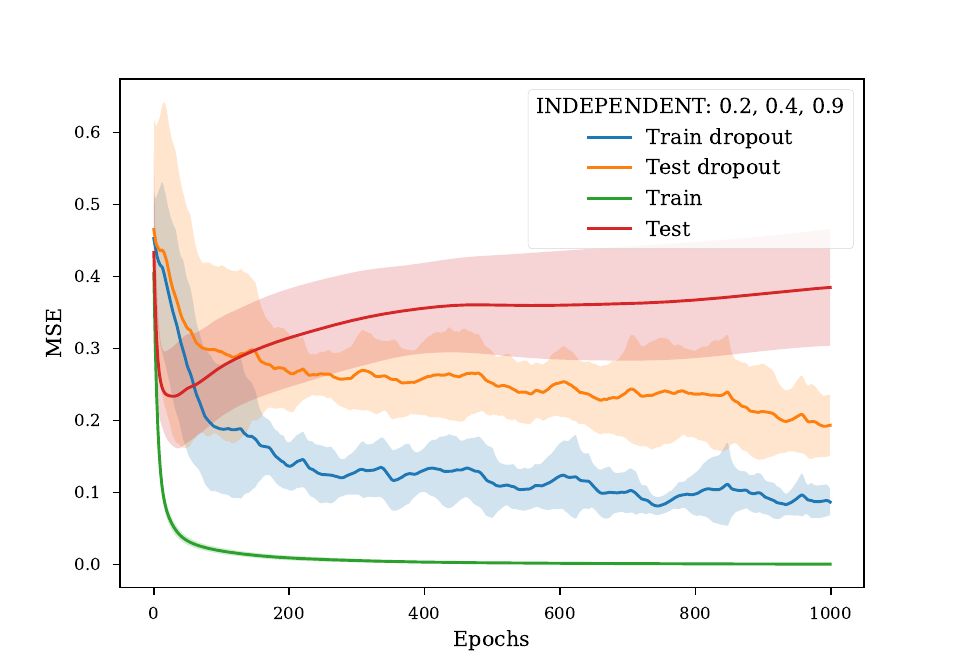}
         \caption{}
     \end{subfigure}
      \caption{\textbf{Regression} The plots illustrate the performances of an overparametrized model employed in a regression task of the \texttt{module} function. \textbf{(a)} The model overfits the noisy data by predicting exactly each of them. \textbf{(b)} The trend of the loss function when training without dropout shows an increasing prediction error typical of an overfitting model, while with dropout the prediction error does not increase. The standard deviation over 10 different runs is shown as a shadow. The sudden deviation of the prediction close to the $x$ limits is due to the Gibbs phenomenon~\cite{introductionFourier} of Fourier series, usually happening in the presence of a jump discontinuity (in this case the end of the $x$ values that we are fitting). }
      \label{fig:module-regr}
\end{figure*}

\begin{figure}[!ht]
    \centering
    \includegraphics[trim={0.cm 0.5cm 0cm 0cm},clip,width=0.5\textwidth]{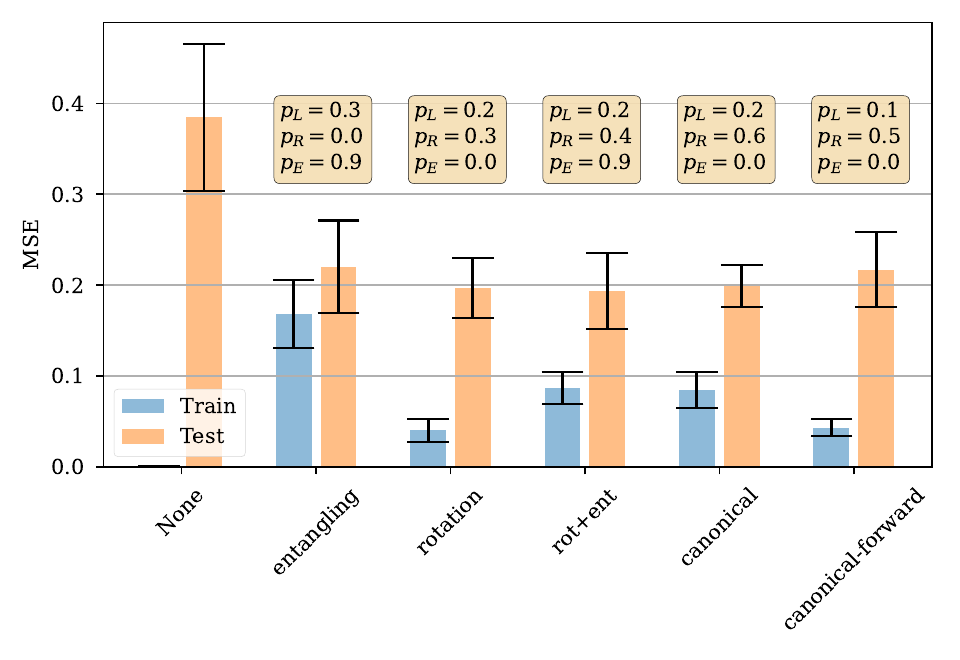}
    \caption{\textbf{Regression} Bar chart comparing the final average performances of all the dropout strategies on the \texttt{module} dataset with their respective optimal hyperparameters. The standard deviation is taken over 10 different runs.}
    \label{fig:histogram module function}
\end{figure}

Here, we focus on the \texttt{module} regression problem and analyze the performances of the models in learning the following function:
\begin{equation}
    y = \lvert x \rvert -\frac{1}{2} + \epsilon\,,
\end{equation}
where $x \in \mathbb{R}$ and $\epsilon$ is still an additive white Gaussian noise with amplitude equal to 0.3, zero mean and a standard deviation of 0.5. 
In Fig.~\ref{fig:module-regr}a we can observe how the trained model approximates the \texttt{module} function with and without quantum dropout. Using quantum dropout helps to achieve better generalization. In this case, we have a sudden deviation of the prediction close to the $x$ limits due to the Gibbs phenomenon~\cite{introductionFourier} of Fourier series, usually happening in the presence of a jump discontinuity (in this case the end of the $x$ values that we are fitting). This is the main cause of the pronounced standard deviation in the generalization performances, as one can see in Fig.~\ref{fig:module-regr}b.
Fig.~\ref{fig:histogram module function} presents a comprehensive analysis of the outcomes for each model with respect to their optimal configurations of $p_L$, $p_R$, and $p_E$. The model without dropout still overfits with almost 0 training MSE and high test error. Conversely, all quantum dropout techniques demonstrate low and comparable errors, both on the training and test sets, indicating their effectiveness in preventing overfitting compared to the model without dropout. The Independent dropout model still gets the best performance in terms of test error, which is 10.2\% lower than the entangling dropout. The standard deviation of all quantum dropout models is lower than the standard deviation of the overfitted model after 10 runs, indicating that the quantum dropout models have better generalization performance.
As in the \texttt{sin} function experiment, also here the gate drop ratios are very low, typically between 6\% and 16\%.

\begin{figure*}
     \centering
     \begin{subfigure}[b]{0.49\textwidth}
         \centering
         \includegraphics[trim={0.cm 0cm 1.2cm 1cm},clip,width=\textwidth]{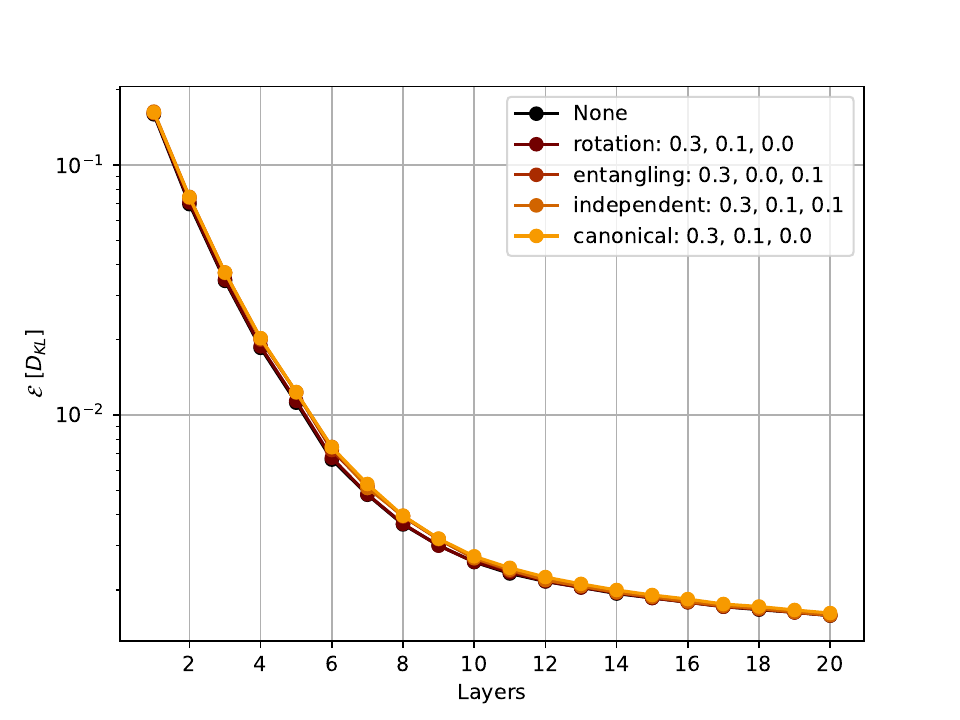}
         \caption{}
     \end{subfigure}
     \begin{subfigure}[b]{0.49\textwidth}
         \centering
         \includegraphics[trim={0.cm 0cm 1.2cm 1cm},clip,width=\textwidth]{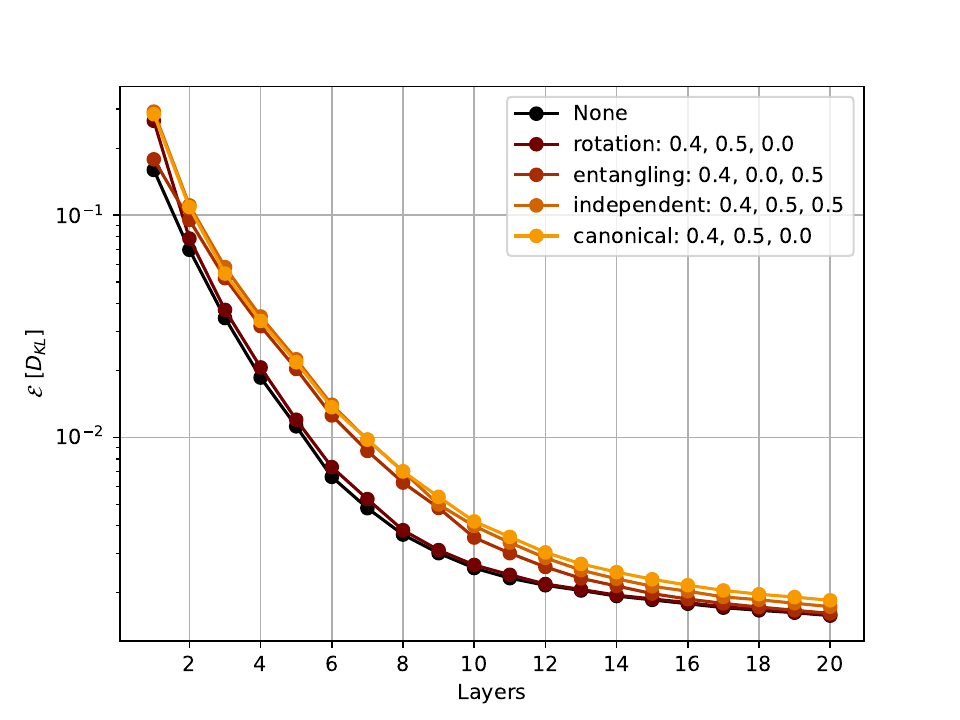}
         \caption{}
     \end{subfigure}
     \begin{subfigure}[b]{0.49\textwidth}
         \centering
         \includegraphics[trim={0.cm 0cm 1.2cm 1cm},clip,width=\textwidth]{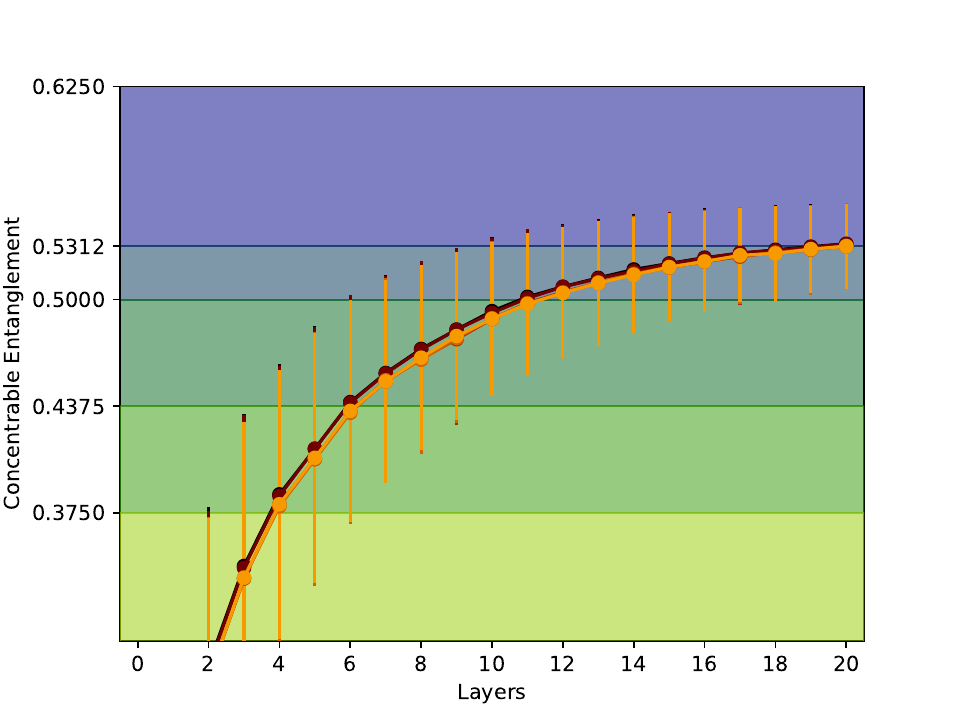}
         \caption{}
     \end{subfigure}
     \begin{subfigure}[b]{0.49\textwidth}
         \centering
         \includegraphics[trim={0.cm 0cm 1.2cm 1cm},clip,width=\textwidth]{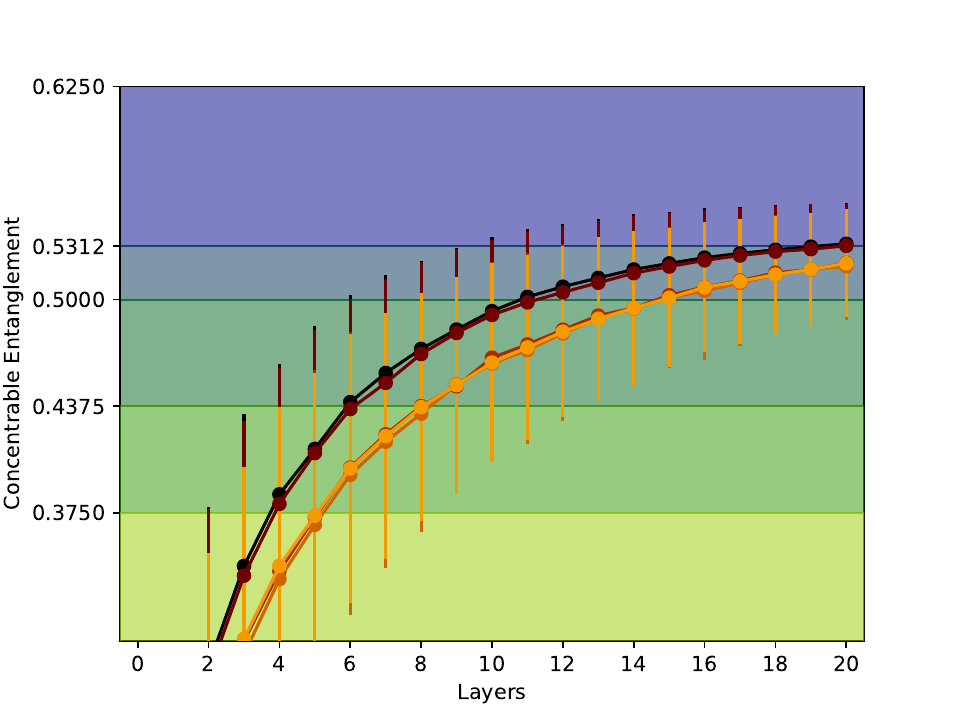}
         \caption{}
     \end{subfigure}
     \begin{subfigure}[b]{0.49\textwidth}
         \centering
         \includegraphics[trim={0.cm 0cm 1.2cm 1cm},clip,width=\textwidth]{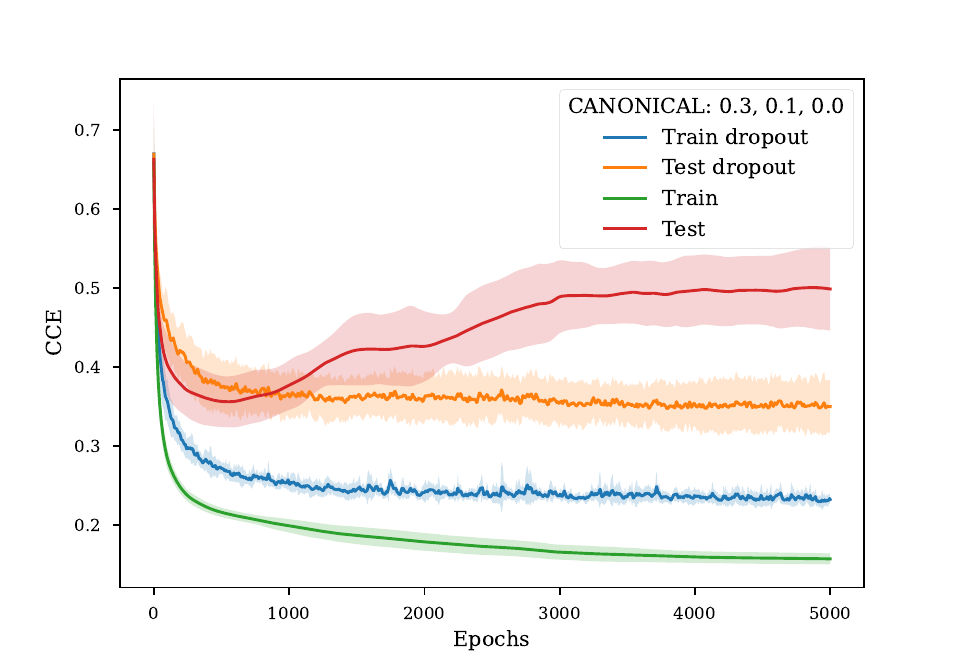}
         \caption{}
     \end{subfigure}
     \begin{subfigure}[b]{0.49\textwidth}
         \centering
         \includegraphics[trim={0.cm 0cm 1.2cm 1cm},clip,width=\textwidth]{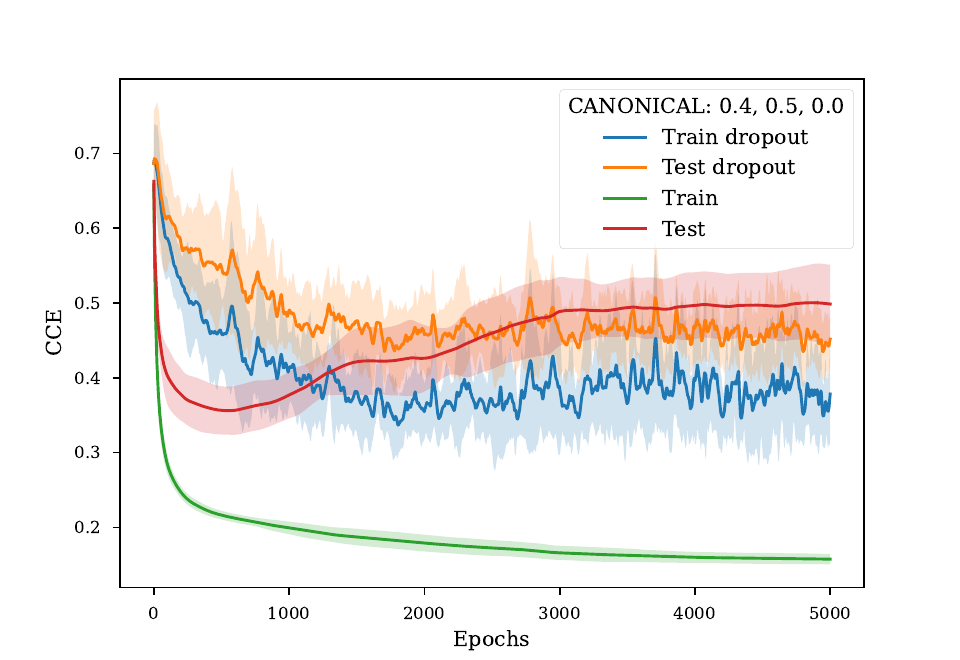}
         \caption{}
     \end{subfigure}
      \caption{Expressibility (in log scale) of QNN w.r.t. the employed layers for drop probability of \textbf{(a)} $3\%$ (good operating regime) and \textbf{(b)} $20\%$. The expressibility is the approximately same for all dropout strategies in the first case while in the second they tend to converge only for more than 20 layers. A similar behaviour can be seen in the degree of concentrable entanglement (CE) produced on average by QNN with the same drop probabilities \textbf{(c)} $3\%$ (good operating regime) and \textbf{(d)} $20\%$. As an example, the performances for the canonical-forward strategy are reported for the two cases in panels \textbf{(e)}  and \textbf{(f)}, respectively. The better-performing strategy is the one not affecting expressibility and entanglement, once again highlighting that this is not why dropout works well.}
      \label{fig:expressib+ent2}
\end{figure*}

\section{Expressibility}
\label{appendix:expressib}

For the definition of expressibility, we refer to Sec.~IIA of~\cite{Sim2019}. Expressibility measures the circuit’s ability to produce states that are well representative of the Hilbert space. To quantify this feature of the PQC, one can compare the distribution of states obtained from the circuit with randomly sampled parameters to the uniform distribution of states, i.e. the ensemble of Haar-random states. This is in practice calculated as the Kullback-Leibler (KL) divergence between the estimated fidelity distribution produced by a PQC and that of the Haar-distributed ensemble:
\begin{equation}
    \mathcal{E }=D_{KL}(\hat{P}_{PQC}(F ; \boldsymbol{\theta}) || P_{Haar}(F))\,,
\end{equation}
where $\hat{P}_{PQC}(F ; \boldsymbol{\theta})$ is the estimated probability distribution of fidelities resulting from sampling states from a PQC. The more the distribution of fidelities overlaps with the Haar one the more the expressibility of the PQC is high. Maximal expressibility corresponds to KL divergence value equal to 0, but in practice, the finite number of samples provides a numerical barrier below which one cannot resolve the
expressibility value for a QNN.

In general, adding more layers to a PQC initially allows better expressibility since more circuit parameters increase the dimension of the manifold of states explored. Nevertheless, sooner or later the expressibility will saturate at a certain value (specific for the chosen ansatz). 

To estimate the expressibility of QNNs we need to investigate the distribution of the produced states. We prepare the states by selecting the first 15 elements of the training dataset, sampling 1000 different parameter vectors and then applying to each embedded data all the parametrized circuits for a total of 15000 different states. When calculating the expressibility for the first QNN with a chosen dropout strategy we employ different dropout layouts per each parameter vector with $p_L=0.7$ and $p_G=0.7$ to have about 50\% of dropout. As displayed in Fig.~\ref{fig:expressib+ent}a the QNN model in Fig.~\ref{fig:QNN circuit} has high expressibility after few layers.

For what concerns the second QNN, it seems to be more sensitive to high dropout rates. This may be related to the particular QNN architecture and to the non-negligible level of redundancy already present before entering the overparametrized regime (see Appendix~\ref{appendix: overparam}). For this reason in Fig.~\ref{fig:expressib+ent2}a~and~b we show the comparison in expressibility between the different dropout strategies in the operating regime ($p_Lp_G<20\%$). One can appreciate that in this case expressibility is slightly reduced by dropout only when removing about $20\%$ of the gates (following a given strategy) but the classification performances are bad (see Fig.~\ref{fig:expressib+ent2}f), once again highlighting that this is not why dropout works well. One possible explanation here is that dropout reduces the overparametrization (while preserving it) making the training harder.

\section{Multipartite entanglement}
\label{appendix: multip ent}

Let's start by giving the definition of entanglement for pure states in systems with only two subsystems (parties):
\begin{definition}[Bipartite Entanglement]
A pure state $\ket{\psi}$ belonging to the Hilbert space $\mathcal{H}=\mathcal{H}_A\otimes\mathcal{H}_B$ is called a product state or \emph{separable} if we can find states $\ket{\psi_A}\in \mathcal{H}_A$ and $\ket{\psi_B}\in \mathcal{H}_B$ such that $\ket{\psi} = \ket{\psi_A} \otimes \ket{\psi_B}$ holds. Otherwise $\ket{\psi}$ is said to be \emph{entangled}.
\end{definition}

Once we have defined the simplest kind of entanglement its generalization to multiple parties is quite straightforward.

\begin{definition}[Multipartite entanglement]
\label{def:multip ent}
Given a pure $N$-partite state $\ket{\psi}$, it is called \emph{fully separable} if it can be written as a product state of all parties. 

We call a pure state \emph{$m$-separable}, with $1 < m < N$, if there exists a splitting of the $N$ parties into $m$ parts $P_1,\dots , P_m$ such that $\ket{\psi} = \bigotimes^m_{i=1}\ket{\psi_i}_{P_i}$ holds.

We call a state genuinely $N$-partite entangled when it is neither fully separable nor $m$-separable, for any $m > 1$.
\end{definition}

Note that if a pure state is not fully separable, it contains some entanglement, then $m$-separable state still may contain some entanglement. For additional general concepts and instruments on bipartite and multipartite entanglement, one can refer to~\cite{Guhne_2009ent}. 

Following Def.~\ref{def:multip ent}, broadly speaking, one can say that a state has Genuine Multipartite Entanglement (GME) if all the qubits (parties) are entangled with each other.

\subsection{Concentrable entanglement}

In this work we quntify the entanglement by means of a recently introduced entanglement measure which is concentrable entanglement~\cite{beckey2021concentrable}. It is directly computable on quantum hardware and allows to analyse different kinds of multipartite entanglement and their possible parties structure~\cite{Schatzki2022hierarchy}.
\begin{definition}[Concentrable Entanglement]
Given a pure $N$-partite state $\ket{\psi}$, the concentrable entanglement of $\ket{\psi}$ is defined as
\begin{equation}
    \mathcal{C}(\ket{\psi})=1-\frac{1}{2^N}\sum_{\alpha\in Q}\emph{\Tr}[\rho_\alpha^2]\,,
\end{equation}
where $\alpha$ is a set of qubit labels, $\rho_\alpha$ is the reduced state of $\ket{\psi}$ on the set of qubits in $\alpha$, and $Q$ is the power set of ${1, 2, \dots , N}$. 
\end{definition}

\begin{table}[]
\begin{tabular}{|c|c|c|}
\hline
{\textbf{ Parties structure }}                   & {\textbf{ min CE }} & \textbf{ max CE } \\ \hline
$5$                                                      & 0.53125                              & 0.625           \\
$3\otimes 2$                                             & 0.5                                  & 0.53125         \\
$4\otimes 1$                                             & 0.4375                               & 0.5             \\
$2\otimes2\otimes1$                                      & 0.375                                & 0.4375          \\
{$3\otimes1\otimes1$}                                    & 0.25                                 & 0.375           \\
{$2\otimes1\otimes1\otimes1$}                            & 0                                    & 0.25            \\
{$1\otimes1\otimes1\otimes1\otimes1$}                    & 0                                    & 0               \\ \hline
\end{tabular}
\caption{Concentrable entanglement intervals for all the different possible parties structures. Intervals are semi-opened $(\min CE, \max CE]$.}
\label{tab:CE}
\end{table}

As mentioned above, this measure allows the detection of different categories of multipartite entanglement and GEM. In particular, in~\cite{Schatzki2022hierarchy} authors calculate the maximal values for different multipartite entanglement combinations useful in our context. For $N=5$ qubits the concentrable entanglement intervals for all the different possible parties structures are reported in Tab.~\ref{tab:CE}. This hierarchy is depicted also in Figs.~\ref{fig:expressib+ent}b and~\ref{fig:expressib+ent2}c~and~d, in color scale, where we report the average concentrable entanglement produced by the QNNs as a function of the number of layers. Here the 15000 states are prepared with the same procedure as in Appendix.~\ref{appendix:expressib}. The first QNN model, on average, produces genuine multipartite entanglement also when quantum dropout is applied. The produced entanglement, after a certain number of layers, reaches a plateau corresponding to states with entanglement distribution analogous to the one expected for random states (Haar states)~\cite{Schatzki2022hierarchy} also for all quantum dropout strategies:
\begin{align*}
    &\langle{\mathcal{C}}\rangle_\mathrm{Haar} = 1-2\frac{3^N}{4^N+2^N} \ \myeq \ 0.53977 \,, \\
    &\mathrm{Var}(\mathcal{C})_\mathrm{Haar}=\mathcal{O}\left(\left(\frac{3}{16}\right)^N\right) \ \myeq \  \mathcal{O}\left(0.00023\right) \,, \\
    &\langle{\mathcal{C}}\rangle_\mathrm{QNN} \approx 0.539\,, \\
    &\mathrm{Var}(\mathcal{C})_\mathrm{QNN} \approx 0.0004\,, 
\end{align*}
This happens because the QNN has high expressibility and thus is producing Haar states. These reported quantities were calculated for the QNN proposed in~\cite{kobayashi_overfitting_2022} with 10 layers with maximal expressibility. 

Since the classification QNN is more sensitive to variations in Fig.~\ref{fig:expressib+ent2}b~and~d we compare the produced entanglement with different dropout strategies in the operating regime ($p_Lp_G<20\%$). As for the expressibility, one can appreciate that, in this case, for dropout of $20\%$ entanglement is reduced and the classification performances are bad (see Fig.~\ref{fig:expressib+ent2}f), showing that this is not the reason why dropout works well in practice. The entanglement produced by the QNN with 20 layers that we employ as QML model for classification reaches GME but on a level slightly below the Haar mean value.\\

\section{Dropout strategies}
\label{appendix:drop strategies}
\begin{figure*}[h!]
\begin{subfigure}[b]{0.48\textwidth}
    \centering
    \includegraphics[trim={3.cm 2cm 6cm 8cm},clip,width=\textwidth]{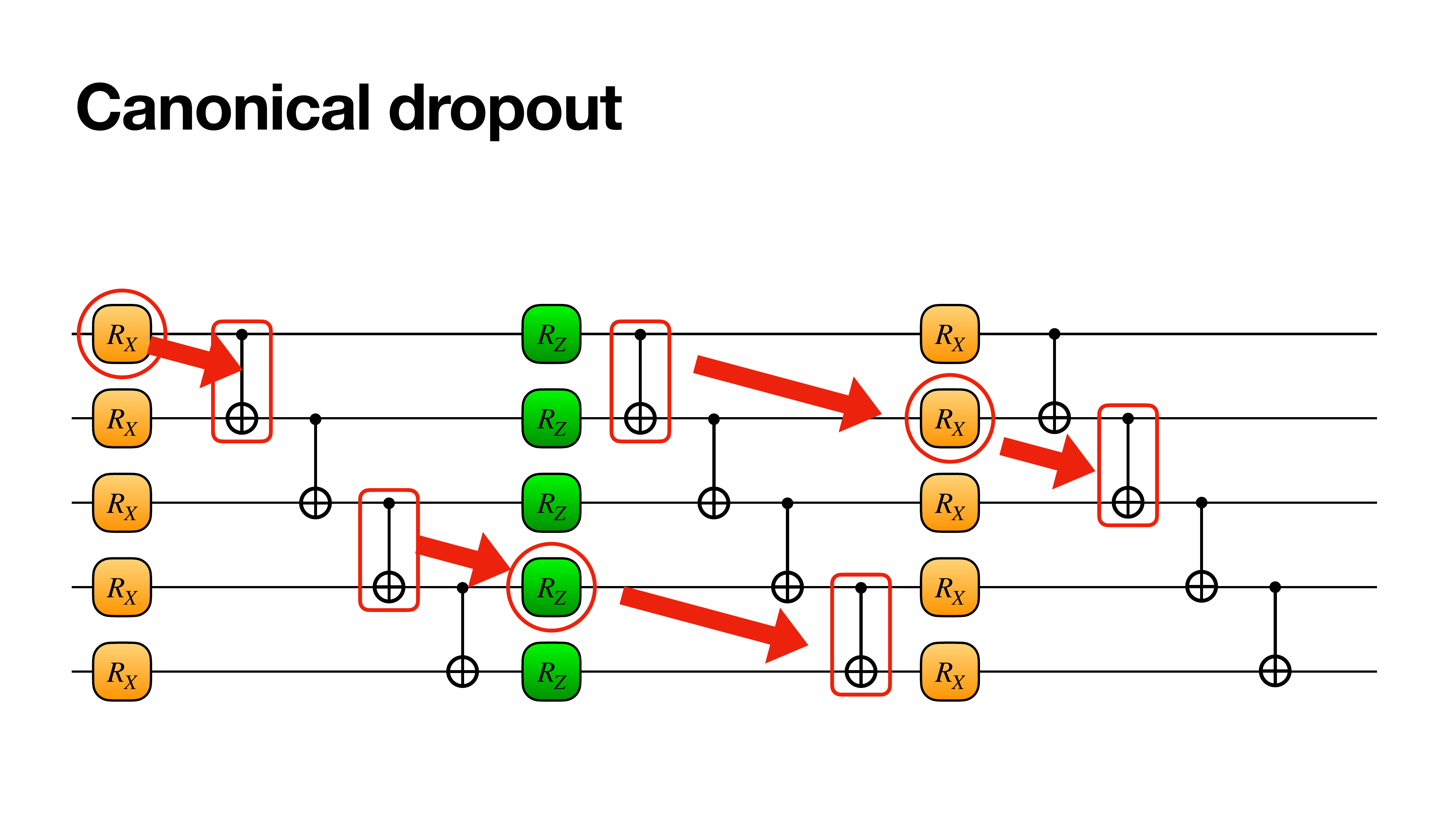}
    \caption{Canonical dropout.  }
    \label{fig:canonical}
\end{subfigure}
\hfill
\begin{subfigure}[b]{0.48\textwidth}
    \centering
    \includegraphics[trim={3.cm 2cm 6cm 8cm},clip,width=\textwidth]{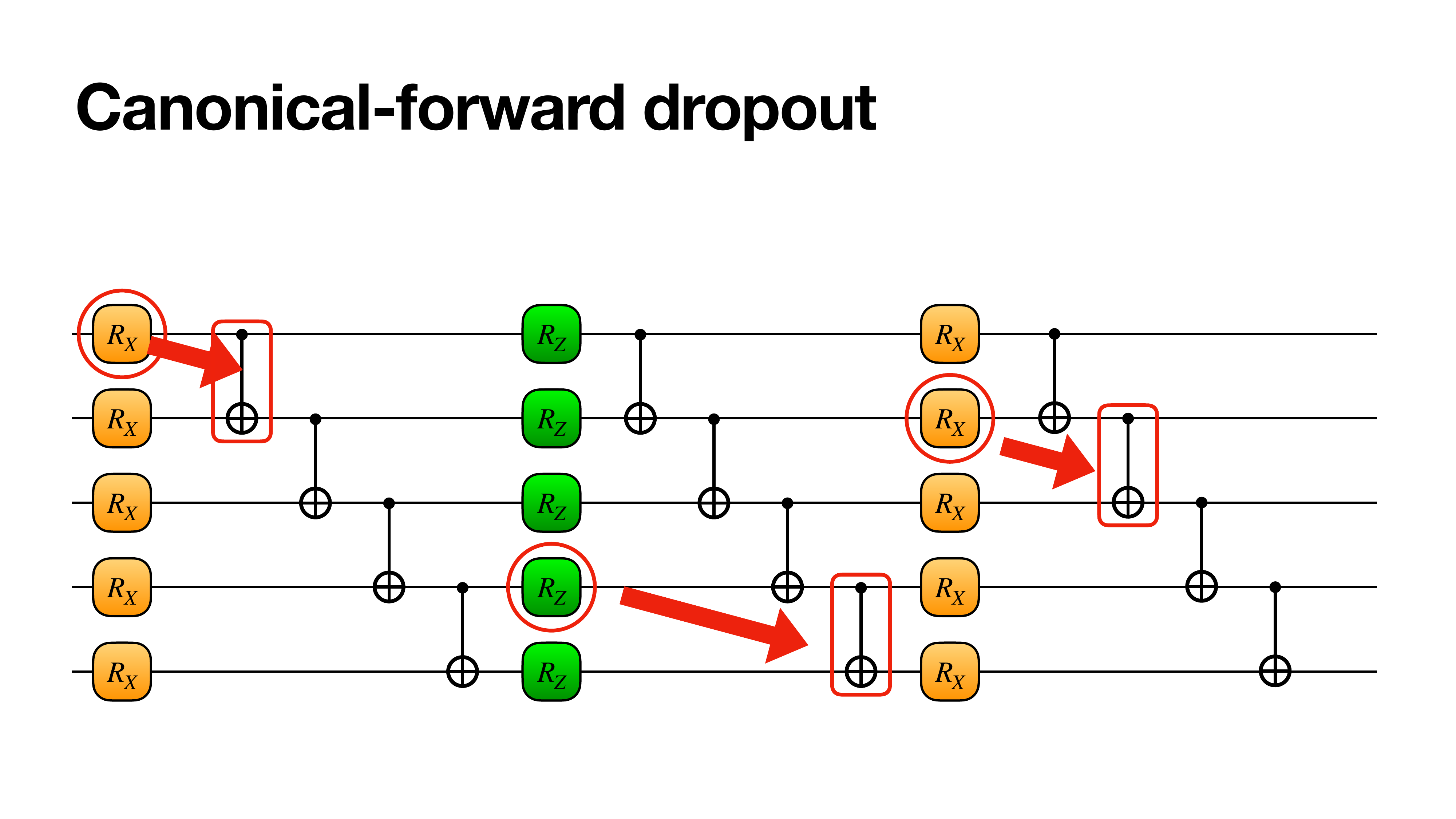}
    \caption{Canonical-forward dropout.  }
    \label{fig:canonical-forward}
\end{subfigure}
\hfill
\begin{subfigure}[b]{0.48\textwidth}
    \centering
    \includegraphics[trim={3.cm 2cm 6cm 8cm},clip,width=\textwidth]{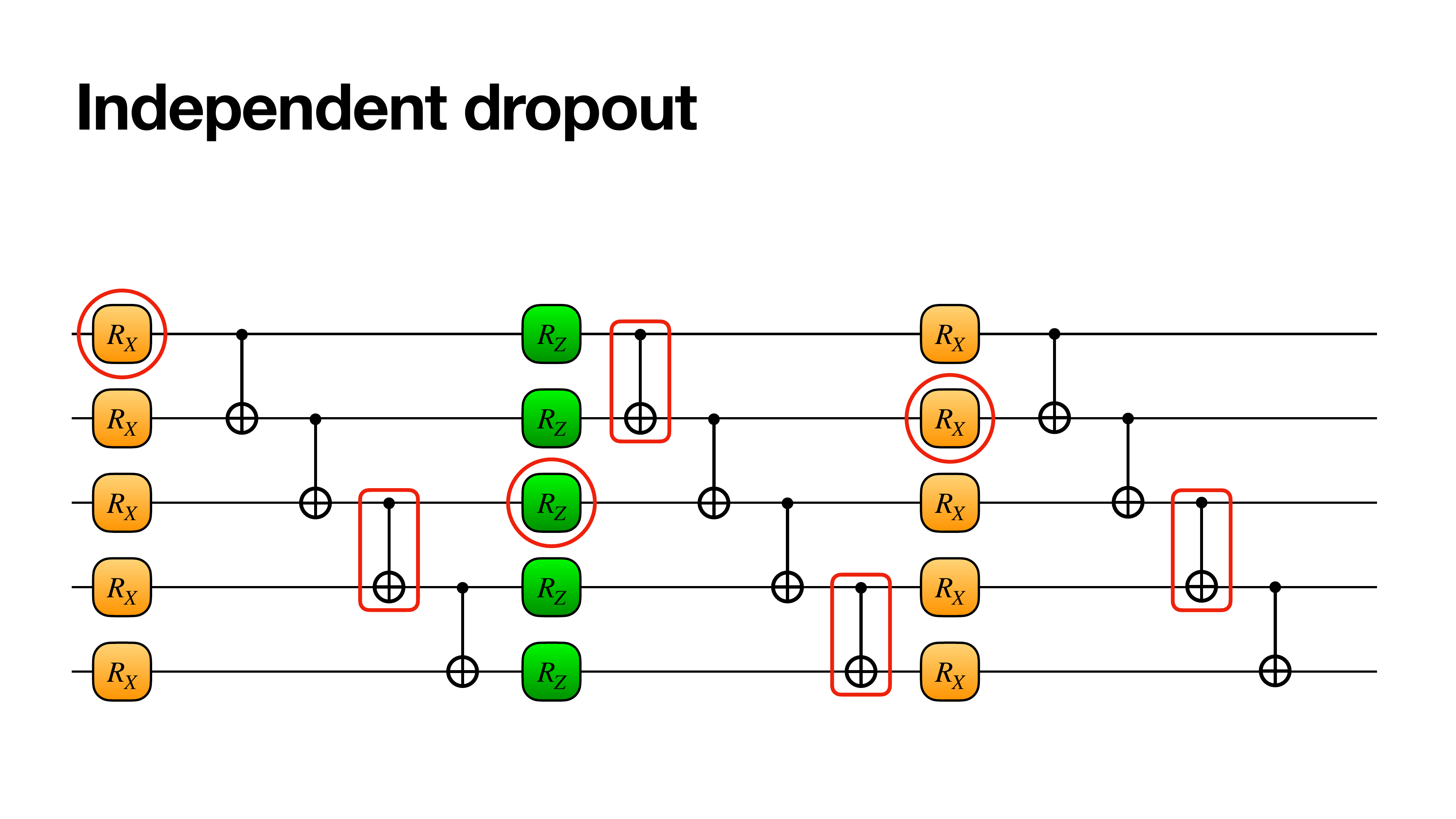}
    \caption{Independent dropout.  }
    \label{fig:independent}
\end{subfigure}
\hfill
\begin{subfigure}[b]{0.48\textwidth}
    \centering
    \includegraphics[trim={3.cm 2cm 6cm 8cm},clip,width=\textwidth]{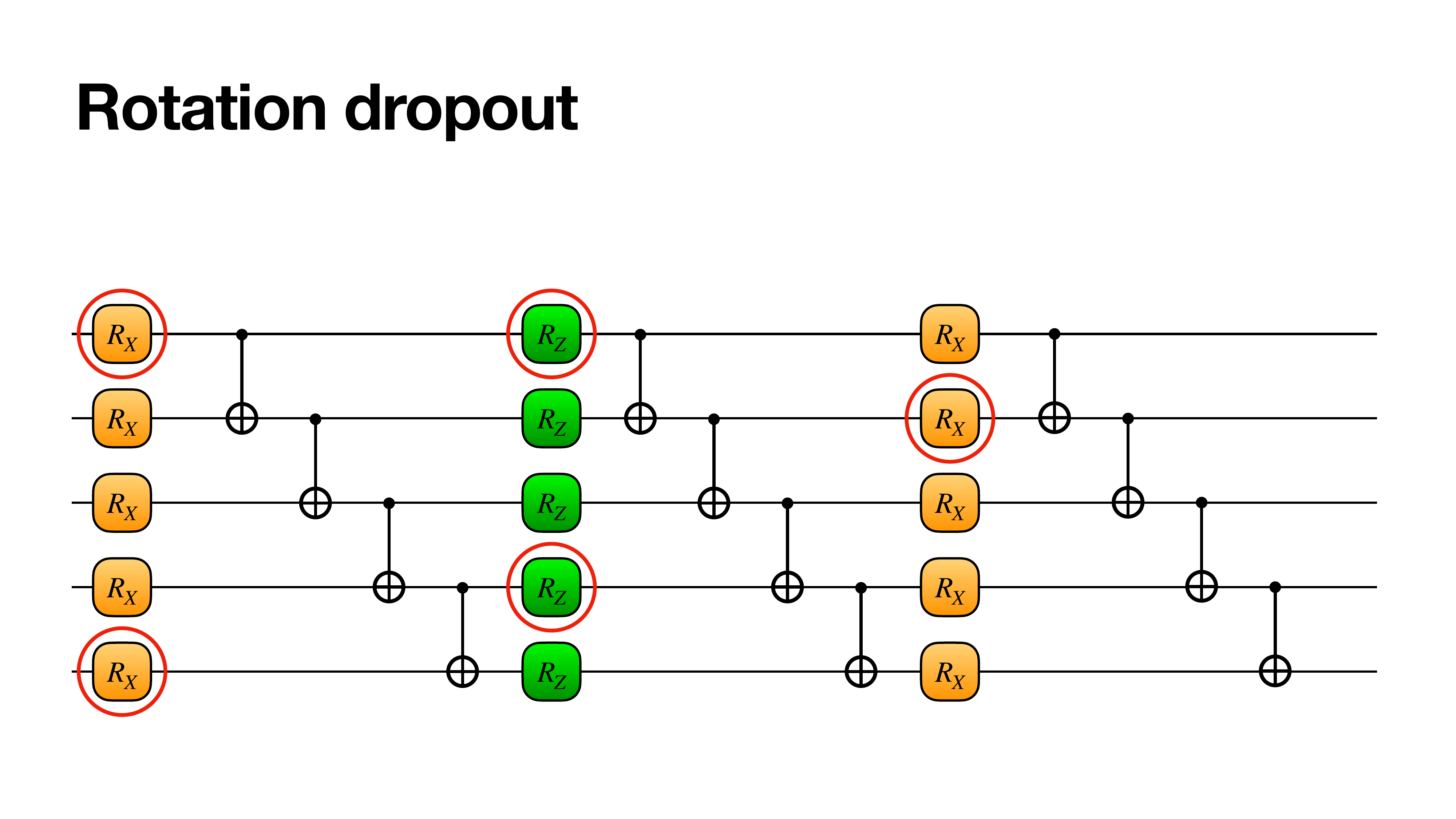}
    \caption{Rotation dropout.  }
    \label{fig:rotation}
\end{subfigure}
\hfill
\begin{subfigure}[b]{0.48\textwidth}
    \centering
    \includegraphics[trim={3.cm 2cm 6cm 8cm},clip,width=\textwidth]{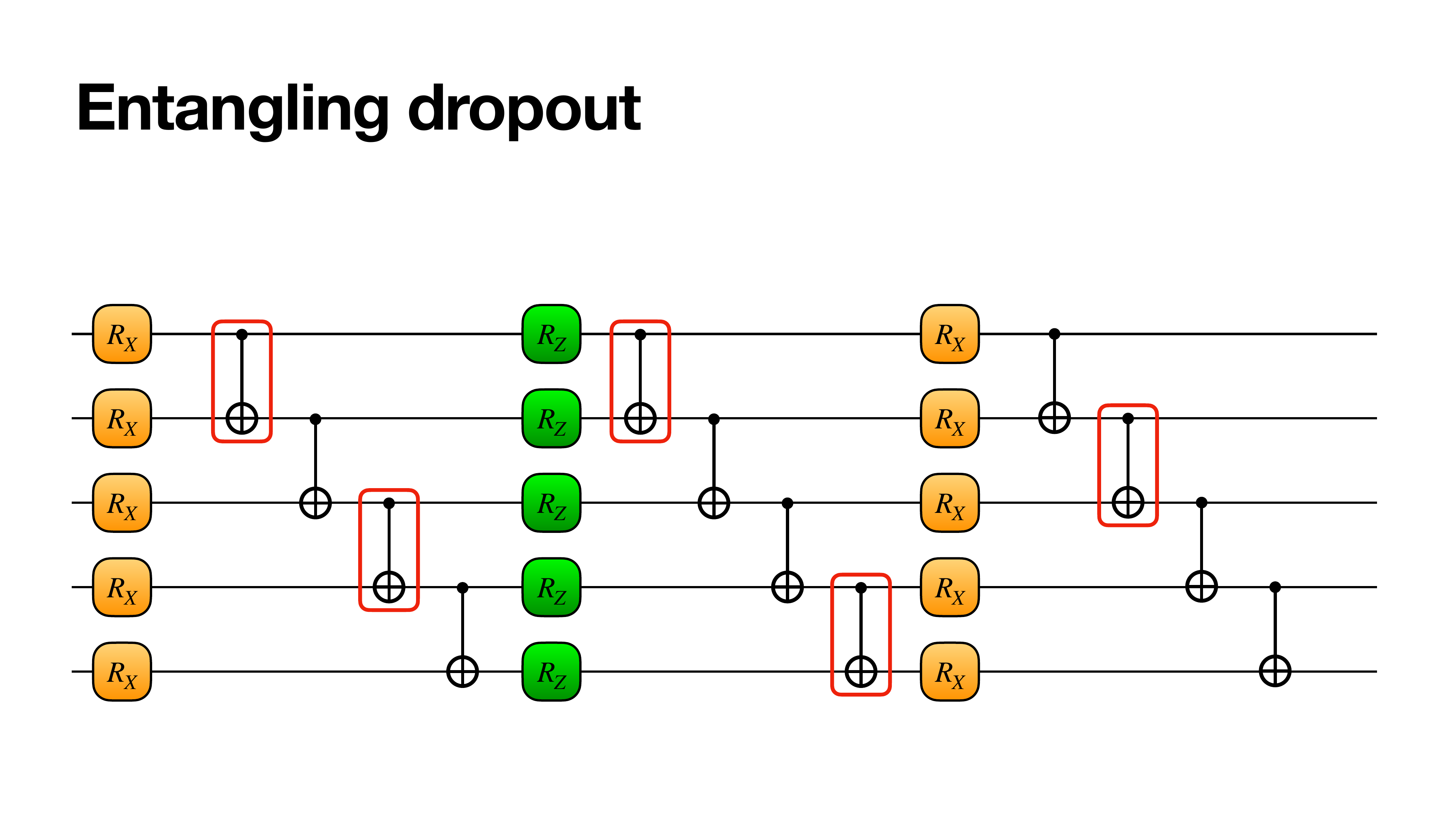}
    \caption{Entangling dropout.  }
    \label{fig:entangling}
\end{subfigure}
\caption{Schematic graphical representations of the different possible quantum dropout strategies}
\label{fig:drop_strategies}
\end{figure*}

Here we report schematic graphical representations of the different possible quantum dropout strategies presented in Sec.~\ref{subsec:strategies} (see Fig.~\ref{fig:drop_strategies}). Single dropped gates are highlighted by a circle/rectangle and the groups are linked by arrows. Fig.~\ref{fig:canonical} illustrates canonical dropout, which involves dropping a single rotation gate along with all the previous entangling gates that targeted that particular qubit and all the subsequent entangling gates that used that qubit as a control. Canonical-forward is shown in Fig.~\ref{fig:canonical-forward}, and involves dropping a single rotation gate along with all the subsequent entangling gates that used that qubit as a control. Independent dropout is represented in Fig.~\ref{fig:independent}, and works by dropping a single rotation gate and a single entangling gate independently one another. Fig.~\ref{fig:rotation} illustrates a simple rotation dropout, where single rotation gates are dropped. Similarly, Fig.~\ref{fig:entangling} shows entangling dropout, which drops single entangling gates.

\end{document}